\documentclass[aps,prl,superscriptaddress]{revtex4}

\usepackage{amssymb}
\usepackage{amsmath}
\usepackage{epsf}
\usepackage{graphicx}

\begin{document}

\bibliographystyle{apsrev}

\title{Hidden Lagrangian coherence and memory effects \\
 in the statistics of
Hamiltonian motions}
\author{Madalina Vlad, Dragos Iustin Palade, Florin Spineanu}
\email{madalina.vlad@inflpr.ro}
\affiliation{National Institute of Laser, Plasma and Radiation Physics, Magurele, Romania}
\maketitle

\section{\protect\bigskip 1. Introduction}

Turbulence is a complex nonlinear process, which appears in many domains as
fluid mechanics, plasma physics, astrophysics, atmosphere and ocean
sciences, chemistry \cite{Falkovich}-\cite{Itoh}. One of the main difficulty
in understanding the dynamics of turbulence is the complicated combination
of stochastic and quasi-coherent aspects that is typical for the strongly
nonlinear regimes.

Quasi-coherence or order appears at the basic level of tracer trajectories
in smooth stochastic velocity fields with finite correlation lengths $%
\lambda $ and times $\tau _{c}.$ Trajectory coherence (or Lagrangian
coherence) is usually a transitory process that lasts during the time of
flight over $\lambda $ with the amplitude $V$ of the stochastic velocity, $%
\tau _{fl}=\lambda /V.$ It exists only for slow time variation of the
velocity with $\tau _{c}>\tau _{fl}$. In the case of two-dimensional
incompressible velocity fields, a much stronger Lagrangian coherence appears
that is due to trajectory eddying or trapping. It generates vortical
structures of trajectories, which produce non-Gaussian statistics of tracer
displacements and strongly modified transport \cite{kraichnan}-\cite{VS17}.
The order of the tracer trajectories determines much more complicated
effects on turbulence evolution, which essentially amplifies the degree of
quasi-coherence. The turbulence that is dominantly two-dimensional has a
self-organizing character \cite{Montgomery}-\cite{Weisse}, which consists of
the generation of quasi-coherent large scale structures \cite{SV}.

This paper is focused on the coherent effects that appear in tracer
statistics in two-dimensional incompressible turbulence in the presence of
an average velocity $\mathbf{V}_{d}$. We show that $\mathbf{V}_{d}$
determines strong modifications of the transport and trajectory statistics,
which are essentially caused by hidden coherent components of the motion.

The results are based on the numerical simulation of the trajectories and
consist of a conditional statistical analysis adapted to the special
properties of the two-dimensional incompressible velocity fields.

The formulation of the problem and the simulation method are presented in
Section 2. The motion is of Hamiltonian type with the Hamiltonian function $%
\phi _{t}$ composed of the stochastic and average potentials. The
trajectories evolve on the contour lines of $\phi _{t}(\mathbf{x})$ in
static (frozen) potentials ($\tau _{c}\rightarrow \infty ),$ and they remain
strongly correlated to these lines for slow time variation (large $\tau
_{c}) $. We consider first (Sections 3-5) frozen potentials. We discuss the
configuration of the contour lines of the potential and the main features of
tracer advection in Section 3. The space of trajectories is organized in two
categories: trapped and free. The statistics of the Lagrangian velocity and
of the trajectories are examined for each category, and their specific
contributions to the global statistics (on the whole set of trajectories)
are identified (Section 4). We show that quasi-coherent Lagrangian
velocities parallel to $\mathbf{V}_{d}$ are generated for both categories.
They are hidden in the global statistics, in the sense that their
contributions compensate each other. However, they provide explanations for
the nonstandard transport produced in these conditions. A deeper examination
of the coherence induced by the average velocity is presented in Section 5,
where we determine the Lagrangian statistics for the trajectories of each
category that evolve on contour lines of the potential with the same value
of $\phi _{t}.$ These results reveal other (hidden) coherent elements of the
motion, and provide important properties that are used for the understanding
of the effects of the average velocity on the statistics of trajectories and
transport. Sections 6 and 7 deal with time dependent potentials. The
Lagrangian statistics conditioned by the value of the potential shows that
the order found in frozen potentials does not decay due to the random time
variation of the potential, as expected. On the contrary, important
quasi-coherent elements significantly increase (Section 6). Explanations are
provided in Section 7. They are essentially related to the constraint of
invariance of the potential, which are approximately valid at large $\tau
_{c},$ and to the slow transition of the trajectories between the two
categories. Long memory effects are identified and their effects is
discussed. A short summary of the results and the conclusions are presented
in Section 8. \ \ 

\section{2. The problem and the simulation method}

Tracer trajectories in two-dimensional stochastic velocity fields are
obtained from%
\begin{equation}
\frac{d\mathbf{x}}{dt}=\mathbf{v}(\mathbf{x,}t)=\widetilde{\mathbf{v}}(%
\mathbf{x,}t)+V_{d}\mathbf{e}_{2},~\ \   \label{eqm}
\end{equation}%
where $\mathbf{e}_{1},$ $\mathbf{e}_{2}$ are the unit vectors in the plane
of the motion $\mathbf{x=(}x_{1},x_{2}),$ $\mathbf{e}_{3}$ is perpendicular
on this plane. The velocity $\mathbf{v}(\mathbf{x,}t)$ has a stochastic
component $\widetilde{\mathbf{v}}(\mathbf{x,}t)$ superposed on a constant
average velocity $\mathbf{V}_{d}=V_{d}\ \mathbf{e}_{2}.$ The
incompressibility condition $\mathbf{\nabla \cdot \widetilde{\mathbf{v}}}(%
\mathbf{x,}t)=0$\ of the velocity field is equivalent with the
representation of $\widetilde{\mathbf{v}}(\mathbf{x,}t)$ by a stochastic
potential (or stream function) $\phi (\mathbf{x,}t)$\ 
\begin{equation}
\widetilde{\mathbf{v}}(\mathbf{x,}t)=-\mathbf{\nabla }\phi (\mathbf{x,}%
t)\times \mathbf{e}_{3}.  \label{pot}
\end{equation}

The equation of motion is of Hamiltonian type, with $x_{1}$ and $x_{2}$\ the
conjugate variables and $\phi _{t}(\mathbf{x,}t)=\phi (\mathbf{x,}%
t)+x_{1}V_{d}$ the Hamiltonian function.

Dimensionless quantities are used in Eq. (\ref{eqm}) with the potential
normalized by its amplitude $\Phi $, the distances by $\lambda _{0}$ that is
of the order of the correlation lengths, the velocities (including $V_{d})$
by $V_{0}=\Phi /\lambda _{0}$ and the time by $\tau _{0}=\lambda _{0}/V_{0}.$
\ \ \ 

The potential is represented by a homogeneous and stationary Gaussian
stochastic field. Its Eulerian correlation (EC) $E(\mathbf{x,}t)\equiv
\left\langle \phi (\mathbf{x}_{0}\mathbf{,}t_{0})~\phi (\mathbf{x}_{0}+%
\mathbf{x,}t_{0}+t)\right\rangle $ in dimensionless quantities is modeled in
the simulations presented here by%
\begin{equation}
E(\mathbf{x,}t)\equiv \exp \left( -\frac{x_{1}^{2}}{2\lambda _{1}^{2}}-\frac{%
x_{2}^{2}}{2\lambda _{2}^{2}}-\frac{t^{2}}{2\tau _{c}^{2}}\right) ,
\label{EC}
\end{equation}%
where $\lambda _{i}$ are the correlation lengths of the 2-dimensional
potential and $\tau _{c}$ is the correlation time. The EC's of the velocity
components $E_{ii}(\mathbf{x,}t)\equiv \left\langle v_{i}(\mathbf{x}_{0}%
\mathbf{,}t_{0})~v_{i}(\mathbf{x}_{0}+\mathbf{x,}t_{0}+t)\right\rangle $ are 
\begin{equation}
E_{11}(\mathbf{x,}t)=-\partial _{2}\partial _{2}E(\mathbf{x,}t),\ E_{22}(%
\mathbf{x,}t)=-\partial _{1}\partial _{1}E(\mathbf{x,}t),  \label{ECv}
\end{equation}%
which determine the normalized amplitudes of the velocity fluctuations $%
V_{1}=$ $\sqrt{E_{11}(\mathbf{0,}0)}=1/\lambda _{2},$ $V_{2}=$ $1/\lambda
_{1}.$ \ 

The statistical properties of the trajectories obtained from Eq. (\ref{eqm})
are numerically analyzed. More precisely, we determine the statistics of the
trajectories and of the Lagrangian velocity, and a class of conditional
Lagrangian correlations that reveal the quasi-coherent components of the
motion and their properties.

We use statistical averages, which consists of generating a large number of
realizations ($r$) of the stochastic Gaussian potential and of determining
the trajectory with the initial condition $\mathbf{x}(0)=0$ in each $r$ by
effectively computing the velocity on the trajectory at each time step, $%
\mathbf{v}(\mathbf{x}(t_{i}),t_{i}).$\ However, the analysis of the results
is connected to the equivalent space averaging procedure. This corresponds
to the statistical ensemble of trajectories obtained in a single typical
realization of the potential by different initial conditions $\mathbf{x}(0)=%
\mathbf{x}_{0}^{r},$ where the points $\mathbf{x}_{0}^{r}$\ are uniformly
distributed in a very large domain.\ \ \ 

We use the simulation code presented in \cite{PV20}, which is based on a
fast generator of Gaussian fields with prescribed spectra. In the present
work, we have implemented the so called FRD representation%
\begin{equation}
\phi (\mathbf{X})=\sum\limits_{i=1}^{N_{c}}\sqrt{S(\mathbf{K}_{i})}\sin
\left( \mathbf{K}_{i}\mathbf{X}+\frac{\pi }{4}\zeta _{i}\right) ,
\label{FRD}
\end{equation}%
where $\mathbf{X}\equiv (\mathbf{x,}t)$ is the three-dimensional space-time
and $\mathbf{K}_{i}\equiv (\mathbf{k}_{\perp }^{i}\mathbf{,}\omega ^{i})$\
are the $N_{c}$ discrete values of the wave numbers $\mathbf{k}_{\perp }^{i}$
and frequencies $\omega ^{i}.$ $S(\mathbf{K})$ is the spectrum of the
stochastic potential, the Fourier transform of the EC (\ref{EC}). This
representation is different of the usual discrete Fourier decomposition by
the set of the values of $\mathbf{K}_{i}$\ that are not the fixed points of
a three-dimensional mesh, but random values with uniform distribution. Also,
the random phases $\zeta _{i}$ have not continuous distributions, but
discrete values $\pm 1$ (with equal probabilities).

Each set of the $N_{c}$ random values of $\mathbf{K}_{i}$\ and $\zeta _{i}$\
determines a realization $r$ of the potential and a trajectory (solution of
Eq. (\ref{eqm}) with initial condition $\mathbf{x}(0)=\mathbf{0}$). The
statistical ensemble $R$ consists of a number $M$ of these sets.

The representation (\ref{FRD}) provides a fast convergence of the Eulerian
properties of the stochastic potential. We have shown that reasonable errors
in the EC and in the probability of the potential are obtained at much
smaller values of $N_{c}$ and $M$ than in the usual fast Fourier
representation (FFR). This leads to the decrease of the computing times by
roughly one order of magnitude compared to the usual FFR method in
two-dimensional potentials \cite{PV20}. Most of the simulations analyzed
here are performed with $N_{c}=500$ and $M=10^{5}.$

\section{3. Main features of tracer advection}

The incompressibility of the two-dimensional velocity field ($\mathbf{\nabla
\cdot \mathbf{v}}(\mathbf{x,}t)=0)$ determines two invariance laws of the
solutions of Eq. (\ref{eqm}). It leads to equations of motion of Hamiltonian
type, with $x_{1}$ and $x_{2}$\ conjugate variables and $\phi _{t}$ the
Hamiltonian function. In the case of time independent (frozen) potentials $%
\phi \left( \mathbf{x}\right) $, the trajectories are linked to the contour
line of the potential $\phi _{t}(\mathbf{x})$, which means that the
Lagrangian potential $\phi _{t}(\mathbf{x}(t))$ is invariant along each
trajectory. The other invariant law is statistical and applies to the motion
in both frozen and time dependent potentials $\phi _{t}(\mathbf{x}(t),t)$
for any value of the correlation time $\tau _{c}$. It concerns the
distribution of the Lagrangian velocity $\mathbf{v}(\mathbf{x(}t),t),$ that
is shown to be time independent, and thus identical with the distribution of
the Eulerian velocity $\mathbf{v}(\mathbf{x},t)$. The Lagrangian potential
is statistically invariant too. This property is trivial in frozen
potentials where $\phi _{t}(\mathbf{x}(t))=\phi _{t}(\mathbf{x}(0))$ and it
is similar to the case of the velocity in time dependent potentials where $%
\phi _{t}(\mathbf{x}(t),t)$ changes in time.

\begin{figure}[tbh]
\centerline{
\includegraphics[height=8cm]{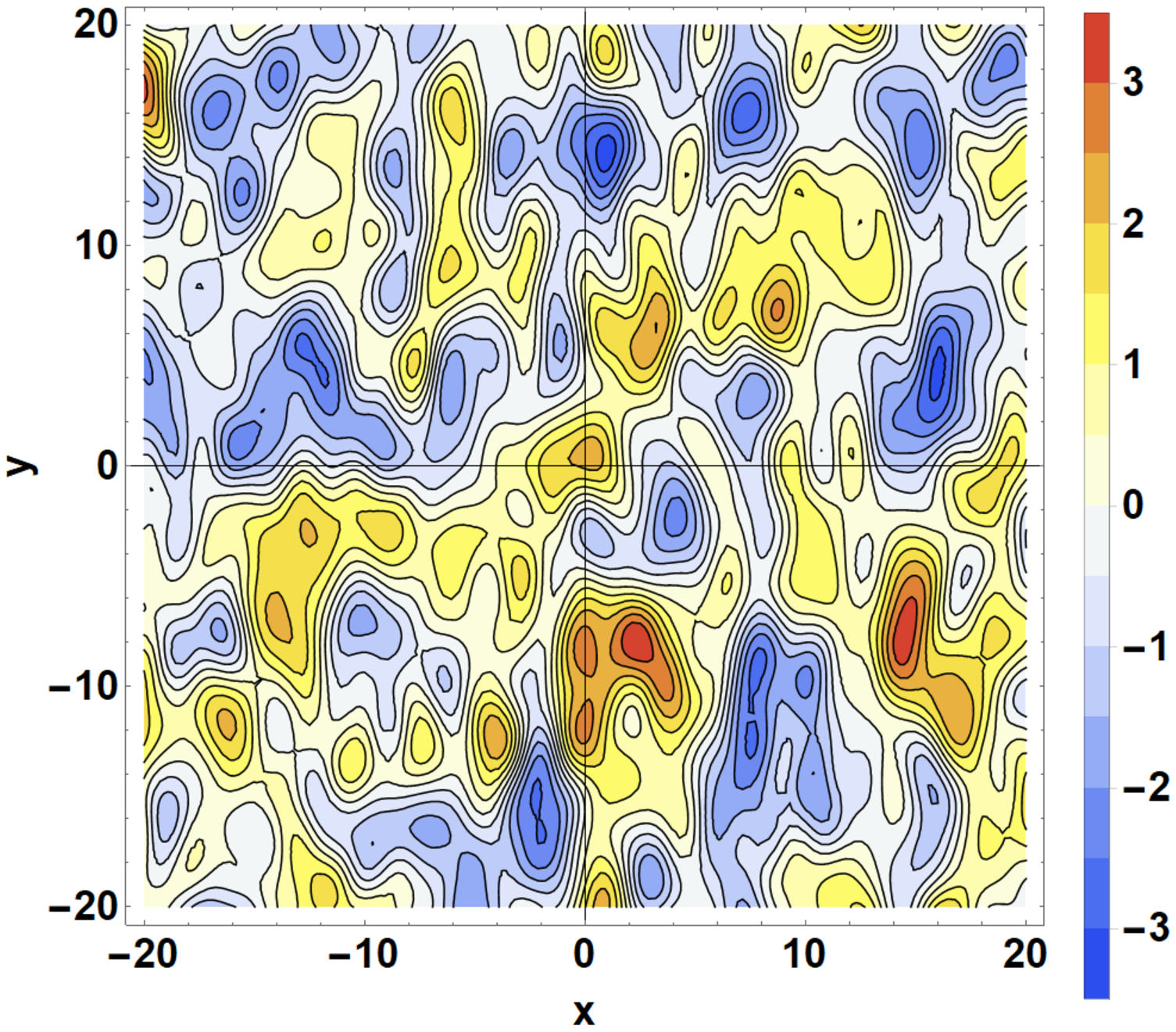}
\includegraphics[height=8cm]{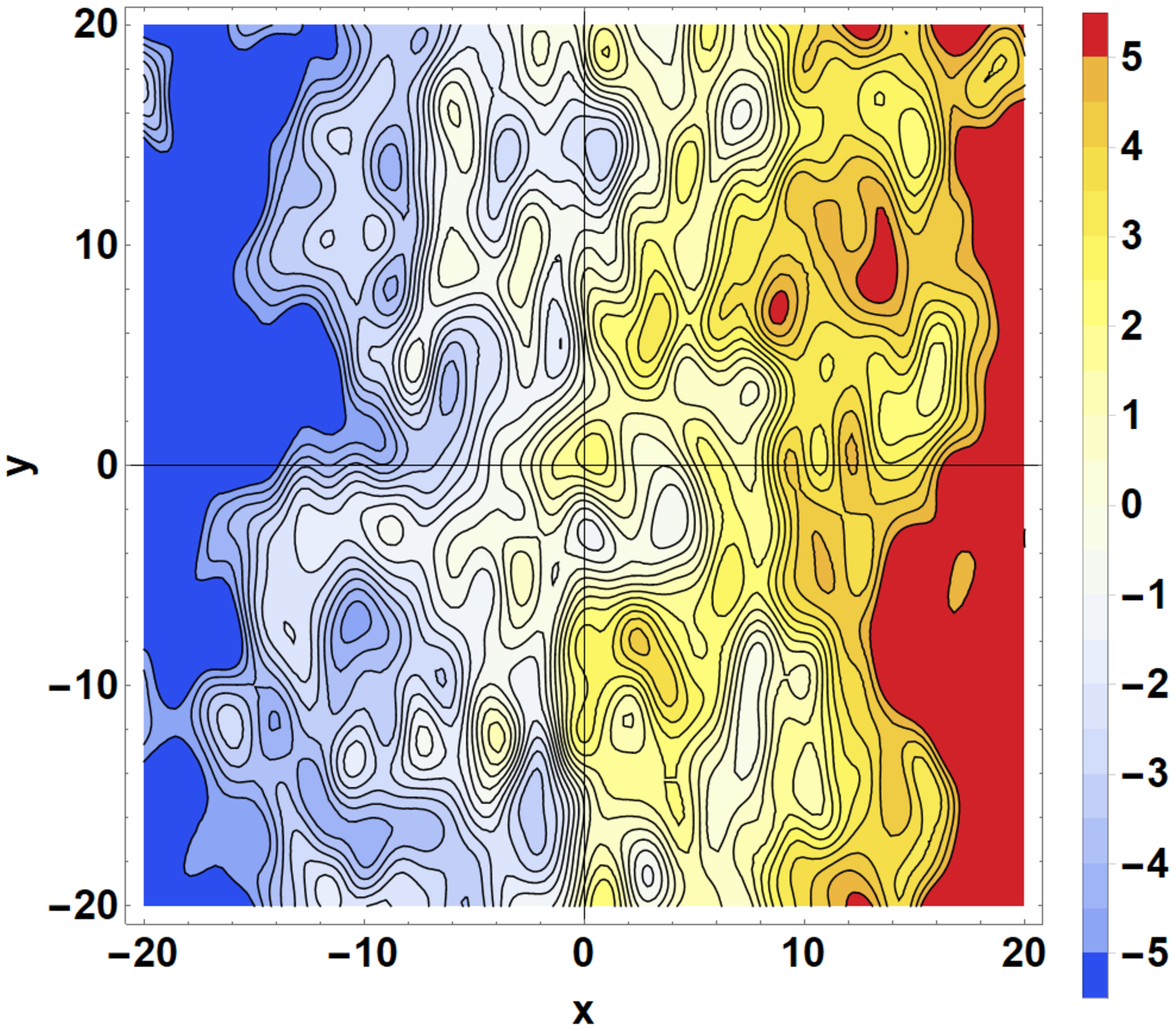}
}
\caption{Typical realization of the potential $\protect\phi _{t}(\mathbf{x})$
for $V_{d}=0$ (left panel) and for $V_{d}=0.3$ (right panel).}
\label{echipot}
\end{figure}

An example of configuration of the contour lines of the potential can be
seen in Fig. \ref{echipot}, where a typical realization of $\phi _{t}(%
\mathbf{x})$ is presented for $V_{d}=0$ (left panel) and for $V_{d}=0.3$
(right panel). The contour lines at $V_{d}=0$ are nested closed curves with
multi-scale sizes that have the dimensions $r_{\max }$ from $r_{\max }\ll
\lambda _{i}$ to $r_{\max }\rightarrow \infty .$ The average velocity $%
\mathbf{V}_{d}$ completely changes the field lines by breaking the large
size contour lines and generating (winding) open paths along its direction.
Islands of closed contour lines remain between the network of open paths,
but their average size decreases as $V_{d}$ increases, and, for $V_{d}$\
much larger than the amplitude of the stochastic velocity, all the lines are
open. The average velocity also determines the limitation of the excursion
of the contour lines perpendicular to $\mathbf{V}_{d}.~$


This configuration of the contour lines of $\phi _{t}(\mathbf{x})$
determines solutions of Eq. (\ref{eqm}) that are, in the presence of an
average velocity, a mixture of localized periodic (or trapped) trajectories
that are closed, and of free trajectories that have unlimited displacements
along $\mathbf{V}_{d}$. The space of trajectories $R$ is organized in two
disjointed subensembles: $tr$ for the trapped trajectories and $fr$ for the
free ones ($R=tr\cup fr,$ $tr\cap fr=\varnothing $). The classification
criterion is the periodicity of the trajectories. A trajectory $r$ with
period $T_{r}$ belongs to $tr$ if $T_{r}$ is finite and to $fr$ otherwise.\ $%
T_{r}$ is defined as the time of the first return in the initial point $%
\mathbf{x}(0)=\mathbf{0}$, and is determined as the first solution of $%
r(t)=0,$ where $r(t)=\sqrt{x_{1}^{2}(t)+x_{2}^{2}(t)}.$ Practically, a
trajectory belongs to the subensemble $tr$ when its period is smaller than
the time of integration $t_{\max }$. The size of each trajectory $r_{\max
}=Max(r(t))$ is also calculated.

\begin{figure}[tbh]
\centerline{\includegraphics[height=6.5cm]{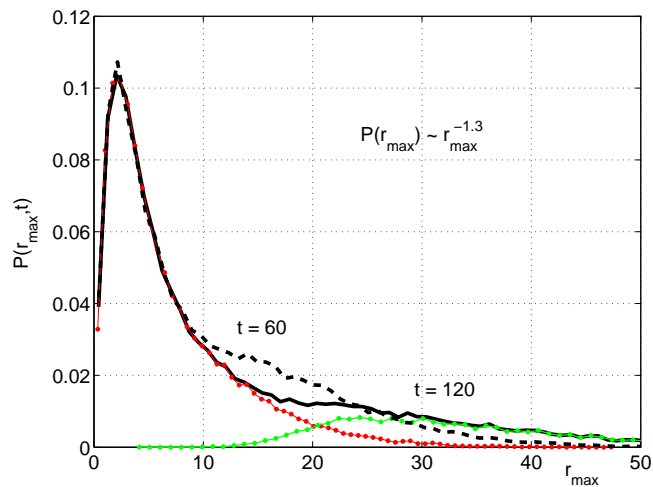}}
\caption{The probability of trajectory sizes $P(r_{max},t)$ for $V_{d}=0$ at 
$t=60$ (dashed black line) and $t=120$ (solid black line). Also shown are
the contributions of the trapped (red points) and free (green points)
trajectories at $t=120$.}
\label{PrmaxVd0}
\end{figure}

For $V_{d}=0,$ all trajectories $\mathbf{x}(t)$ are closed, periodic
functions of time when $t\rightarrow \infty .$ At finite time $t$, open
trajectories are found, which correspond to large periods $T_{r}>t$ (and to
large size contour lines). As $t$ increases the fraction of free
trajectories decreases, and, in the limit $t\rightarrow \infty ,$ all
trajectories are trapped ($tr=R$ and $fr=\varnothing )$. The probability of
trajectory sizes $P(r_{\max },t)$ is represented in Fig. \ref{PrmaxVd0} at
two time moments, $t=60$ (dashed line) and $t=120$\ (solid line). One can
see that the time evolution of $P(r_{\max },t)$ affects only the large
distances, while the small $r_{\max }$ domain\ has invariant probability.
The contributions of the closed and open trajectories to $P(r_{\max },t)$
are also represented in the figure. The closed trajectories (red points)
determine the invariant part of $P(r_{\max },t).$ The open trajectories
(green points) have large sizes and they determine the time variation of $%
P(r_{\max },t).$ Their contribution move toward larger $r_{\max }$ as time
increases and it decays, such that $P(r_{\max },t)$ is determined only by
closed trajectories in the limit $t\rightarrow \infty $. It is a decaying
function of $r_{\max }$ that scales as $P(r_{\max },t)\sim r_{\max }^{-1.3}$
at large $t.$ The slow algebraic decay of the asymptotic probability shows
that the sizes of the trajectories cover multiple scales from $r_{\max }\ll
\lambda _{i}$ to $r_{\max }\rightarrow \infty .$

Thus, the invariance of the Lagrangian potential determines a process of
trajectory trapping manifested by eddying in the structure of $\phi \left( 
\mathbf{x}\right) .$

\begin{figure}[tbh]
\centerline{
\includegraphics[height=5.5cm]{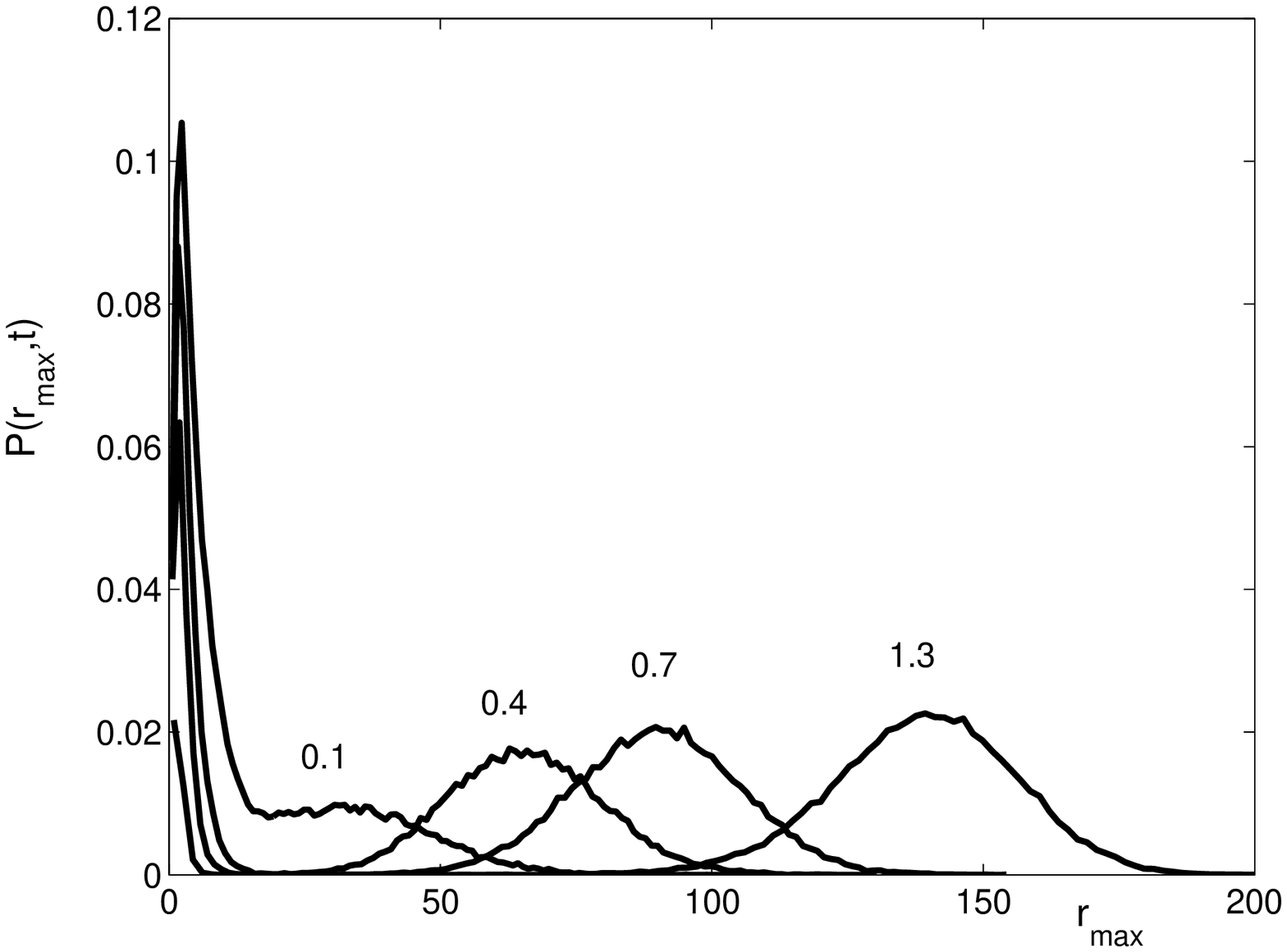}
\includegraphics[height=5.5cm]{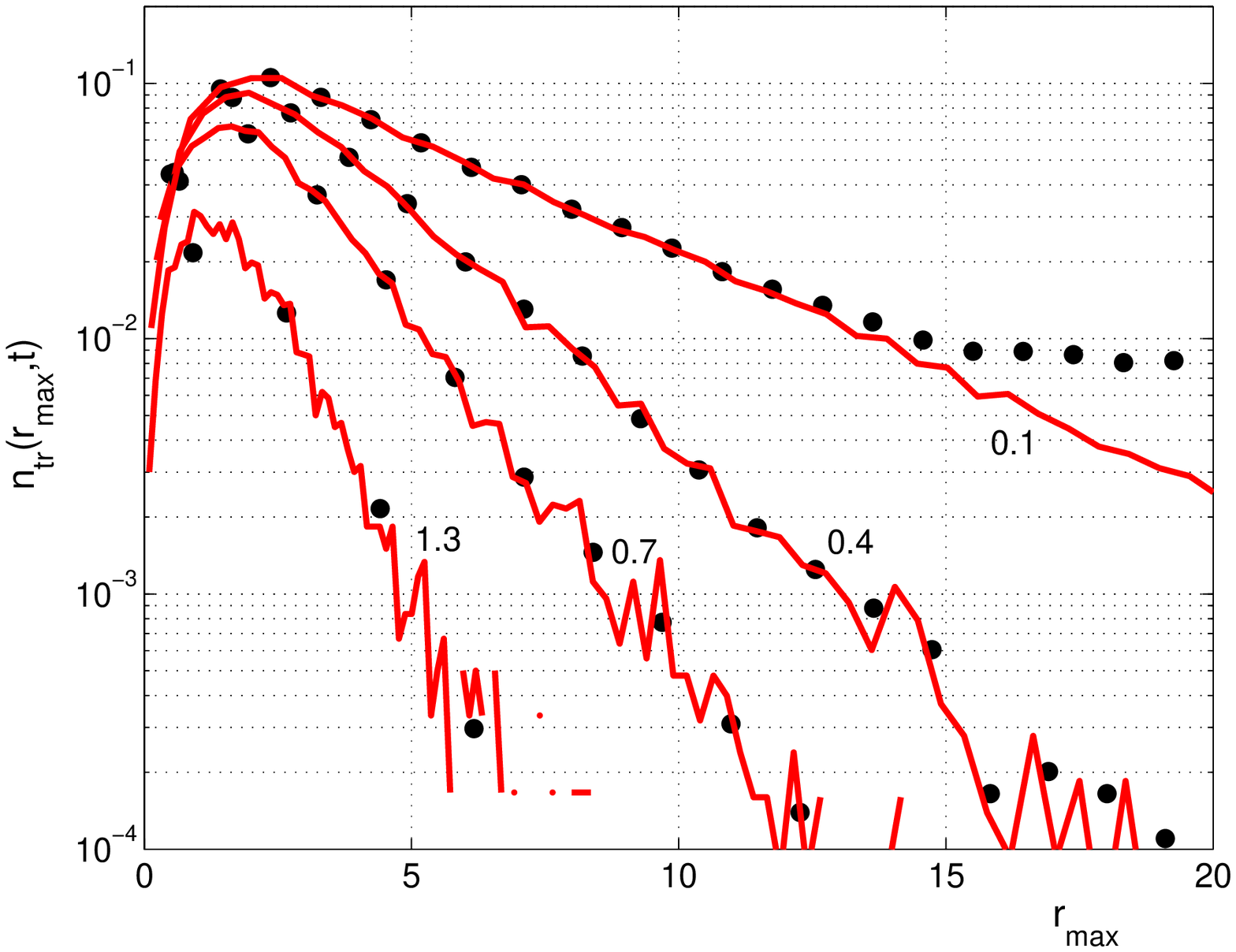}
}
\caption{The probability $P(r_{max},t)$ for several average velocities $V_d$
that label the curves, at $t=60$ as function of $r_{max}$ (the curves in the
left panel and black points in the right panel) and the contribution of the
trapped trajectories $n_{tr}(r_{max},t)$ (right panel, red lines).}
\label{PrmaxVd}
\end{figure}

The average velocity $V_{d}$ that strongly modifies the structure of the
field lines of $\phi _{t}\left( \mathbf{x}\right) $\ determines a
significant change of the probability of trajectory sizes. Two categories of
trajectories coexist for $V_{d}\lesssim V:$ periodic, closed trajectories
situated on the islands of closed contour lines of $\phi _{t}\left( \mathbf{x%
}\right) $ and non-periodic trajectories along the open paths generated by
the average potential $xV_{d}.$ The latter are free trajectories that make
large displacements along $\mathbf{V}_{d}.$\ The probability $P(r_{\max },t)$
can be written as the sum of the contributions of these two types of
trajectories%
\begin{equation}
P(r_{\max },t)=n_{tr}(r_{\max },t)+n_{fr}(r_{\max },t),  \label{ntfr}
\end{equation}%
where $n_{tr},$ $n_{fr}$\ are determined in the subensembles $tr$ and $fr$
at time $t.$ $P(r_{\max },t),$\ shown in Fig. \ref{PrmaxVd}\ (left panel)
has a second maximum. It appears at a large value of $r_{\max }$ \ that
increases with the increase of $V_{d}.$\ Also, the amplitude and the width
of this peak increase with $V_{d}.$\ It is determined by the free
trajectories. The narrow peak $n_{tr}(r_{\max },t)$ at small $r_{\max }$\ is
the contribution of the trapped, periodic trajectories. It is represented in
the right panel of Fig. \ref{PrmaxVd}, which shows that both the maximum
size and the amplitude of the trapped trajectories decrease as $V_{d}$
increases. The average velocity hinders and eventually eliminates the
trapping process. The contribution $n_{tr}(r_{\max },t)$ in Eq. (\ref{ntfr})
decreases with $V_{d}$ and become negligible at $V_{d}\gg 1.$ The
contribution of the free trajectories is negligible in this range of small
sizes, at any $V_{d}$, as shown in Fig. \ref{PrmaxVd} (right panel) where
the black points for $P(r_{\max },t)$ are superposed on the red curves for $%
n_{tr}(r_{\max },t).$ The two contributions in Eq. (\ref{ntfr}) separates at
large time.

\begin{figure}[tbh]
\centerline{
\includegraphics[height=6.0cm]{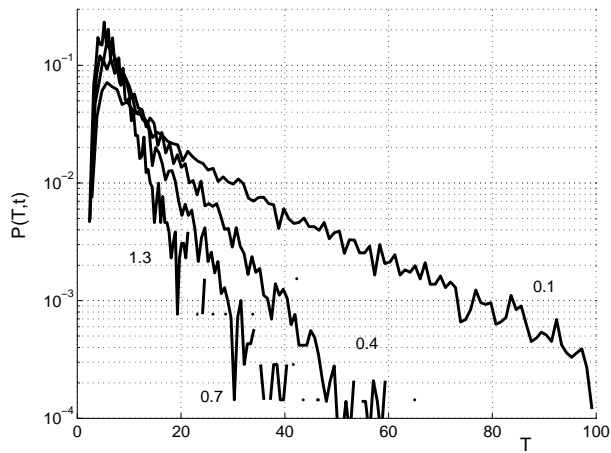}}
\caption{The probability of the periods of the trapped trajectories $P(T,t)$
as functions of $T$ at $t=60$ and at the values of $V_{d}$ that label the
curves.}
\label{PT-Vd}
\end{figure}

The probability of the periods of the closed trajectories $P(T,t)$
calculated from the trajectories $\mathbf{x}(t)$ at $t=60$ is shown in Fig. %
\ref{PT-Vd}. One can see that, at small $V_{d},$ this probability extends to
large values of $T\lesssim 100$ and it has a weak decay. As $V_{d}$\
increases, the width of $P(T,t)$ decreases and its decay is steeper. This
behavior is in agreement with the decay of the trajectory sizes at large $%
V_{d}.$\ An average velocity can be defined for the trapped trajectories as
the maximum displacement over the period, $v^{eff}=r_{\max }/T.$ Its
probability is weakly dependent on the average velocity. \ \ 


\begin{figure}[tbh]
\centerline{
\includegraphics[height=5.5cm]{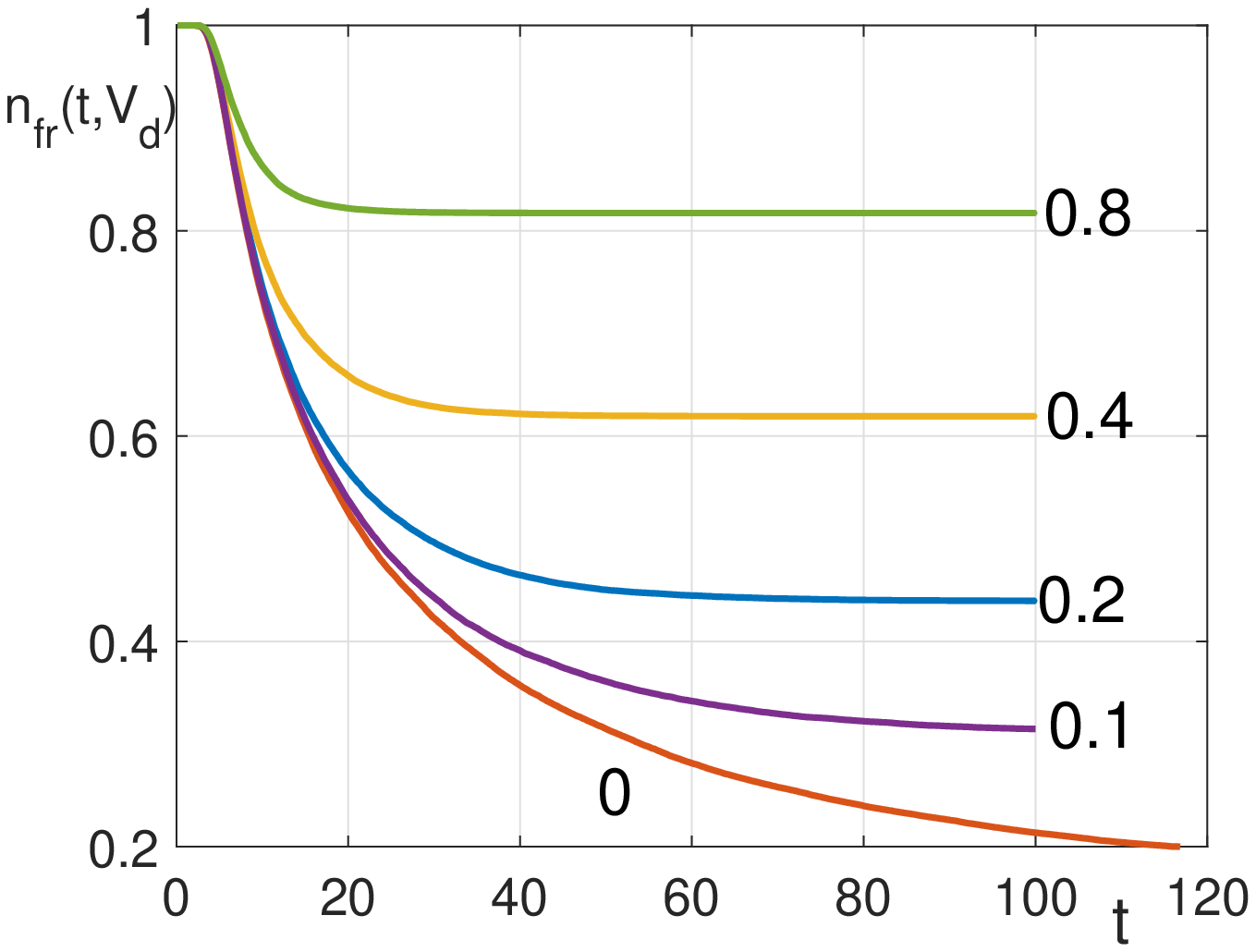}
\hspace{0.5cm}
\includegraphics[height=5.5cm]{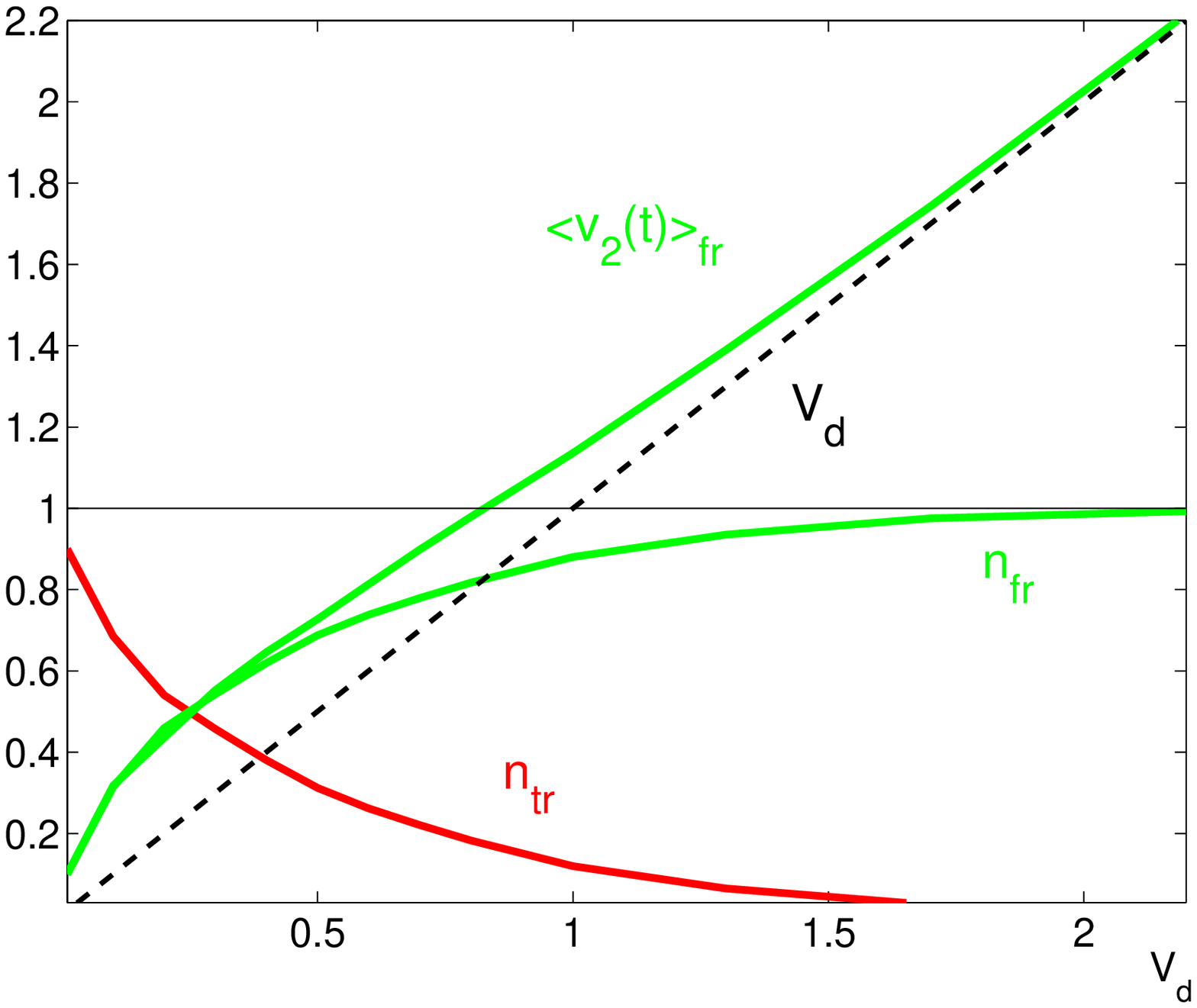}}
\caption{Left panel: the fractions of free trajectories as function of time
for the values of $V_d$ that label the curves. Right panel: the asymptotic
values $n_{fr}$ and $n_{tr}$ and the average velocity of the free
trajectories (see next Section) as functions of $V_d$. }
\label{ntrfrvm}
\end{figure}



\begin{figure}[tbh]
\centerline{
\includegraphics[height=5.5cm]{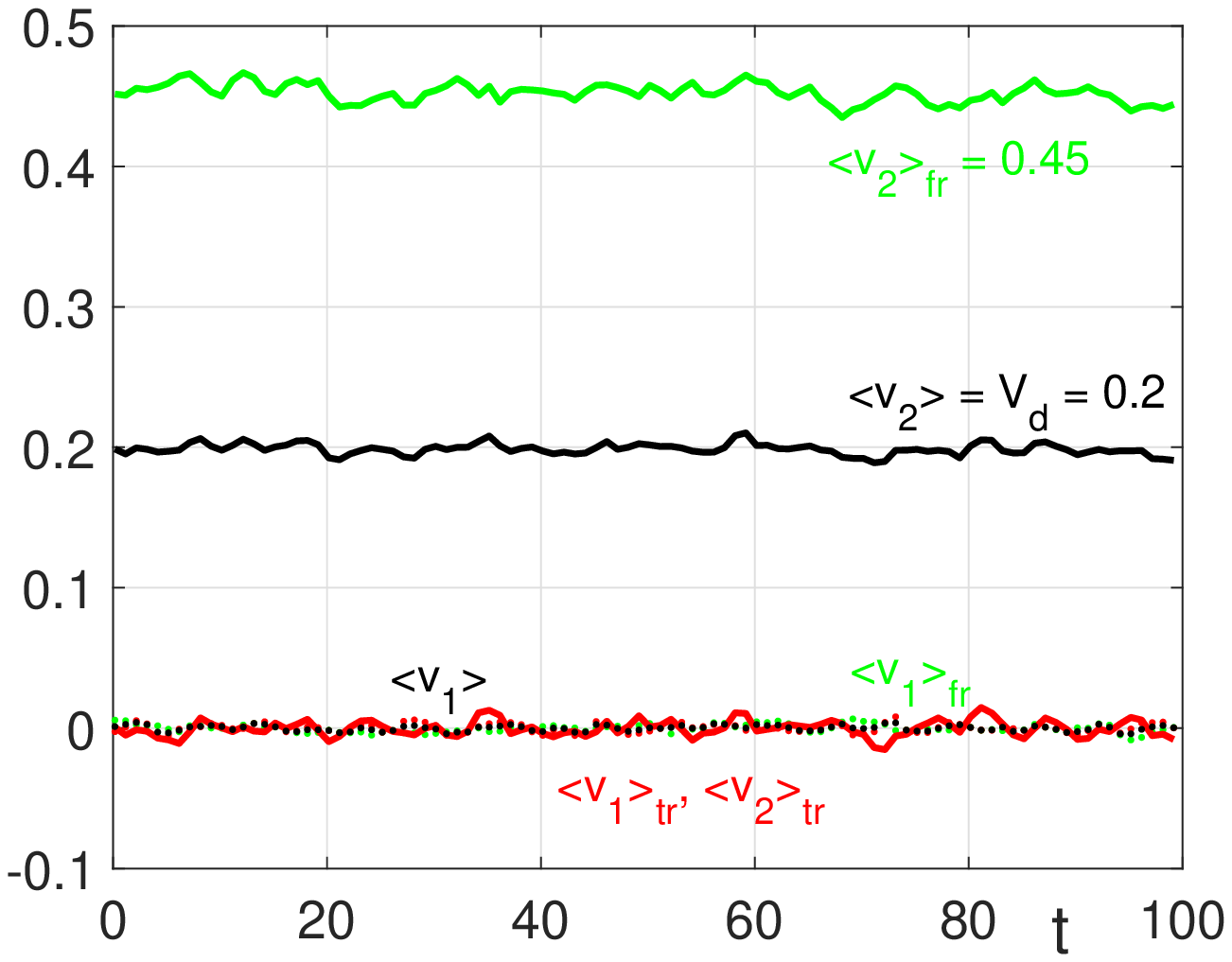}
\hspace{0.5cm}
\includegraphics[height=5.5cm]{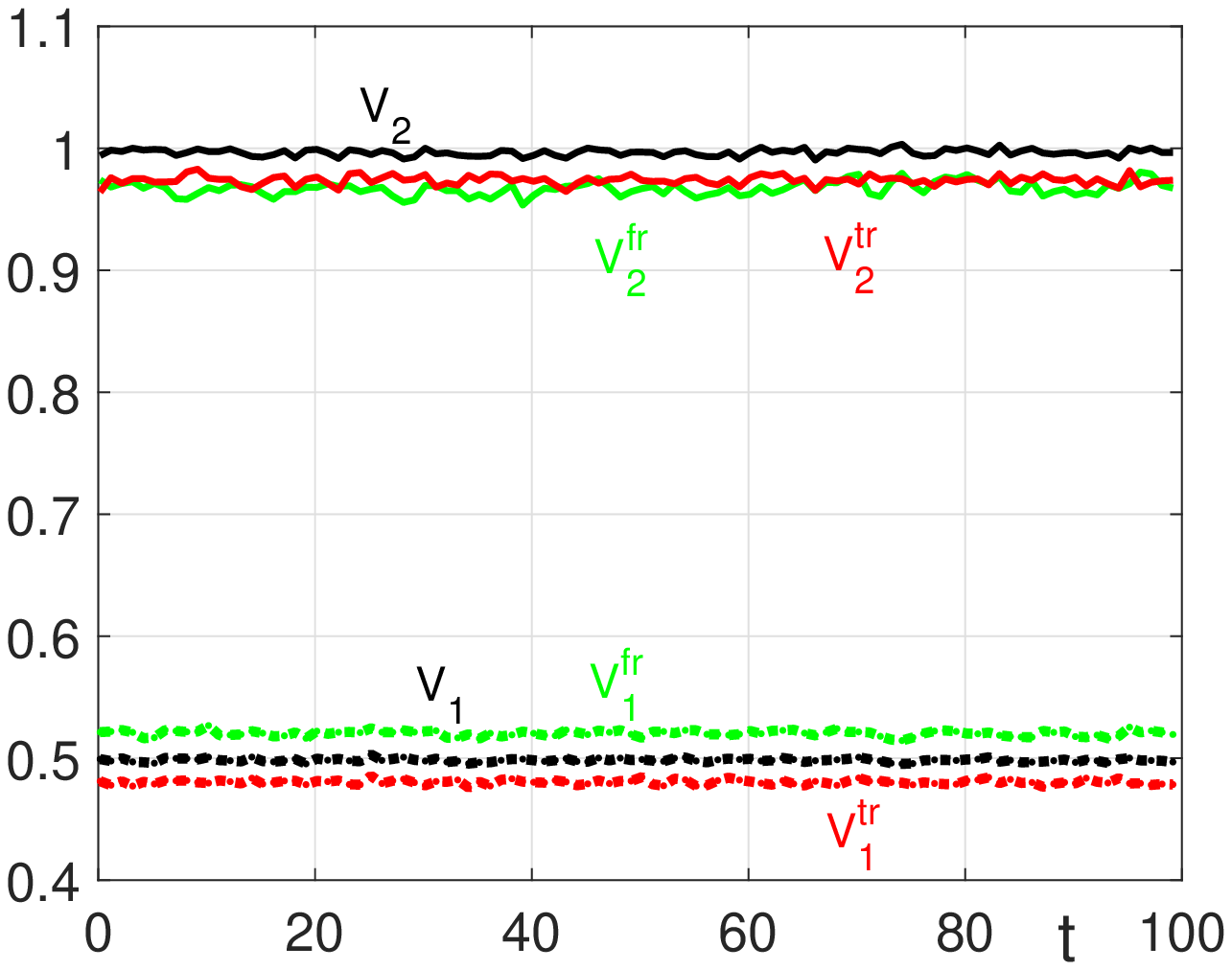}
}
\caption{The average Lagrangian velocities (left panel) and the fluctuations
of the Lagrangian velocities (right panel) as functions of time. The dashed
lines are for the $v_1$ and the solid lines for the $v_2$. The green lines
are for the free trajectories and the red lines for the trapped
trajectories, while the black are averages on the whole statistical ensemble 
$R$. $V_d=0.2$.}
\label{vymed}
\end{figure}


\begin{figure}[tbh]
\centerline{
\includegraphics[height=5.5cm]{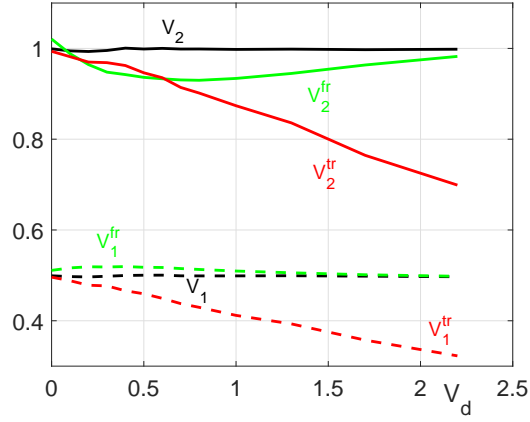}
}
\caption{The amplitudes of fluctuations of the Lagrangian velocities as
functions of $V_d$. The dashed lines are for the $v_1$ and the solid lines
for the $v_2$. The green lines are for the free trajectories and the red
lines for the trapped trajectories, while the black are averages on the
whole statistical ensemble $R$.}
\label{vitfluct}
\end{figure}


\begin{figure}[tbh]
\centerline{
\includegraphics[height=5.5cm]{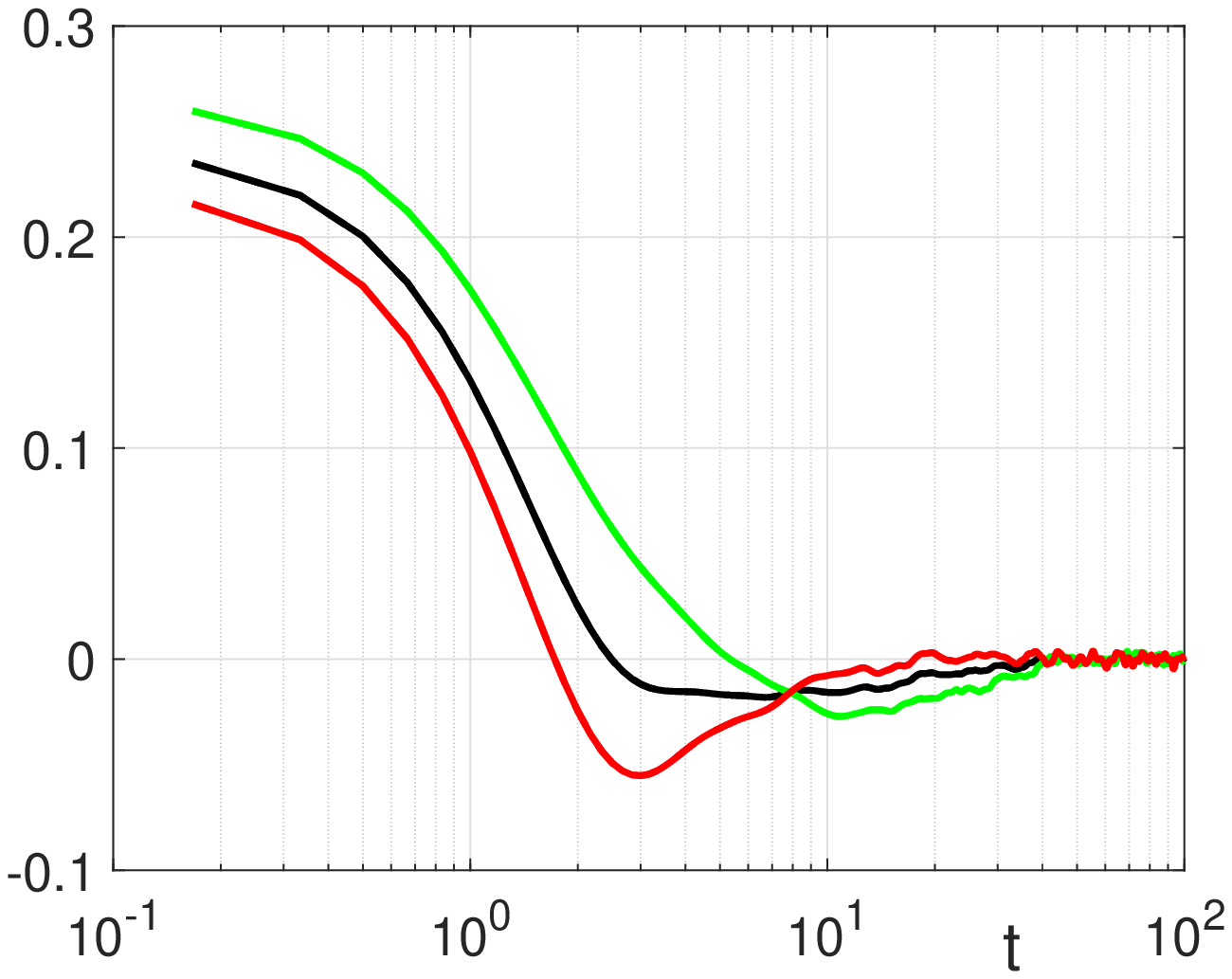}
\hspace{0.5cm}
\includegraphics[height=5.5cm]{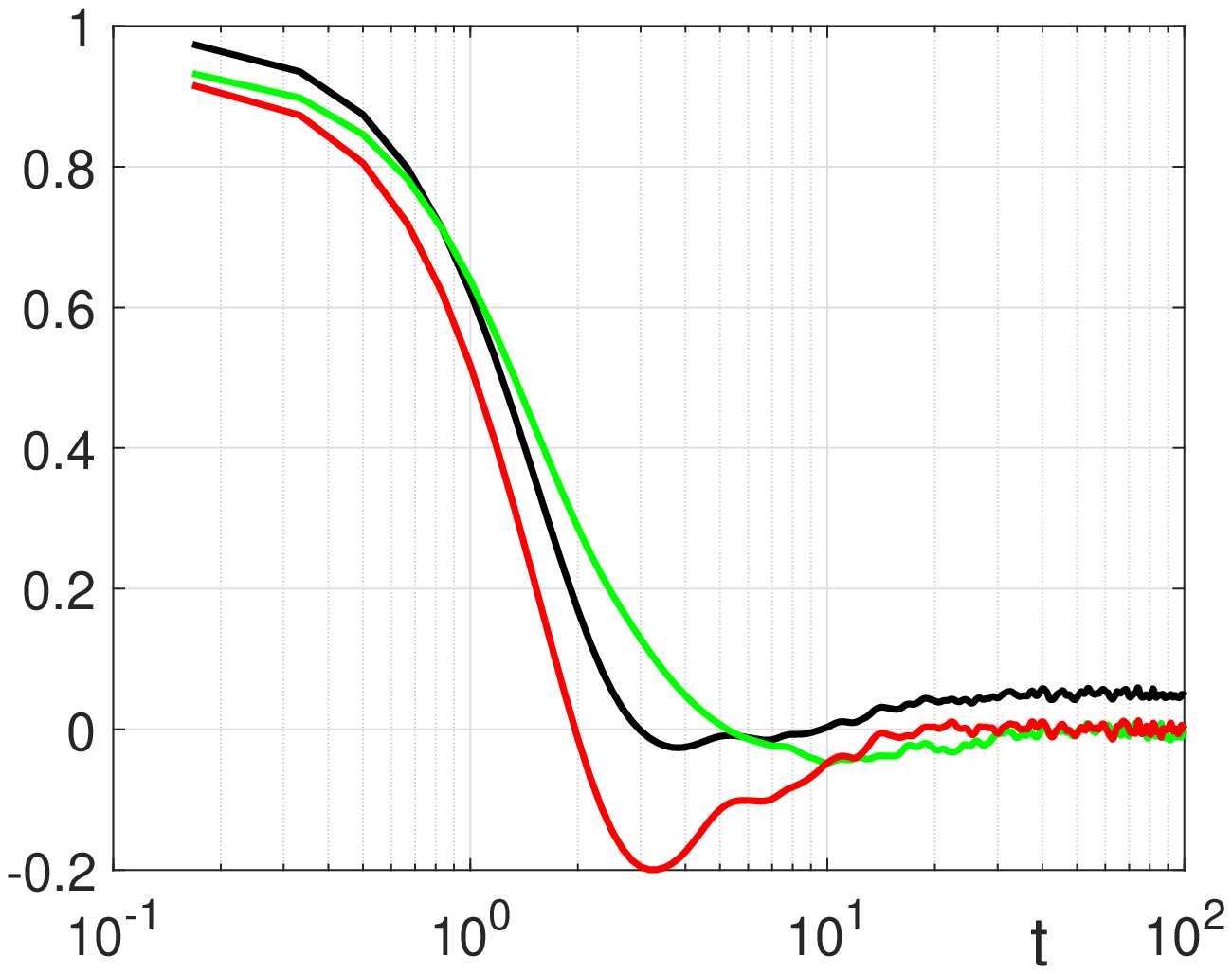}
}
\caption{The correlations of the Lagrangian velocity $v_1(t)$ (left panel)
and $v_2(t)$ (right panel). The correlations on the whole statistical
ensemble $L_i(t)$ (black lines) are compared to the subensemble correlations 
$L_i^{tr}(t)$ (red lines) and $L_i^{fr}(t)$ (green lines). $V_d=0.2.$}
\label{Lvx}
\end{figure}



\begin{figure}[tbh]
\centerline{
\includegraphics[height=5.5cm]{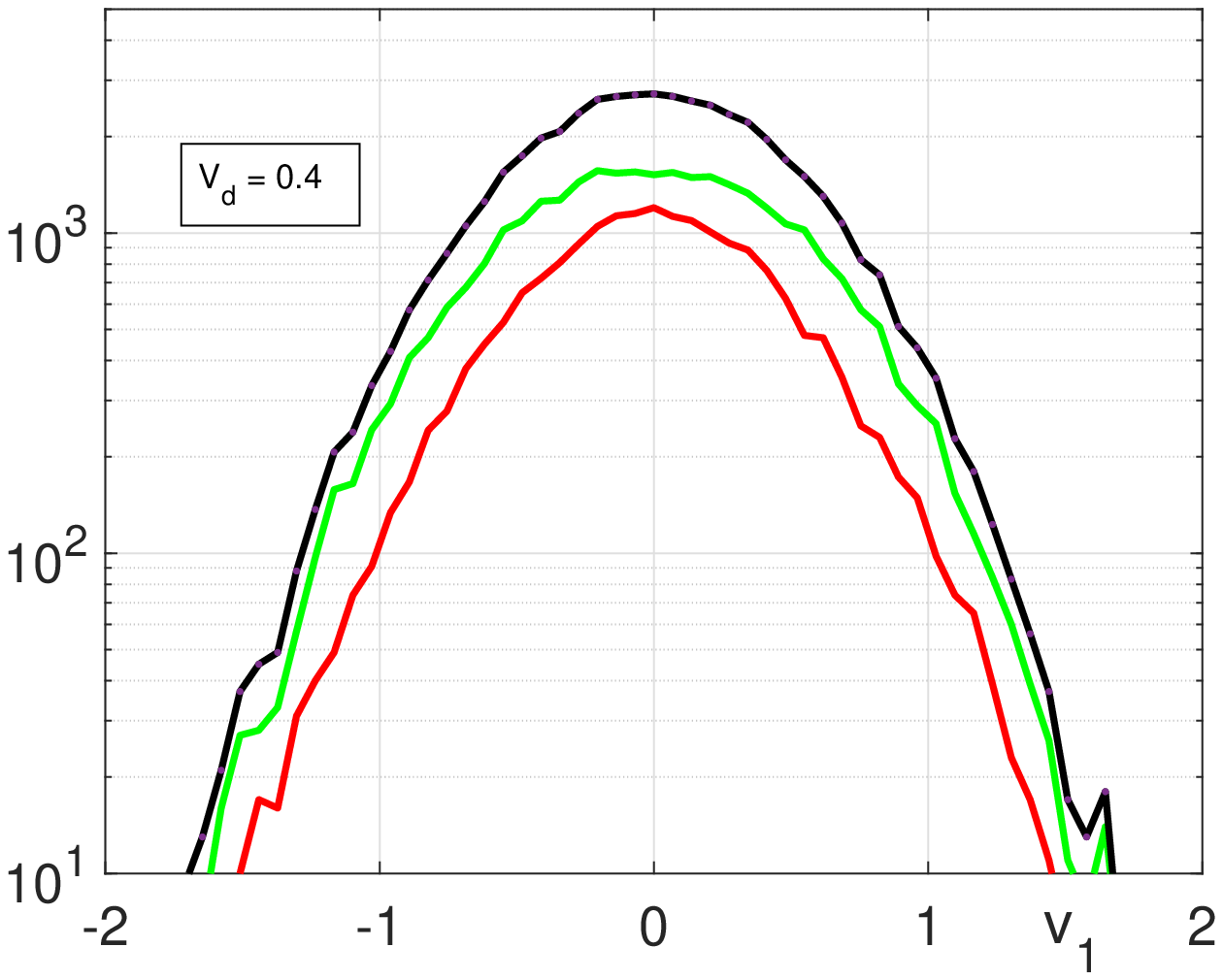}
\hspace{0.5cm}
\includegraphics[height=5.5cm]{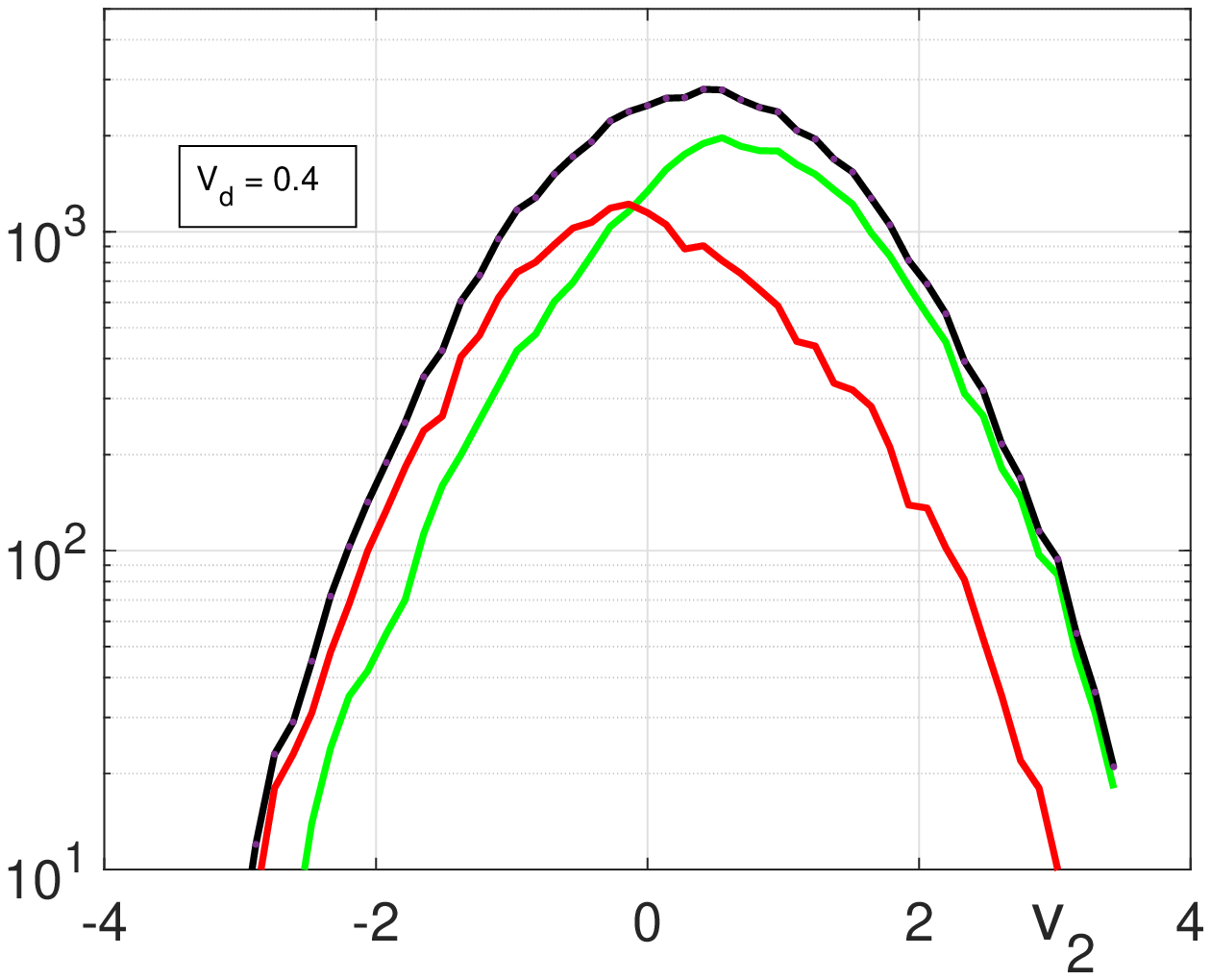}
}
\caption{Histograms of the Lagrangian velocities $v_1$ (left panel) and $v_2$
(right panel) represented by the black curves and the contributions
determined by the free (green curves) and trapped (red curves) trajectories. 
$V_d=0.4$.}
\label{Pvx}
\end{figure}


The fraction of trajectories that are not closed at time $t,$ $%
n_{fr}(t,V_{d})$\ is obtained from the probability of the periods of the
closed trajectories (calculated at the time of integration, $t_{\max })$ 
\begin{equation}
n_{fr}(t,V_{d})=1-n_{tr}(t,V_{d}),\ \ \
n_{tr}(t,V_{d})=\int_{0}^{t}P(T,t_{\max })~dT.  \label{nfrtT}
\end{equation}%
This function decreases in time from $n_{fr}(0,V_{d})=1,$ as seen in Fig. %
\ref{ntrfrvm} (left panel). In the case $V_{d}\neq 0,$ \ $n_{fr}(t,V_{d})$\
saturates at a value $n_{fr}(V_{d})$ in a time that becomes shorter at
larger $V_{d}$. In the case of $V_{d}=0,$\ the decay is not limited, and it
scales as $n_{fr}(t,0)\sim t^{-0.6}.$\ 

The results obtained for the asymptotic fraction of free trajectories $%
n_{fr}(V_{d})\equiv \underset{t\rightarrow \infty }{\lim }n_{fr}(t,V_{d}),$
presented in Fig. \ref{ntrfrvm} (right panel), are well approximated by 
\begin{equation}
n_{fr}(V_{d})=\left[ 1-\exp \left( -V_{d}^{2}\right) \right] ^{1/4}.
\label{nfr}
\end{equation}%
The fraction of trapped trajectories is $n_{tr}(t,V_{d})=1-n_{fr}(t,V_{d})$
at any time, with the asymptotic value $n_{tr}(V_{d})=1-n_{fr}(V_{d})$ that
is also represented in Fig. \ref{ntrfrvm} (right panel).

\section{4. Lagrangian statistics in static potentials}

Thus, the trajectories obtained in the stochastic potential $\phi _{t}(%
\mathbf{x})$\ were divided into two categories: trapped and free. They have
different topologies and different sizes, which suggests that their
contributions to the global statistical properties of the trajectories are
qualitatively different.

We analyze here the statistics of the Lagrangian velocity and of the
displacements of each category of trajectories. For any Lagrangian quantity $%
A(\mathbf{x}(t)),$ we determine $\left\langle A(\mathbf{x}(t))\right\rangle
_{tr}$ and $\left\langle A(\mathbf{x}(t))\right\rangle _{fr}$ that are
conditional averages restricted to the trapped and free trajectories,
respectively. These are statistical averages calculated over the subspaces $%
tr$ and $fr.$ The contribution of each subensemble to the global average
(over $R$) is the product of the probability that a trajectory belongs to
the subensemble multiplied by the statistical average over the subensemble, $%
n_{c}(t,V_{d})\left\langle A(\mathbf{x}(t))\right\rangle _{c},$ where $%
c=tr,~fr.$ It yields

\begin{equation}
\left\langle A(\mathbf{x}(t))\right\rangle =n_{tr}(t,V_{d})\ \left\langle A(%
\mathbf{x}(t))\right\rangle _{tr}+n_{fr}(t,V_{d})\ \left\langle A(\mathbf{x}%
(t))\right\rangle _{fr}.  \label{cond-type}
\end{equation}%
The separation of the trajectories in these categories is performed at a
large time such that $n_{fr}(t,V_{d})$ is saturated (see Fig. \ref{ntrfrvm},
left panel).

\subsection{4.1 Statistics of the Lagrangian velocity}

The statistical parameters of the Lagrangian velocity $\mathbf{v}\left( 
\mathbf{x}(t)\right) \equiv \mathbf{v}(t)$ are shown in Fig. \ref{vymed}\
for a stochastic potential with $\lambda _{1}=1,$ $\lambda _{2}=2$ and $%
V_{d}=0.2.$ The average Eulerian velocity and fluctuation amplitudes are in
this case $\left\langle v_{1}\right\rangle =0,~\left\langle
v_{2}\right\rangle =V_{d},$ $V_{1}=0.5$ and $V_{2}=1,$\ where $V_{i}=\sqrt{%
\left\langle \widetilde{v}_{i}^{2}\right\rangle }$\ are obtained from Eq. (%
\ref{ECv}).

The Lagrangian quantities maintain the Eulerian values at any time,\ as
stated by Lumley theorem. Besides this, the conditional average velocity and
fluctuation amplitudes are time invariant, as seen in Fig. \ref{vymed}, but
their values depend on the category.

It is interesting to note that the average velocity is determined only by
the free trajectories, while the trapped trajectories do not contribute ($%
\left\langle v_{2}(t)\right\rangle _{tr}=0$ at any time). The average
velocity of the free trajectories is larger than $V_{d},$ and it can be
approximated with%
\begin{equation}
\left\langle v_{2}(t)\right\rangle _{fr}=\frac{V_{d}}{n_{fr}}>V_{d}
\label{vfr}
\end{equation}%
It is $\left\langle v_{2}(t)\right\rangle _{fr}=0.45$ for the example
presented in Fig. \ref{vymed}, left panel, obtained for $V_{d}=0.2$. The
conditional average velocity $\left\langle v_{2}(t)\right\rangle _{fr}$ is
also shown in Fig. \ref{ntrfrvm} (right panel) as function of $V_{d}.$ One
can see that this average velocity is significantly larger than $V_{d}$ only
in the presence of trajectory trapping (for $V_{d}\lesssim 1).$

This result shows that a supplementary ordered component of the Lagrangian
velocity appears for the free trajectories that exactly compensates the
missing contribution of the trapped particles, such that $\left\langle
v_{2}(t)\right\rangle =n_{fr}\left\langle v_{2}(t)\right\rangle _{fr}=V_{d}.$
It seems to be a trivial consequence of $\left\langle v_{2}(t)\right\rangle
_{tr}=0$, but the underlying physical process is rather complex. It
essentially consists of generation of ordered motion from the stochastic
velocity $\widetilde{\mathbf{v}}(\mathbf{x,}t)$ for both types of
trajectories 
\begin{equation}
\left\langle \widetilde{v}_{2}(t)\right\rangle _{tr}=-V_{d},\ \ \left\langle 
\widetilde{v}_{2}(t)\right\rangle _{fr}=V_{d}\frac{n_{tr}}{n_{fr}}.
\label{vsupl}
\end{equation}%
The supplementary average velocity of the trapped trajectories is opposite
to $\mathbf{V}_{d}$ and exactly compensates it. The supplementary average
velocity of the free trajectories is along $\mathbf{V}_{d}$ and it
contributes to the increase of the Lagrangian over the Eulerian velocity.

Equations (\ref{vsupl}) are valid at any time, including $t=0.$ It can be
interpreted as the condition for the separation of the trajectories in the
free and trapped categories. The trapped trajectories start from the
geometric locus for which $\left\langle \widetilde{v}_{2}(\mathbf{x}%
)\right\rangle =-V_{d}$ and they remain in this domain, while the free
trajectories are confined in the complement of this domain. These ordered
components of the motion are hidden, in the sense that they are not "seen"
in the average velocity calculated on the whole ensemble $R$ ($\left\langle
v_{2}(t)\right\rangle =V_{d}).$ However, as shown below, they have strong
effects on the transport along $\boldsymbol{V}_{d}$ through the modification
of the correlation of the Lagrangian velocity.

The amplitudes of velocity fluctuations around the average velocity are
shown in Fig. \ref{vymed} (right panel). They are different for the two
types of trajectories. It is interesting to underline that the supplementary
order that characterizes trapped and free trajectories appears in the
fluctuations of the velocity in $R$. The average of the square velocity
decomposed on $tr$ and $fr$\ subensembles according to Eq. (\ref{cond-type})
for large time \ \ \ 
\begin{equation}
\left\langle v_{i}^{2}(t)\right\rangle =n_{tr}\left\langle
v_{i}^{2}(t)\right\rangle _{tr}+n_{fr}\left\langle v_{i}^{2}(t)\right\rangle
_{fr},  \label{vy2}
\end{equation}%
leads to%
\begin{equation}
V_{2}^{2}=n_{tr}(V_{2}^{tr})^{2}+n_{fr}(V_{2}^{fr})^{2}+\frac{n_{tr}}{n_{fr}}%
V_{d}^{2},  \label{ord-dez}
\end{equation}%
where%
\begin{equation}
(V_{i}^{c})^{2}\equiv \left\langle \left( v_{i}(t)-\left\langle
v_{i}(t)\right\rangle _{c}\right) ^{2}\right\rangle _{c}  \label{cond-ampl}
\end{equation}%
\ are the amplitudes of the fluctuations of the velocity $\delta
v_{i}(t)\equiv v_{i}(t)-\left\langle v_{i}(t)\right\rangle _{c},$ $i=1,2,$\
conditioned by the category of trajectories $c=tr,\ fr$ (on the subensembles 
$tr$ and $fr).$ Thus, a contribution produced by the ordered motion appears
(the last term of Eq. (\ref{ord-dez})) besides the direct contributions of
the conditional fluctuations. It is determined by the ordered motion (\ref%
{vsupl}) generated by $V_{d}$ in the presence of trapping (for $%
V_{d}\lesssim 1).$ The results presented in Fig. \ref{vymed} (right panel)
show values $V_{2}^{tr}<V_{2}$ and $V_{2}^{fr}<V_{2},$ which reproduce Eq. (%
\ref{ord-dez}).

The conditioned amplitudes of the velocity fluctuations $V_{i}^{c}$ depend
on the average velocity $V_{d}.$ As seen in Fig. \ref{vitfluct}, the
amplitudes of both components of the trapped trajectory velocity (red
curves) are continuously decreasing functions of $V_{d}$. This is the effect
of the decrease of the size of the islands of closed contour lines of the
potential, which, as $V_{d}$ increases, shrink around the maxima and minima
of $\phi (\mathbf{x})$ where the gradients are small. In the case of free
trajectories (green lines), the amplitudes of the Lagrangian velocities are
different of $V_{i}$ only in the range of $V_{d}$ that corresponds to the
existence of islands of closed contour lines of the potential. The
perpendicular amplitude is increased ($V_{1}^{fr}>V_{1}),$ while the
parallel amplitude is decreased ($V_{2}^{fr}<V_{2})$ such that the
supplementary parallel velocity is compensated (Eq. (\ref{ord-dez})).

One can deduce from these results that the EC defined on the geometric locus
of the free trajectories is different of the EC (\ref{EC}). As shown below
(Section 6), the amplitude of the stochastic potential of the free
trajectories $\Delta $ is smaller than in the whole space ($\Delta <\Phi $).
The correlation lengths are evaluated using the amplitudes of velocity
fluctuations of the free trajectories%
\begin{equation}
\lambda _{1}^{fr}\sim \frac{\Delta }{V_{2}^{fr}}=\lambda _{1}\frac{\Delta }{%
\Phi }\frac{V_{2}}{V_{2}^{fr}},\ \lambda _{2}^{fr}\sim \frac{\Delta }{%
V_{1}^{fr}}=\lambda _{2}\frac{\Delta }{\Phi }\frac{V_{1}}{V_{1}^{fr}},
\label{lambdafr}
\end{equation}%
where $\lambda _{1}=\Phi /V_{2}$ and $\lambda _{2}=\Phi /V_{1}.$\ Thus, the
correlation lengths on the domain of free trajectories decrease with the
factor $\Delta /\Phi $\ on both directions and are modified by the velocity
amplitudes (decreased along $\mathbf{V}_{d}$ and increased across\ $\mathbf{V%
}_{d}$).\ 

The correlations of the Lagrangian velocity are shown in Fig. \ref{Lvx},
where the notations are%
\begin{equation}
L_{i}(t)=\left\langle \delta v_{i}(0)\ \delta v_{i}(t)\right\rangle ,\
L_{i}^{c}(t)=\left\langle \delta v_{i}(0)\ \delta v_{i}(t)\right\rangle
_{c},\ \ c=tr,fr.\   \label{defLC}
\end{equation}%
One can see that all the conditional correlations (for both categories and
both components of the velocity) decay to zero at large $t.$ However, the
correlation of the velocity along $\mathbf{V}_{d}$ calculated on all
trajectories, $L_{2}(t),$ has a finite asymptotic value. It is determined by
the ordered components of motion produced in subensembles $tr$ and $fr.$ An
equation similar to (\ref{ord-dez}) can be obtained from (\ref{cond-type})
written for $A=v_{i}(0)\ v_{i}(t)$ 
\begin{equation}
L_{2}(t)=n_{tr}L_{2}^{tr}(t)+n_{fr}L_{2}^{fr}(t)+\frac{n_{tr}}{n_{fr}}%
V_{d}^{2},  \label{LCsup}
\end{equation}%
\ which shows that $L_{2}(t)$\ has a finite asymptotic tail in spite of the
decay to zero of $L_{2}^{tr}(t)$\ and $L_{2}^{fr}(t).$ It is determined by
the presence of trapped trajectories at small average velocity $V_{d}.$\ \ \
\ \ \ \ 

The histograms for the Lagrangian velocity components are time invariant for
all statistical ensembles $R,$ $tr$ and $fr.$ The histogram for all
trajectories (in $R)$ is shown in Fig. \ref{Pvx} together with the
contributions of the trapped and free trajectories (that include the
fractions of trajectories). One can see that the distribution is Gaussian in 
$R$, while significant departures from Gaussianity appear in the
subensembles $tr$ and $fr,$\ especially for the velocity component $v_{2}$\
(right panel). The domain of large positive velocities is dominated by the
free trajectories, while the trapped trajectories have the main contribution
for the large negative $v_{2}.$ The most probable value of $v_{2}$ on $tr$
(that is slightly negative) is compensated by a longer tail at positive $%
v_{2}.$

The non-Gaussian distribution of the Lagrangian velocity of the trapped
trajectories provides additional information on the geometric locus of this
category of trajectories. It shows that the space average of the parallel
velocity $v_{2}(\mathbf{x})$ on this locus (that is zero for any $V_{d})$\
results from the elimination of the regions with large, positive $v_{2}(%
\mathbf{x}).$ In other words, the regions where the stochastic velocity is
oriented parallel to $\mathbf{V}_{d}$\ belong to the geometric locus $fr.$

\subsection{4.2 Transport and statistics of trajectories}

The statistics of the displacements is strongly non-Gaussian, in spite of
the Gaussian Lagrangian velocity. Moreover, the average and mean square
displacements (calculated for all trajectories and in the subensembles $tr$
and $fr)$ have asymptotic regimes that can be linear, quadratic or saturated
functions of time, which shows that the transport has anomalous aspects.

The average displacements are in agreement with the average Lagrangian
velocities%
\begin{eqnarray}
\left\langle x_{1}(t)\right\rangle &=&\left\langle x_{1}(t)\right\rangle
_{tr}=\left\langle x_{1}(t)\right\rangle _{fr}=0,  \label{depmed} \\
\left\langle x_{2}(t)\right\rangle &=&V_{d}t,\ \left\langle
x_{2}(t)\right\rangle _{tr}=0,\ \left\langle x_{2}(t)\right\rangle _{fr}=%
\frac{V_{d}}{n_{fr}}t.  \notag
\end{eqnarray}

\begin{figure}[tbh]
\centerline{
\includegraphics[height=5.5cm]{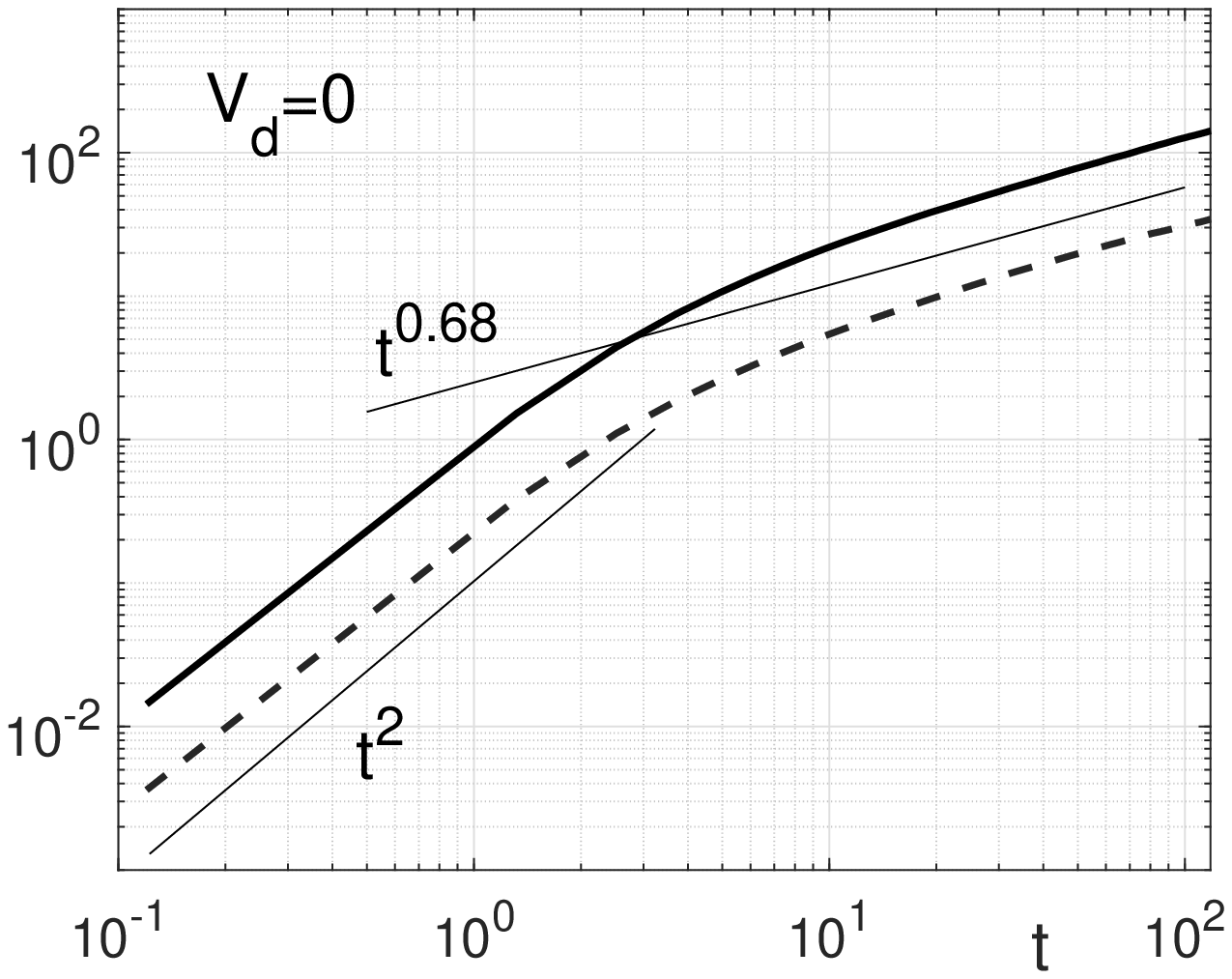}
\hspace{0.5cm}
\includegraphics[height=5.5cm]{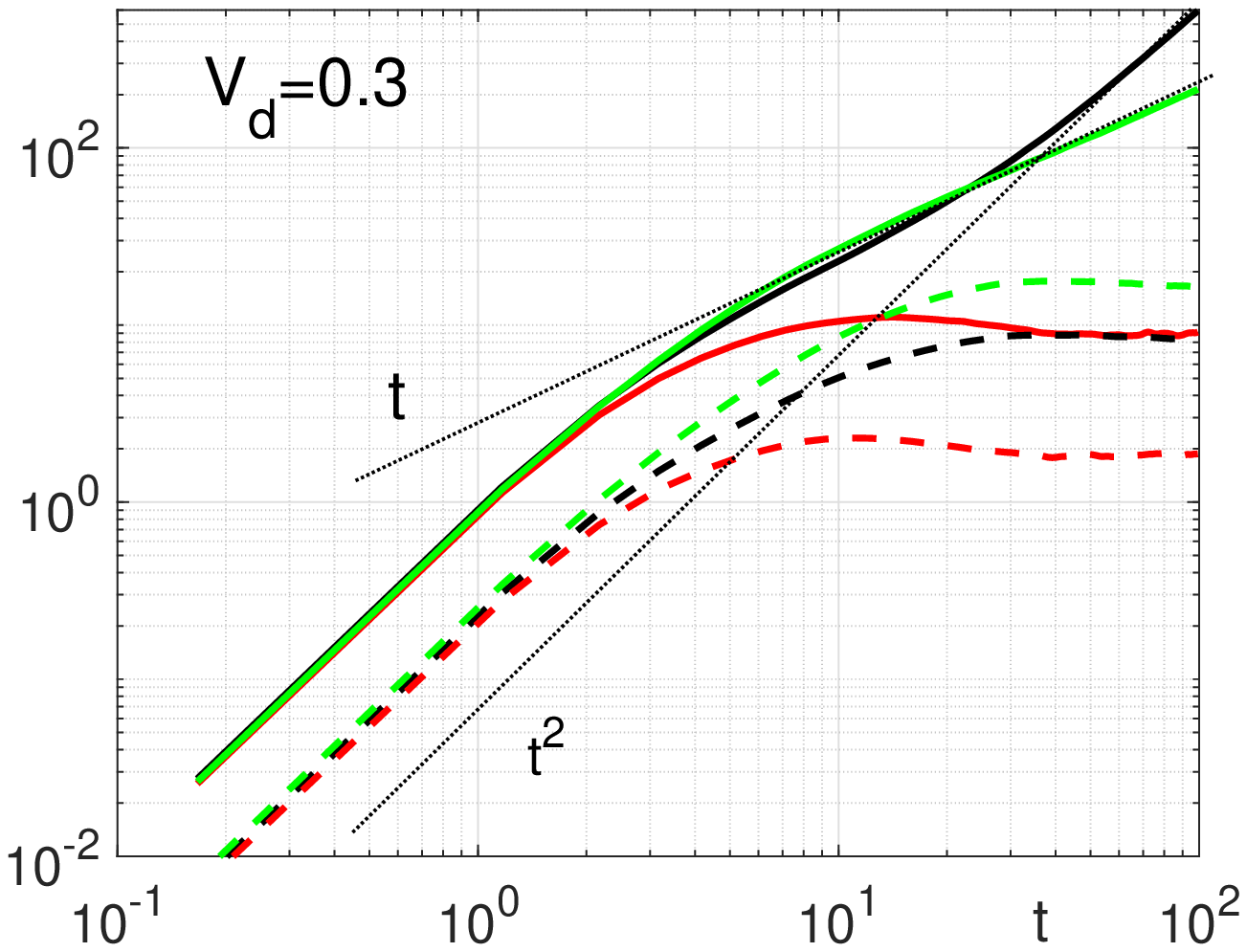}
}
\caption{ The dispersions of the trajectories on the whole statistical
ensemble (black line) for $V_d=0$ (left panel) and $V_d=0.2$ (right panel)
as functions of time. The dashed lines are for $x_1(t) $ and the solid lines
for $x_2(t)$. The conditional dispersions for trapped (red) and free (green)
trajectories are also shown in the right panel.}
\label{x2}
\end{figure}

The dispersion $\left\langle \left( \delta x_{i}(t)\right) ^{2}\right\rangle
,$ where $\delta x_{i}(t)=x_{i}(t)-\left\langle x_{i}(t)\right\rangle $\ are
shown in Fig. \ref{x2} for $V_{d}=0$ (left panel) and for $V_{d}=0.2$ (right
panel), as functions of time.

In the absence of the average velocity ($V_{d}=0),$ the dispersions are
similar along the two directions. The curves in the left panel of Fig. \ref%
{x2} are only translated due to the different amplitudes of the stochastic
velocities $V_{1}=0.5,~V_{2}=1.$ The dispersions are sub-diffusive, with
time increase that is slower than linear $\left\langle \left( \delta
x_{i}(t)\right) ^{2}\right\rangle \sim t^{0.68}.$\ The reason is\ the
progressive saturation of the contributions of the trajectories with small
periods. At a time $t,$ all the trajectories with $T<t$\ have saturated
dispersion and only the free trajectories (that are still not closed)
determine the time variation of $\left\langle \left( \delta x_{i}(t)\right)
^{2}\right\rangle .$ The latter results from two factors with opposite
effects: the fraction of free trajectories at time $t$ and their average
size. As seen in Fig. \ref{ntrfrvm}, left panel, $n_{fr}(t,0)$ is a
decreasing function of time, $n_{fr}(t,0)\sim t^{-0.6}.$ The average size of
the closed trajectories is an increasing function of $t,$ because it is an
increasing function of the average period.

The average velocity $V_{d}$ makes trajectory dispersion strongly
non-isotropic, as seen in Fig. \ref{x2}, right panel. The dispersion $%
\left\langle \left( \delta x_{i}(t)\right) ^{2}\right\rangle $ for all
trajectories (black lines) are compared to the results obtained for the
trapped $\left\langle \left( \delta x_{i}(t)\right) ^{2}\right\rangle _{tr}$
(red lines) and free $\left\langle \left( \delta x_{i}(t)\right)
^{2}\right\rangle _{fr}$ (green lines) trajectories. The dispersions across $%
\mathbf{V}_{d}$ (of $x_{1}(t)$) for the whole set of trajectories and for
the subensembles $tr$ and $fr$ are all saturated (the dashed curves in Fig. %
\ref{x2}, right panel), which corresponds to the minimum sub-diffusive
transport. This means that the average velocity completely hinders the
perpendicular transport in the case of static stochastic potentials. The
contrary happens to the transport parallel to $\mathbf{V}_{d}$: the
dispersion of the trajectories has a very fast time-increase,\ $\left\langle
\left( \delta x_{2}(t)\right) ^{2}\right\rangle \sim t^{2},$\ which
correspond to the maximum super-diffusive transport that is of the ballistic
type. It appears in spite of the much weaker transport of the trapped and
free trajectories ($\left\langle \left( \delta x_{2}(t)\right)
^{2}\right\rangle _{tr}$ saturates and $\left\langle \left( \delta
x_{2}(t)\right) ^{2}\right\rangle _{fr}\sim t$\ is diffusive).\ 

This super-diffusive parallel transport is the effect of the coherent
parallel motion generated by $V_{d}$, as demonstrated using Eq. (\ref%
{cond-type})\ for $A=x_{i}^{2}(t).$ The relations between the dispersion of
all trajectories (in $R$) and the subensemble $tr$\ and $fr$\ dispersions are%
\begin{equation}
\left\langle \left( \delta x_{1}(t)\right) ^{2}\right\rangle
=n_{tr}\left\langle \left( \delta x_{1}(t)\right) ^{2}\right\rangle
_{tr}+n_{fr}\left\langle \left( \delta x_{1}(t)\right) ^{2}\right\rangle
_{fr},  \label{deltx1}
\end{equation}%
\begin{equation}
\left\langle \left( \delta x_{2}(t)\right) ^{2}\right\rangle
=n_{tr}\left\langle \left( \delta x_{2}(t)\right) ^{2}\right\rangle
_{tr}+n_{fr}\left\langle \left( \delta x_{2}(t)\right) ^{2}\right\rangle
_{fr}+\frac{n_{tr}}{n_{fr}}V_{d}^{2}\ t^{2}.  \label{deltx2}
\end{equation}%
The last term in Eq. (\ref{deltx2}) is dominant at large time and it makes
the asymptotic regime superdiffusive of ballistic type. This term is
determined by the supplementary average velocity generated from the
stochastic components for the free and trapped trajectories, Eq. (\ref{vsupl}%
). It leads to the "concentration" of the average velocity along the free
trajectories Eq. (\ref{vfr}). Thus, the super-diffusive parallel transport
is determined by the average velocity ($V_{d}\neq 0$) only in the presence
of the islands of trapped trajectories ($n_{tr}\neq 0$), which corresponds
to $V_{d}\lesssim 1.$

\ \ \ 


\begin{figure}[tbh]
\centerline{
\includegraphics[height=5.5cm]{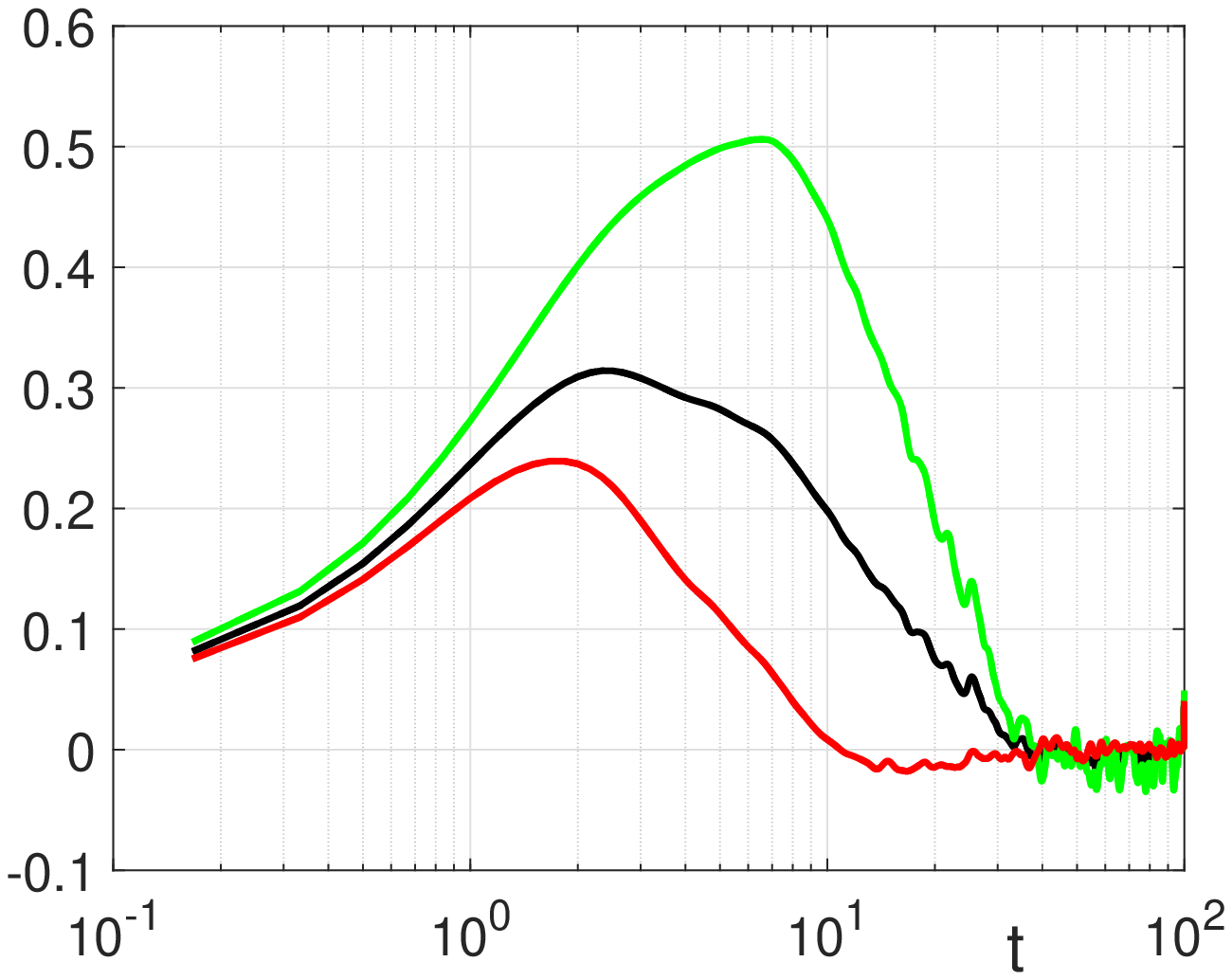}
\hspace{0.5cm}
\includegraphics[height=5.5cm]{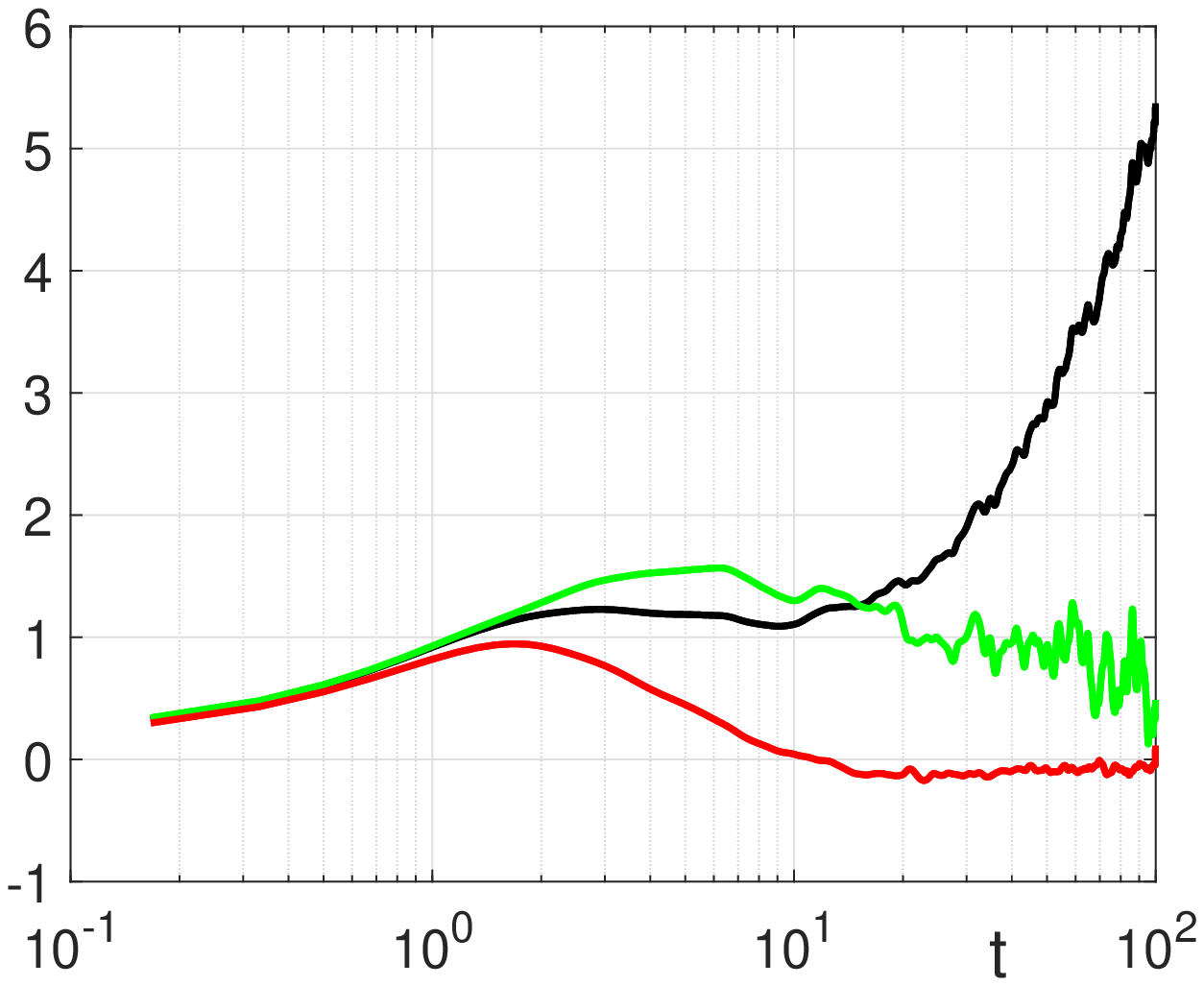}
}
\caption{The time dependent diffusion coeficients in the direction
perpendicular (left panel) and parallel (right panel) to the average
velocity for the whole statistical ensemble $R$ (black lines) and restricted
to the $tr$ (red lines) and $fr$ (green lines) subensembles. $V_d=0.2.$}
\label{Dx}
\end{figure}

The dispersions of the trajectories (Fig. \ref{x2}) are connected to the
correlations of the Lagrangian velocity (Fig. \ref{Lvx}) and to the time
dependent diffusion coefficients, defined by $2D_{i}(t)=d\left\langle \left(
\delta x_{i}(t)\right) ^{2}\right\rangle /dt,$ by Taylor formulas \cite%
{Taylor} 
\begin{equation}
D_{i}(t)=\int_{0}^{t}L_{i}(\tau )~d\tau ,\ \ \left\langle \left( \delta
x_{i}(t)\right) ^{2}\right\rangle =2\int_{0}^{t}\left( t-\tau \right)
~L_{i}(\tau )~d\tau .  \label{T}
\end{equation}%
Similar equations can be written for each category of trajectories (trapped
or free). Figure \ref{Dx} presents the time dependent diffusion coefficients
compared to their restrictions to the trapped and free trajectories for the $%
x_{1}$ (left panel) and $x_{2}$ (right panel) directions. This confirms that
the perpendicular diffusion is completely hindered (even for the free
trajectories). The time integral of $L_{1}(t)$ vanishes for all categories
at a finite time. The parallel transport is ballistic $D_{2}(t)\sim t,$\ in
spite of the normal diffusion of the free trajectories and of the total
confinement of the trapped ones.\ It is the result of the ordered parallel
motion, as seen by performing the time derivative in Eq. (\ref{deltx2}).


\begin{figure}[tbh]
\centerline{
\includegraphics[height=5.5cm]{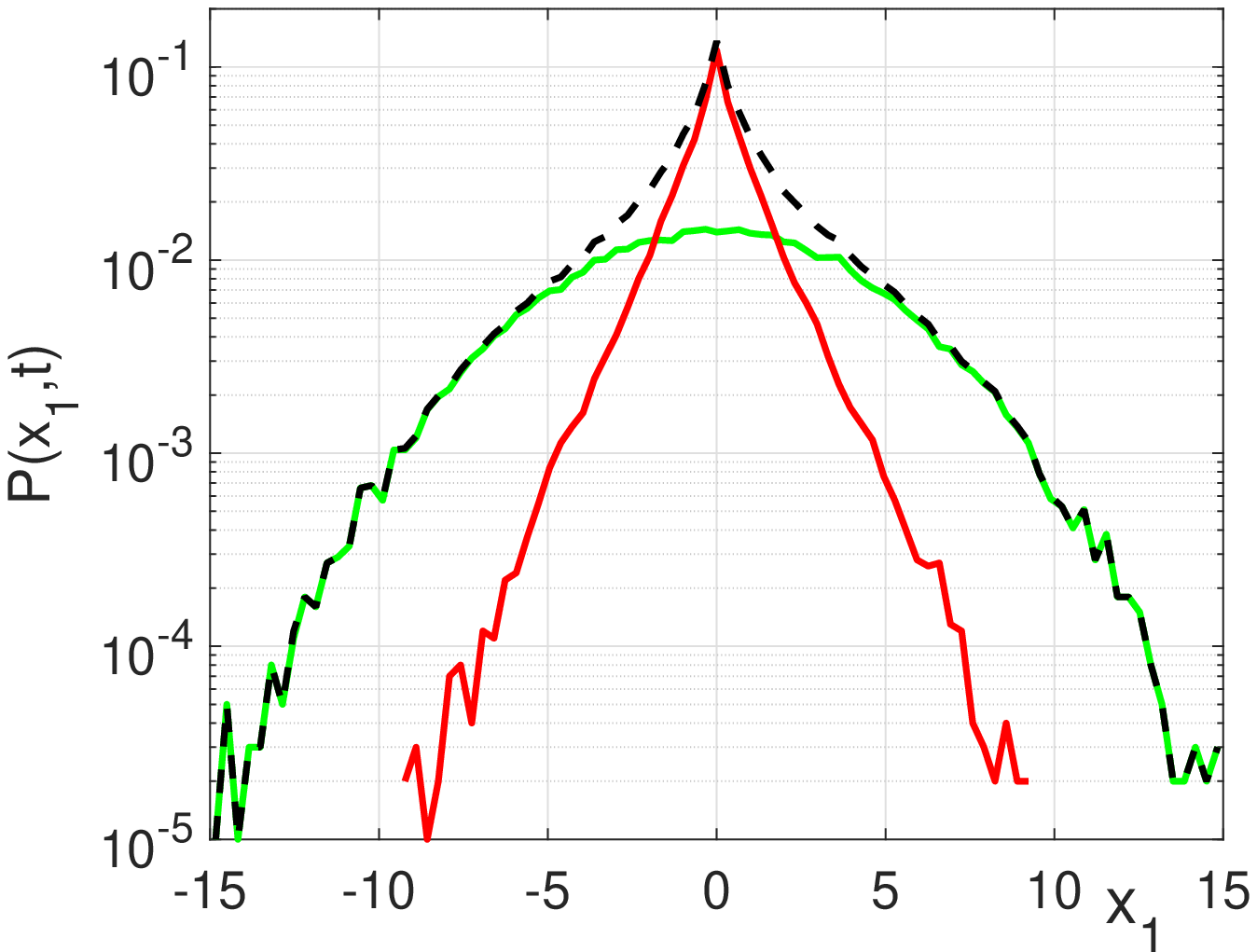}
\hspace{0.5cm}
\includegraphics[height=5.5cm]{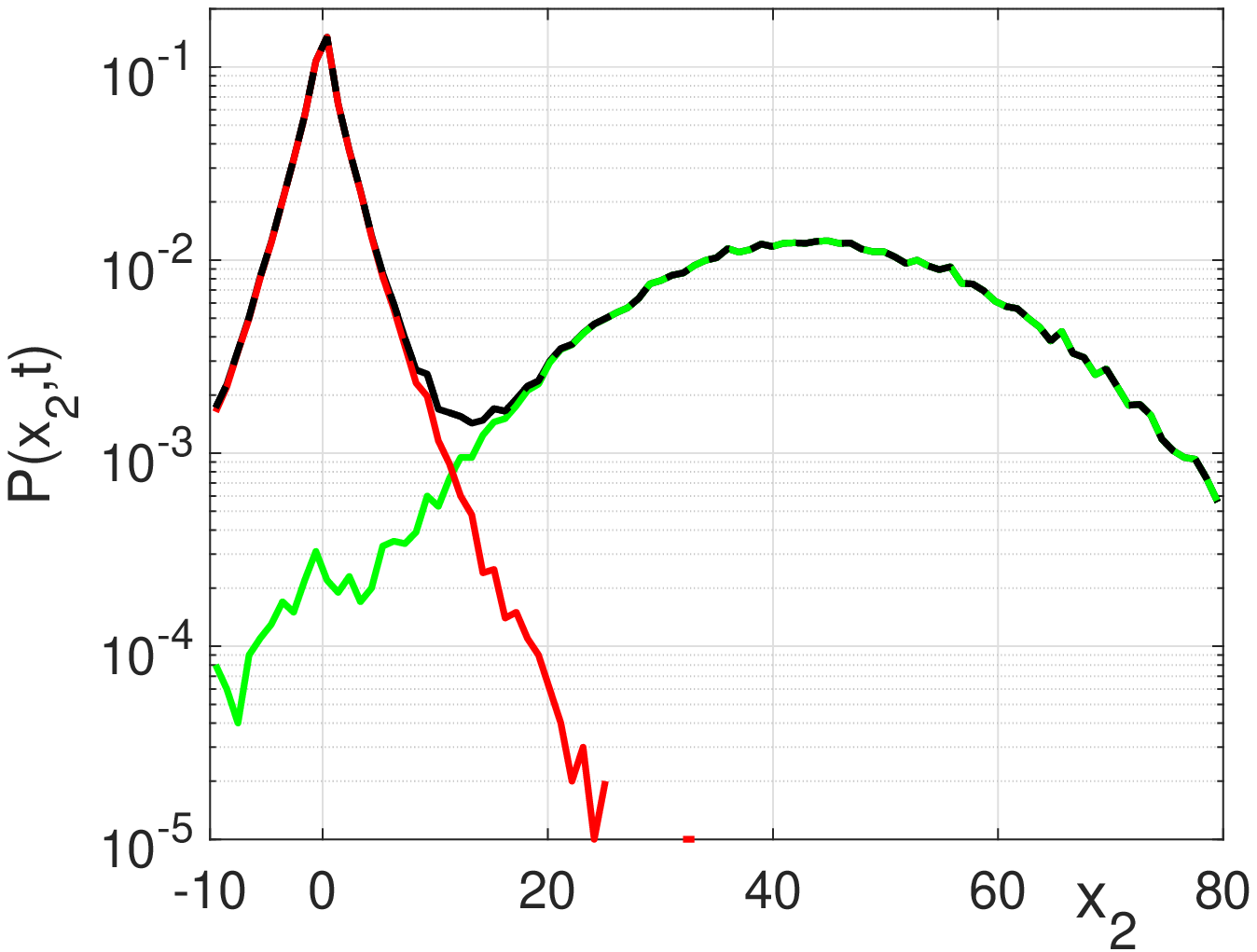}
}
\caption{The probabilities of the $x_1$ (left panel) and $x_2$ (right panel)
for the whole set of realizations (black lines) compared to the
contributions of the free (green lines) and trapped (red lines) trajectories
for $V_d=0.2$ and $t=97.$ }
\label{Px}
\end{figure}


The probability of the displacements $P(\mathbf{x},t)$ is strongly
non-Gaussian, as seen in in Fig. \ref{Px} (black curves). The contributions
of the two categories of trajectories are completely different: the trapped
trajectories determine the steep peak in $\mathbf{x}=\mathbf{0}$, and the
free ones have a large Gaussian distribution with the average parallel
displacement $\left\langle x_{2}(t)\right\rangle _{fr}=V_{d}t/n_{fr}.$ We
note that the transport, which is essentially produced by the free
trajectories, results from a Gaussian distribution.

\section{5. Coherence induced by an average velocity}

The Hamiltonian structure of equation (\ref{eqm}) is the origin of the order
that characterizes the two-dimensional incompressible turbulence. It
determines the strong connection between trajectories and the contour lines
of the potential, which are paths of the motion. The order (quasi-coherence)
of the motion is essentially represented by the existence of correlations
between the potential and the trajectories. They are represented by nonzero
average displacements or velocities conditioned by the (initial) potential.

\bigskip

Significant quasi-coherent characteristics of the transport process can be
found by analyzing statistical Lagrangian quantities restricted on the
contour lines with given potential $\phi ^{0}.$ The trajectories that belong
to this class correspond to solutions of Eq. (\ref{eqm}) that start (in $%
\mathbf{x}(0)=\mathbf{0}$) from points with $\phi (\mathbf{0})=\phi ^{0}.$
The invariance of the total potential in this class gives%
\begin{equation}
\phi _{t}(\mathbf{x}(t))=\phi (\mathbf{x}(t))+x_{1}(t)V_{d}=\phi ^{0}.
\label{fitinv}
\end{equation}

The fractions of trajectories that evolve on the $\phi ^{0}$ potential
lines, the average and the amplitude of fluctuations of their displacements
and Lagrangian velocities are determined below for each type of trajectories
using conditional averages.

\bigskip

The analysis starts from the representation (\ref{cond-type}) and introduces
a supplementary condition for the trajectories, namely that the initial
potential is $\phi ^{0}$ $[\phi (\mathbf{x}(0))=\phi ^{0}].$ Defining the
fraction of these trajectories by $n(\phi ^{0})$ and the corresponding
conditional average by $\left\langle {}\right\rangle _{\phi ^{0}},$ the
average\ $\left\langle A(\mathbf{x}(t))\right\rangle $ is the sum of the
contributions from each value $\phi ^{0}$\ 
\begin{equation}
\left\langle A(\mathbf{x}(t))\right\rangle =\int_{-\infty }^{\infty
}\left\langle A(\mathbf{x}(t))\right\rangle _{\phi ^{0}}\ P(\phi ^{0})~d\phi
^{0},  \label{fi0cond}
\end{equation}%
where $P\left( \phi ^{0}\right) $ is the Gaussian distribution of the
(normalized) potential 
\begin{equation}
P\left( \phi ^{0}\right) =\frac{1}{\sqrt{2\pi }}\exp \left( -\frac{\left(
\phi ^{0}\right) ^{2}}{2}\right) .  \label{nfi0}
\end{equation}%
Similar equations can be written for the contributions of the free and
trapped trajectories%
\begin{equation}
n_{c}~\left\langle A(\mathbf{x}(t))\right\rangle _{c}=\int_{-\infty
}^{\infty }\left\langle A(\mathbf{x}(t))\right\rangle _{\phi
^{0},c}~n^{c}(\phi ^{0})~d\phi ^{0},  \label{fi0condc}
\end{equation}%
where $n^{c}(\phi ^{0})$\ is the fraction of trajectories that evolve on
contour lines $\phi ^{0}$\ and are in the category $c=tr,\ fr,$\ and $%
\left\langle {}\right\rangle _{\phi ^{0},c}$\ is the conditional average
taken on the subensemble of these trajectories. $n^{c}(\phi ^{0})$ is
related to $n_{tr},$ $n_{fr}$ (defined in Section 3) 
\begin{equation}
n_{c}=\int_{-\infty }^{\infty }n^{c}(\phi ^{0})\ d\phi ^{0}.  \label{n-fi0}
\end{equation}%
One obtains using Eq. (\ref{cond-type})

\begin{equation}
\left\langle A(\mathbf{x}(t))\right\rangle _{\phi ^{0}}P(\phi
^{0})=\left\langle A(\mathbf{x}(t))\right\rangle _{\phi ^{0},tr}\
n^{tr}(\phi ^{0})+~\left\langle A(\mathbf{x}(t))\right\rangle _{\phi
^{0},fr}\ n^{fr}(\phi ^{0}),  \label{cond-fi0}
\end{equation}%
which connects the contribution of all trajectories $\left\langle
A\right\rangle _{\phi ^{0}}P(\phi ^{0})~$ to the contributions of each
category $\left\langle A\right\rangle _{\phi ^{0},c}n^{c}(\phi ^{0}).$

\ 


\begin{figure}[tbh]
\centerline{\includegraphics[height=6.0cm]{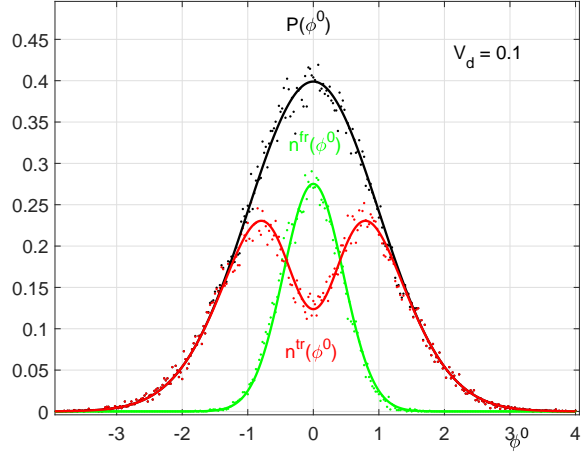}}
\caption{The fraction of the trajectories that evolve on the contour lines
with potential $\protect\phi^0$ for the free (green points), trapped (red
points) and for all trajectories (black points) obtained from the numerical
simulations compared to $P(\protect\phi^0)$ (solid black line), with Eq. (%
\protect\ref{nfifr}) (solid green line) and Eq. (\protect\ref{nfit}) (solid
red line).}
\label{nfi}
\end{figure}


\begin{figure}[tbh]
\centerline{\includegraphics[height=6.0cm]{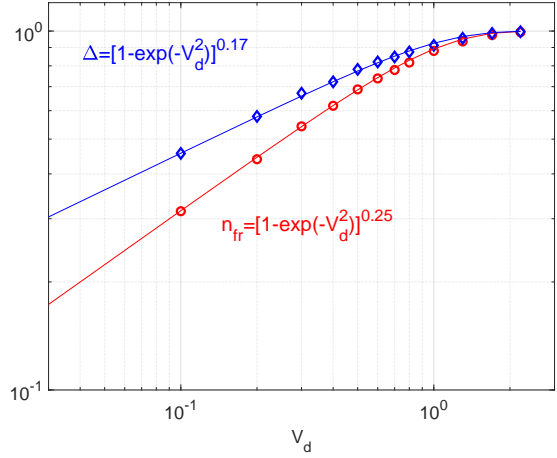}} .
\caption{The width $\Delta$ of the initial potential of the free
trajectories (diamonds) and the fraction of free trajectories (circles) as
functions of the average velocity $V_d$. The numerical results are
interpolated by Eqs. (\protect\ref{Delta}) and (\protect\ref{nfr})}
\label{nfrDelta}
\end{figure}


The fractions of trajectories fulfil the equation%
\begin{equation}
P(\phi ^{0})=n^{tr}(\phi ^{0})+n^{fr}(\phi ^{0}),  \label{P-nfi0}
\end{equation}%
\ which is obtained from Eq. (\ref{cond-fi0}) for $A=1.$ The numerical
results obtained for $n^{fr}(\phi ^{0})$ and $n^{tr}(\phi ^{0})$\
(represented by points), are compared to analytical approximations (solid
lines) in Fig. \ref{nfi}. One can see that the fraction of trajectories that
evolve on $\phi ^{0}$ contour lines (black line and points in Fig. \ref{nfi}%
) reproduces Eq. (\ref{nfi0}). The fraction of the free trajectories is
narrower, but it is still Gaussian. We have found, as seen in Fig. \ref{nfi}
(green curve), a good approximation of the data by 
\begin{equation}
n^{fr}(\phi ^{0})=n_{fr}~G(\phi ^{0};\Delta ),  \label{nfifr}
\end{equation}%
where $G(\phi ^{0};\Delta )$ is the Gaussian distribution 
\begin{equation}
G(\phi ^{0};\Delta )=\frac{1}{\sqrt{2\pi }\Delta }\exp \left( -\frac{\left(
\phi ^{0}\right) ^{2}}{2\Delta ^{2}}\right)  \label{G}
\end{equation}%
with a width $\Delta $ that depends on the average velocity $V_{d}.$ The
fraction of the trapped trajectories (red curve in Fig. \ref{nfi}), which
according to Eq. (\ref{P-nfi0}) is 
\begin{equation}
n^{tr}(\phi ^{0})=P(\phi ^{0})-n_{fr}~G(\phi ^{0};\Delta ),  \label{nfit}
\end{equation}%
provides a good representation of the numerical results (red points).

The width $\Delta $ as function of the average velocity $V_{d}$\ is shown in
Fig. \ref{nfrDelta} together with the fraction of free trajectories $n_{fr}.$
The numerical results for $\Delta (V_{d})$ (diamonds) are well approximated
by%
\begin{equation}
\Delta (V_{d})=\left[ 1-\exp \left( -V_{d}^{2}\right) \right] ^{0.17}.
\label{Delta}
\end{equation}%
Both functions saturate at large $V_{d}$ ($V_{d}>V_{1},V_{2}),$ and they
have power law dependence for small $V_{d}$ 
\begin{equation}
n_{fr}\sim V_{d}^{0.5},\ \Delta \sim V_{d}^{0.34}.  \label{scaling}
\end{equation}%
The asymptotic value $n_{fr}\rightarrow 1$ for $V_{d}\rightarrow \infty $
corresponds to the complete elimination of the islands of close contour
lines of the potential ($n_{tr}=0)$. The limit $\Delta \rightarrow 1$
confirms that all trajectories are free, because $G(\phi ^{0};1)=P(\phi
^{0}),$ where $P(\phi ^{0})$ is the probability of the Eulerian potential.

Thus, the free trajectories are localized on the contour lines with small
values of $\left\vert \phi ^{0}\right\vert \lesssim \Delta .$ The potential
on the geometrical locus of free trajectories is Gaussian with an amplitude
that is smaller than in the whole space.

The trapped (periodic) trajectories mainly have large $\left\vert \phi
^{0}\right\vert $ : they completely occupy the range of large potential $%
\left\vert \phi ^{0}\right\vert \gg \Delta ,$ but also have significant
presence at small potential $\left\vert \phi ^{0}\right\vert \lesssim \Delta 
$ that correspond to free trajectories.\ \ 

\bigskip

\begin{figure}[tbh]
\centerline{
\includegraphics[height=5.5cm]{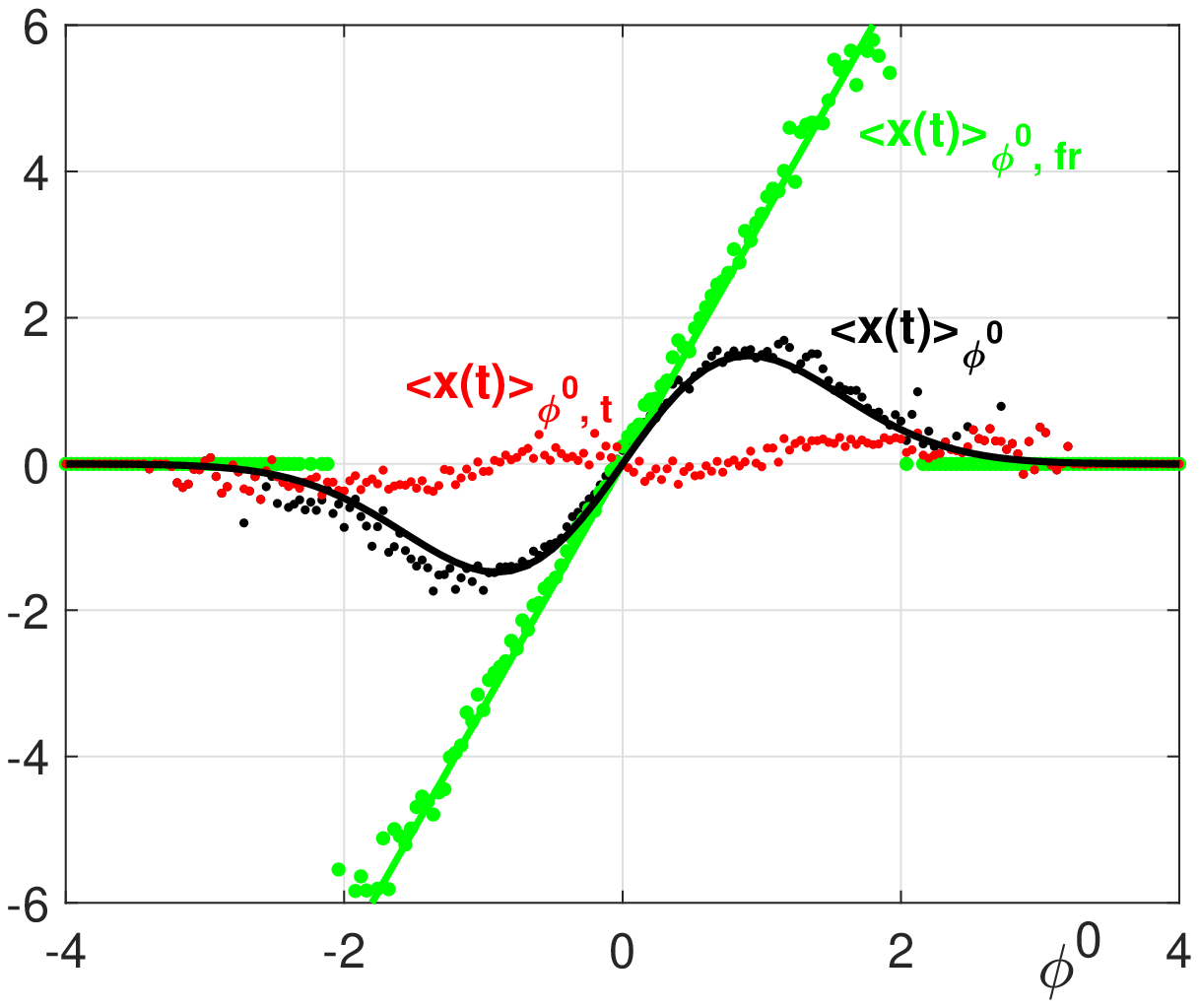} 
\hspace{0.5cm}
\includegraphics[height=5.5cm]{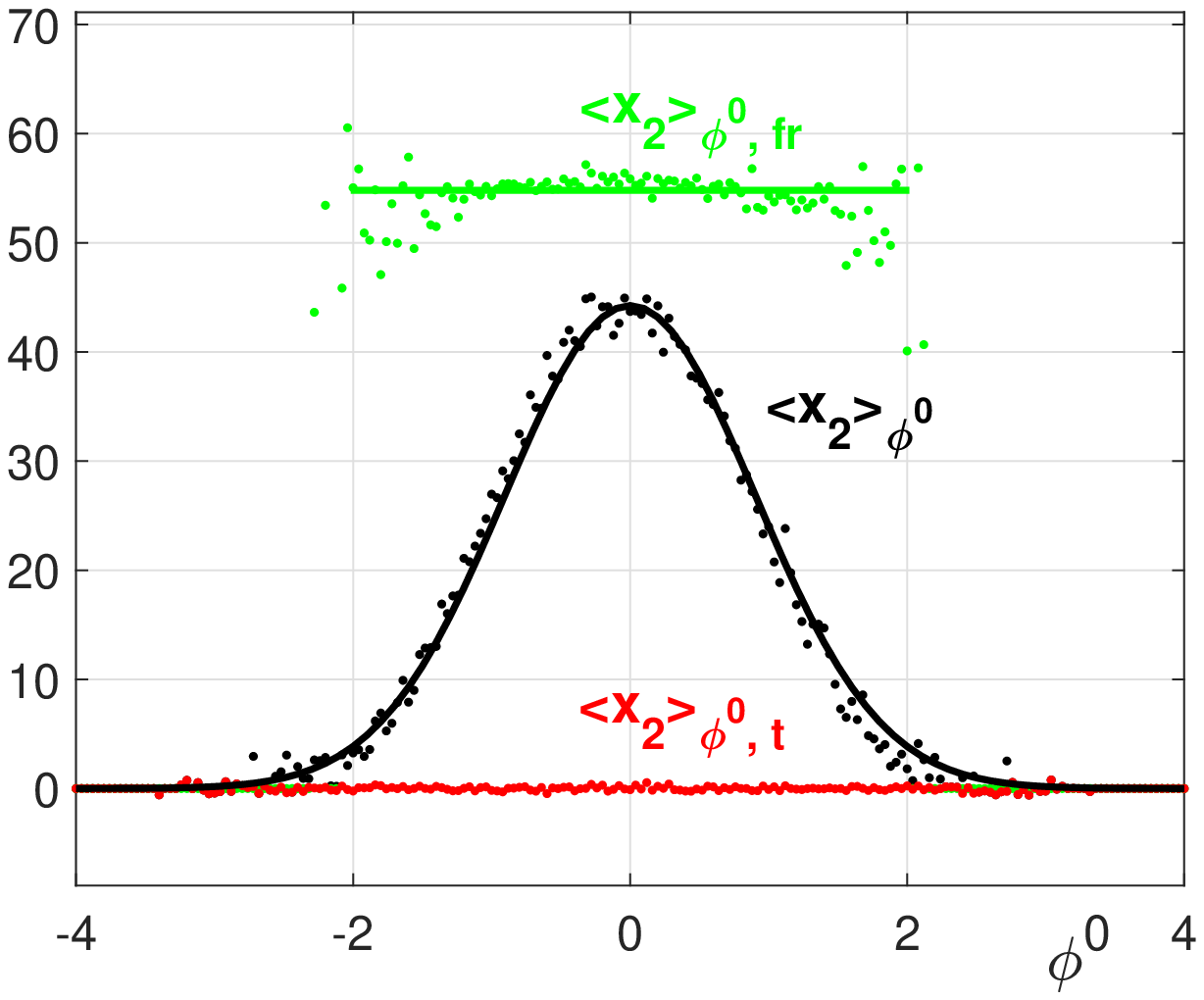}
}
\caption{The conditional average displacements along $x_1$ axis (left panel)
and along $x_2$ axis (right panel) as functions of $\protect\phi^0$ for the
trapped (red points), free (green points) and all (black points)
trajectories, compared to the approximations (\protect\ref{medcondx})-(%
\protect\ref{medcondy}) (green lines) and (\protect\ref{medcondxy}) (black
lines). $V_d=0.3.$ }
\label{xycond}
\end{figure}

The average displacements conditioned by the value of the initial potential
and by the category of the trajectories are shown in Fig. \ref{xycond} for
free (green), trapped (red) and all (black) trajectories. The perpendicular
(left panel) and the parallel (right panel) displacements are shown at a
large time $t=97,$ larger than the saturation time of $n_{fr}(t,V_{d})$
(seen in Fig. \ref{ntrfrvm}, left panel).\ These represent quasi-coherent
components of the motion, and appear only in the presence of an average
velocity. One can see that the average conditional displacements are small
for the trapped trajectories (red points), and that significant values
appear for the free trajectories in both directions (green points). As shown
in Fig. \ref{xycond}, these quantities can be approximated by%
\begin{equation}
\left\langle x_{1}(t)\right\rangle _{\phi ^{0},fr}=\frac{\phi ^{0}}{V_{d}},\
\ \left\langle x_{1}(t)\right\rangle _{\phi ^{0},tr}\cong 0,\ \ 
\label{medcondx}
\end{equation}%
\begin{equation}
\left\langle x_{2}(t)\right\rangle _{\phi ^{0},fr}=\frac{V_{d}t}{n_{fr}},\ \
\left\langle x_{2}(t)\right\rangle _{\phi ^{0},tr}=0,  \label{medcondy}
\end{equation}%
which are represented by red and green lines respectively. The black lines
have the equations%
\begin{equation}
\left\langle x_{1}(t)\right\rangle _{\phi ^{0}}=\frac{\phi ^{0}}{V_{d}}%
F(\phi ^{0}),\ \ \left\langle x_{2}(t)\right\rangle _{\phi ^{0}}=\frac{V_{d}t%
}{n_{fr}}F(\phi ^{0}),  \label{medcondxy}
\end{equation}%
where%
\begin{equation}
F(\phi ^{0})=\frac{n^{fr}(\phi ^{0})}{P(\phi ^{0})}=\frac{n_{fr}}{\Delta }%
\exp \left( -\frac{\left( \phi ^{0}\right) ^{2}}{2}\frac{1-\Delta ^{2}}{%
\Delta ^{2}}\right) .  \label{F}
\end{equation}%
They result from Eq. (\ref{cond-fi0}) with $A=x_{i}(t)$ using (\ref{medcondx}%
) and (\ref{medcondy}), and provide, as seen in Fig. \ref{xycond}, good
approximations of the data for $\left\langle x_{i}(t)\right\rangle _{\phi
^{0}}$ (black points).


\begin{figure}[tbh]
\centerline{
\includegraphics[height=5.5cm]{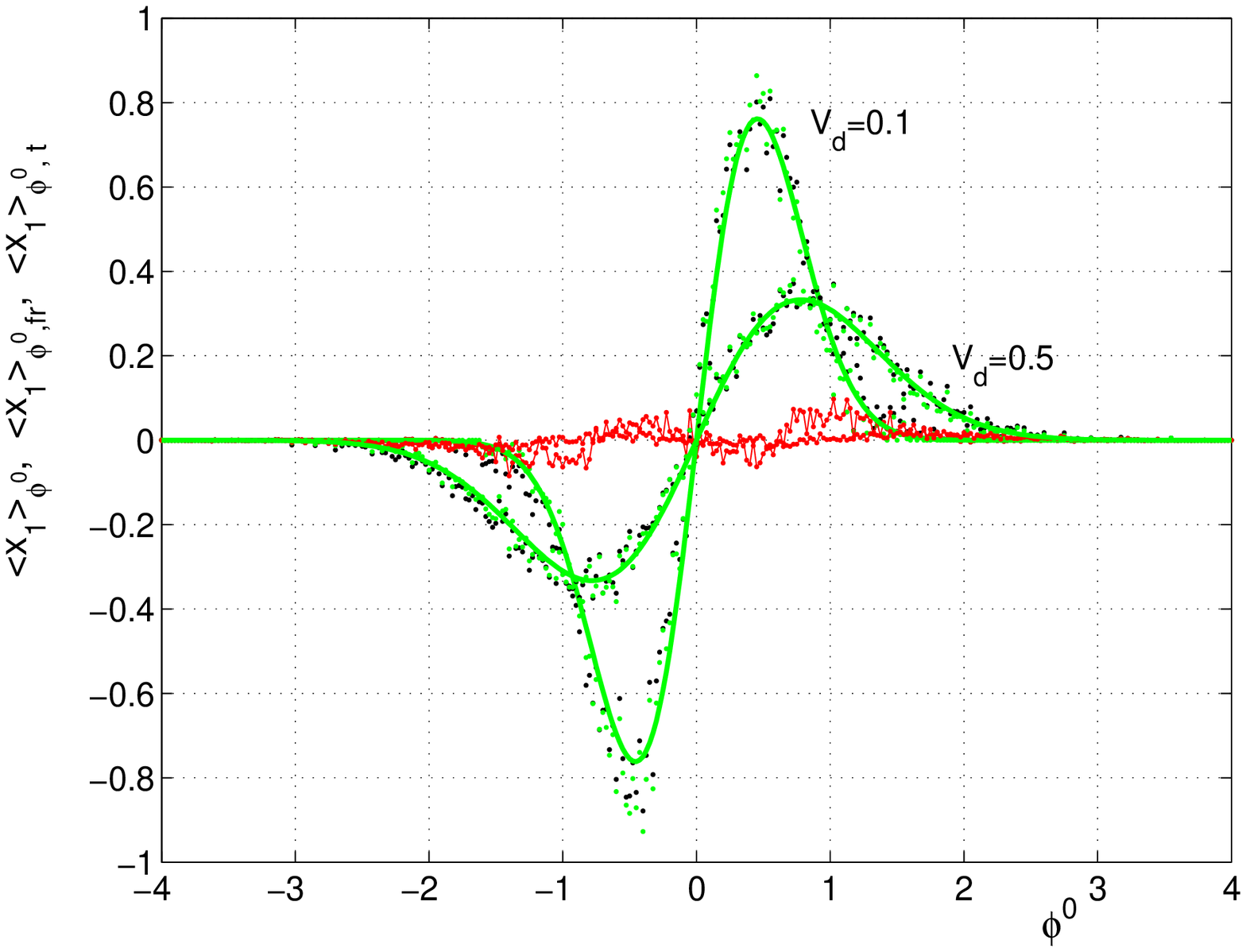} 
\hspace{0.5cm}
\includegraphics[height=5.5cm]{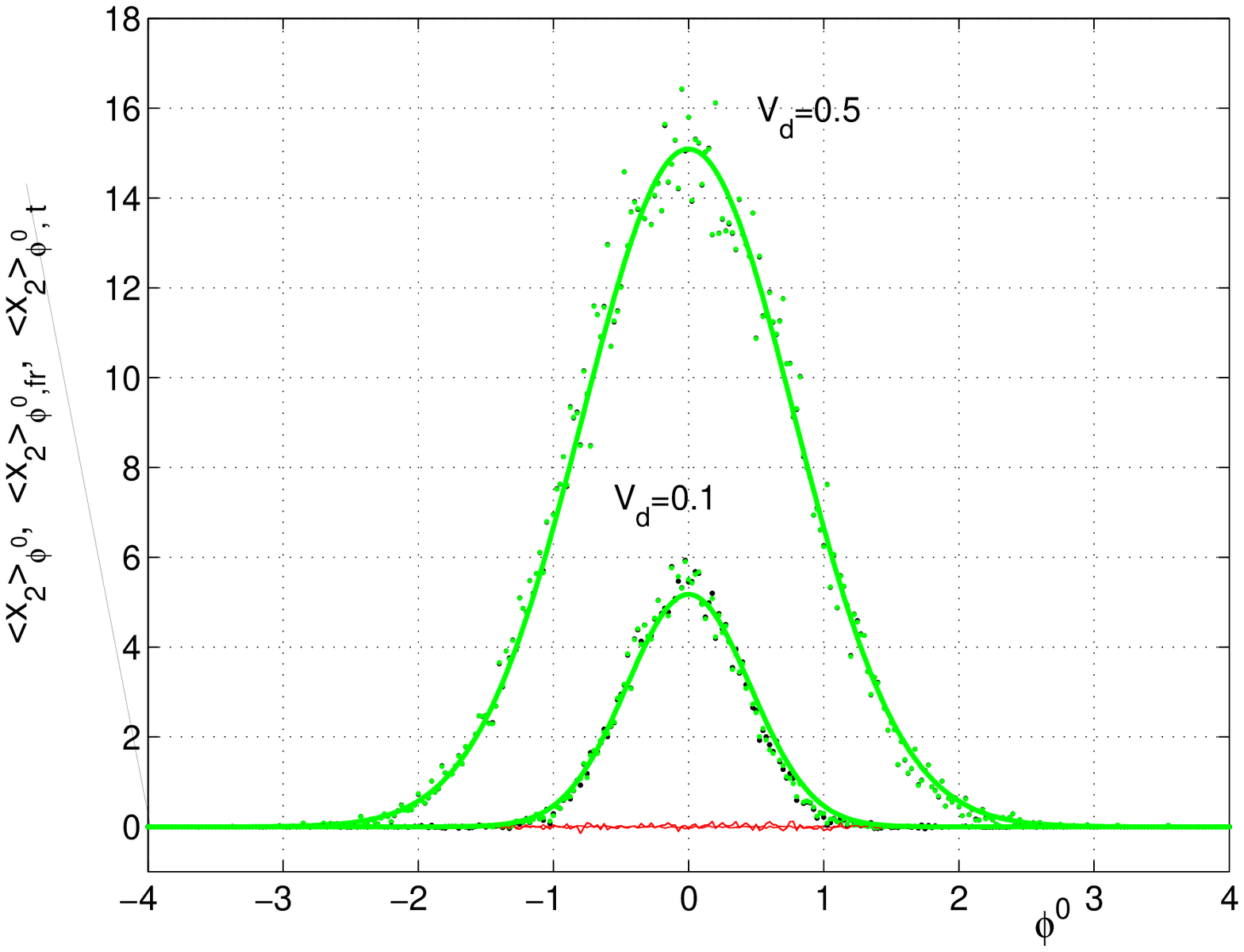}
}
\caption{ The contributions of the trapped (red points) and free (green
points) trajectories, to the average displacements (black points) along $x_1$
(left panel) and $x_2$ (right panel) directions as functions of $\protect\phi%
^0$ for the values of $V_d$ label the curves. The approximations for the
free trajectories are represented by the green lines.}
\label{xycontrib}
\end{figure}


The contributions to the average displacements, obtained by multiplying the
conditional averages with the corresponding fractions of trajectories ($%
P(\phi ^{0}),$ $n^{fr}(\phi ^{0})$ or $n^{tr}(\phi ^{0})$), are shown in
Fig. \ref{xycontrib}. It appears more clearly that the coherent
displacements are produced only by the free trajectories. The black points
for all trajectories are practically superposed on the green points and they
are well approximated by the green lines that represent the contributions of
the free trajectories ($\phi ^{0}/V_{d}\ n^{fr}(\phi ^{0})$ in the left
panel and $V_{d}~t/n_{fr}\ n^{fr}(\phi ^{0})$ in the right panel). The
dependence on $\phi ^{0}$ is different across and along the average velocity 
$\mathbf{V}_{d}.$ In the first case, it is an anti-symmetrical function of $%
\phi ^{0}$ that saturates in time, and, in the second case, it is a
symmetrical Gaussian function that increases linearly in time.

The parallel average displacement on the contour lines with initial
potential $\phi ^{0},$ which increases linearly with $t$ (\ref{medcondy}),
leads to an average Lagrangian velocity 
\begin{equation}
\left\langle v_{2}(t)\right\rangle _{\phi ^{0},fr}=\frac{V_{d}}{n_{fr}}.
\label{v2med}
\end{equation}%
It is important to note that this velocity does not depend on $\phi ^{0},$
and it equals the average velocity (\ref{vfr}). The contribution of the
conditional average velocity is determined only by the free trajectories
since $\left\langle v_{2}(t)\right\rangle _{\phi ^{0},tr}=0.$ The
perpendicular average displacement of the free trajectories also determines
an average velocity, but it is transitory since $\left\langle
x_{1}(t)\right\rangle _{\phi ^{0},fr}$ saturates in time.

\begin{figure}[tbh]
\centerline{
\includegraphics[height=5.5cm]{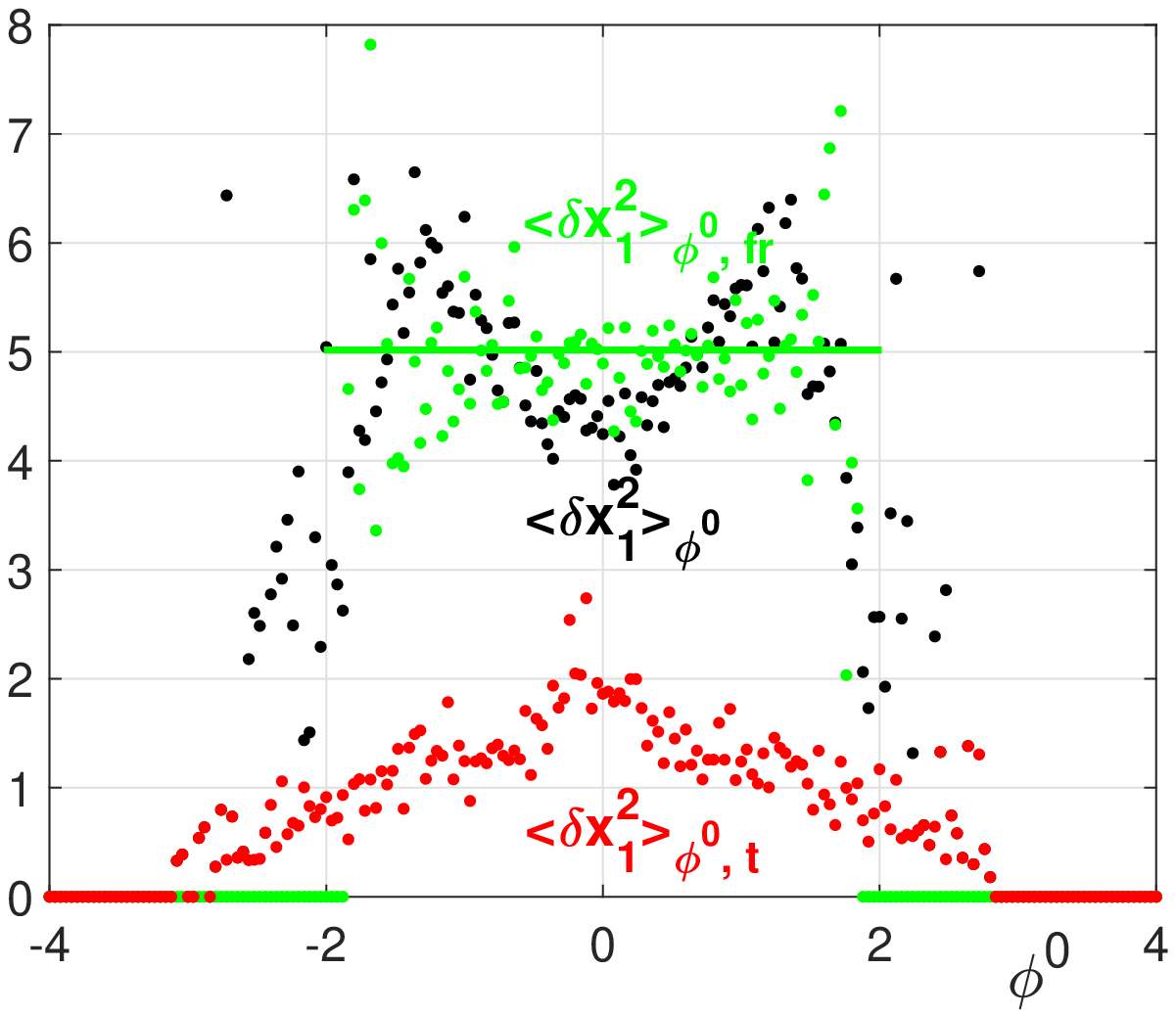} 
\hspace{0.5cm}
\includegraphics[height=5.5cm]{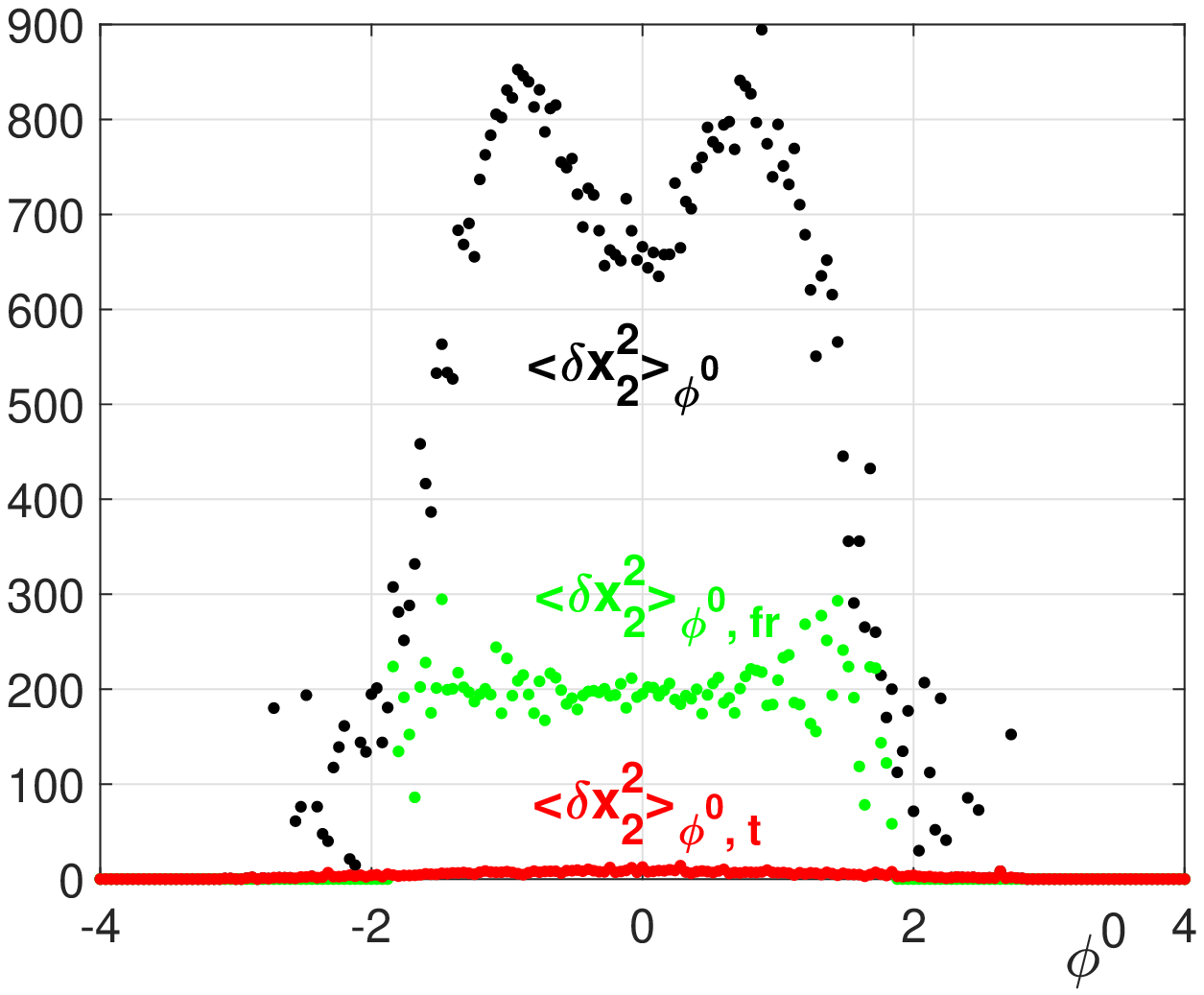}
}
\caption{The conditional trajectory dispersion along $x_1$ axis (left panel)
and along $x_2$ axis (right panel) as functions of $\protect\phi^0$ for the
trapped (red points), free (green points) and all (black points)
trajectories, compared to the approximations (\protect\ref{dispfr}) (green
lines). $V_d=0.3$ and $t=97$. }
\label{dx2dy2cond}
\end{figure}

The dispersion of the trajectories conditioned by the value of the initial
potential and by the category are shown in Fig. \ref{dx2dy2cond} for free
(green), trapped (red) and all (black) trajectories, in the perpendicular
(left panel) and parallel (right panel) directions. One can see that the
dispersion of the free trajectories along both directions are not dependent
on the initial potential $\phi ^{0},$ and can be approximated by 
\begin{equation}
\left\langle \delta x_{1}^{2}\right\rangle _{\phi ^{0},fr}=\frac{\Delta ^{2}%
}{V_{d}^{2}},\ \ \left\langle \delta x_{2}^{2}\right\rangle _{\phi
^{0},fr}=2\ D_{2}^{fr}t  \label{dispfr}
\end{equation}%
represented by the green lines. On the contrary, the trapped trajectories
have dispersions that decay with the increase of $\phi ^{0}$ (the red
points). The dispersion of the trajectories conditioned by the potential is
obtained using Eq. (\ref{cond-fi0}) for $A=x_{i}^{2}(t)$ 
\begin{equation}
\left\langle \delta x_{i}^{2}\right\rangle _{\phi ^{0}}=\left\langle \delta
x_{1}^{2}\right\rangle _{\phi ^{0},fr}F+\left\langle \delta
x_{1}^{2}\right\rangle _{\phi ^{0},tr}(1-F)+\left\langle
x_{i}(t)\right\rangle _{\phi ^{0},fr}^{2}F(1-F),  \label{dispfi0}
\end{equation}%
which depends on $\phi ^{0}$ as seen in Fig. \ref{dx2dy2cond} (black points).


\emph{\ }

Thus, the analysis of the Lagrangian statistics conditioned by the initial
potential reveals a coherent component of motion perpendicular to $\mathbf{V}%
_{d}$. It consists of average displacements that have the sign correlated
with the sign of $\phi ^{0}.$ They appear for the free trajectories and are
hidden in the sense that the contributions of all contour lines (with all
values of $\phi ^{0})$ mix to zero. The amplitude of the ordered motion is
defined by the displacements conditioned by the sign of $\phi ^{0},$ $%
\left\langle x_{1}(t)\right\rangle _{+}$\ obtained by integration over $\phi
^{0}$\ on the interval $[0,\infty )$ and $\left\langle x_{1}(t)\right\rangle
_{-}$\ that is the integral over $(-\infty ,0].$\ These are symmetrical
quantities since $\left\langle x_{1}(t)\right\rangle =\left\langle
x_{1}(t)\right\rangle _{+}+\left\langle x_{1}(t)\right\rangle _{-}=0.$ The
time derivatives of these functions determine a pair of average velocities
with opposite directions that exactly compensate each other (the hidden
drifts, HDs) that are oriented across $\mathbf{V}_{d}$. The HDs were first
found in \cite{VS18} using an approximate theoretical approach, the
decorrelation trajectory method \cite{V98}. In the presence of components of
the motion that introduce a small compressibility, an average velocity can
be generated by breaking the equilibrium of the HDs. Such effects were found
in magnetically confined plasmas \cite{VS18b}-\cite{VPS21}. The HDs are
transitory in frozen potentials because the average displacements saturate $%
\left\langle x_{1}(t)\right\rangle _{\phi ^{0},fr}\rightarrow \phi
^{0}/V_{d}.$

The analysis also shows that the parallel motion is similar on the contour
lines with different $\phi ^{0},$ and that it depends only on the category
of trajectories. The asymptotic value $\left\langle x_{1}(t)\right\rangle
_{\phi ^{0},fr}\rightarrow \phi ^{0}/V_{d}$ represents the centrum of the
space domain that contains the trajectories, which start from points on the
line $x_{1}=0$ where the potential is $\phi ^{0}.$ It is limited by the
lines $x_{1}^{-}=(\phi ^{0}-\Delta )/V_{d},$ $x_{1}^{+}=(\phi ^{0}+\Delta
)/V_{d},$ and has infinite dimension along $\mathbf{V}_{d}.$ Reported at
this centrum, the free trajectories are statistically identical for all
values of the initial potential $\phi ^{0}.$

\section{6. Lagrangian statistics in time dependent potentials}

The general conclusion of the analysis of trajectory statistics in frozen
potentials is the existence of a high degree of coherence, which reflects
the structure of the contour lines of $\phi _{t}(\mathbf{x})$ on which the
trajectories are bounded.

The time-dependence of the potential determines the variation of the
Lagrangian potential and the decorrelation from its initial value $\phi ^{0}$%
. It is expected to strengthen the random aspects of the trajectories and to
cause the elimination of the Lagrangian coherence in a time of the order of
the decorrelation time $\tau _{c}$. More precisely, the random
time-variation of the potential should vanish the averages and correlations
conditioned by $\phi ^{0},$ which show the existence of hidden order. It is
thus expected that the order found in static potential is in this case only
a transitory processes with life-time $\tau _{c}.$

The trajectories are more complex than in static potentials. Closed periodic
trajectories do not exist in time dependent potentials, but trapping events
represented by almost closed eddying segments appear on all trajectories
when the decorrelation time $\tau _{c}$ is large compared to the time of
flight $\tau _{fl}=\lambda /V,$ ($\lambda =(\lambda _{1}^{2}+\lambda
_{2}^{2})^{1/2}$ and $V=(V_{1}^{2}+V_{2}^{2})^{1/2}),$\ and the integration
time is much longer than $\tau _{c}.$ The trapping events are separated by
long jumps, which are similar with the free trajectories.

The separation of the trajectories in the categories $c=tr,~fr$ has no
meaning in time-dependent potentials. However one can define related
quantities that are not properties of the trajectories but of the contour
lines of the potential. The latter are geometric objects. The fraction of
free/trapped trajectories can be defined using the number of trajectories
that stay on open/closed contour lines of the potential at time $t.$ These
fractions do not depend on time for stationary stochastic potentials,
because the amplitude, the space correlation and the structure of the
contour lines are statistically time-invariant. They equal the asymptotic
values of the time dependent fractions $n_{c}(t,V_{d})$ obtained in static
potentials from the trajectories (Sections 3)%
\begin{equation}
n_{c}(V_{d})=\ \underset{t\rightarrow \infty }{\lim }\ n_{c}(t,V_{d}),\ 
\label{fractd}
\end{equation}%
\ for any $\tau _{c}$ and $c=tr,fr.$ In a similar way, the fraction of
trajectories that stay at time $t$ on contour lines of category $c$ with the
potential $\phi ^{t}=\phi (\mathbf{x}(t))$ is a time-independent function of 
$\phi ^{t}$ and $c,$ which is the asymptotic value $n^{c}(\phi ^{t})$ of the
fractions obtained in static potential ($\tau _{c}\rightarrow \infty )$ in
Eqs. (\ref{nfifr}-\ref{nfit}). \ The physical meaning of these quantities
will be clarified after analyzing the significant modifications of the
statistics of trajectories produced by the time-variation of the potential.

We underline that, in time-dependent potentials, one can define the
statistics conditioned by the initial potential, but not by the category.

Our aim is to see if the special order determined by the average velocity
survives at finite $\tau _{c}.$ We analyze here the statistics on the whole
set of trajectories (in $R$), while, in the next section, the statistics
conditioned by the initial potential will be used for understanding the
direct effects of time variation on the coherent elements found here.

\bigskip

The time variation of $\phi $ represents a decorrelation mechanism, because
it determines the stochastic change of the Lagrangian velocity, which
vanishes the Lagrangian correlations $L_{i}(t)$ at times $t\gg \tau _{c.}$
Usually, this produces the saturation of the time dependent diffusion
coefficients $D_{i}(t)\rightarrow D^{i}$\ and diffusive transport with $%
\left\langle \delta x_{i}^{2}(t)\right\rangle \rightarrow 2D^{i}t$\ (as
obtained from Eq. (\ref{T})). The memory of the initial velocity is lost in
a time $\tau _{c},$ which means that the displacements at large time $t\gg
\tau _{c}$ are sequences of non-correlated random steps that yield a
Gaussian distribution.

We show that this general behavior is not at all observed in the presence of 
$\mathbf{V}_{d}$ at large correlation times (weak time variation). A strong
non-standard influence appears both on the transport and on the probability
of the displacements.\ 

The Lagrangian velocity is, as expected, Gaussian at any time and for any $%
\tau _{c}$, as in the static case. The time variation influences only its
correlation.

\begin{figure}[tbh]
\centerline{\includegraphics[height=6.0cm]{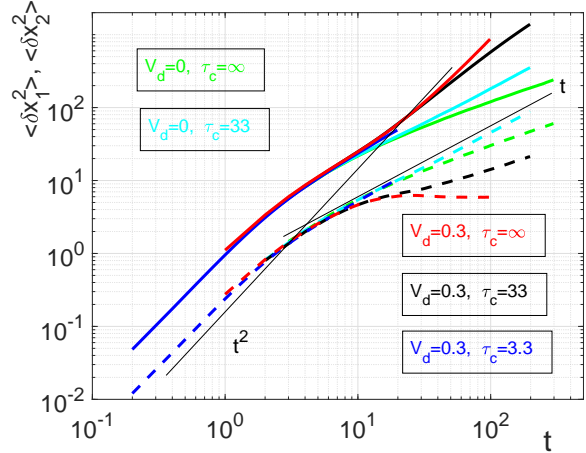}} .
\caption{Comparision of the dispersions of the trajectories in
time-dependent potential with the static cases along $x_1$ (dashed lines)
and $x_2$ (solid lines) directions. The values of $V_d$ and $\protect\tau_c$
label the curves. }
\label{MSD}
\end{figure}

\begin{figure}[tbh]
\centerline{
\includegraphics[height=5.5cm]{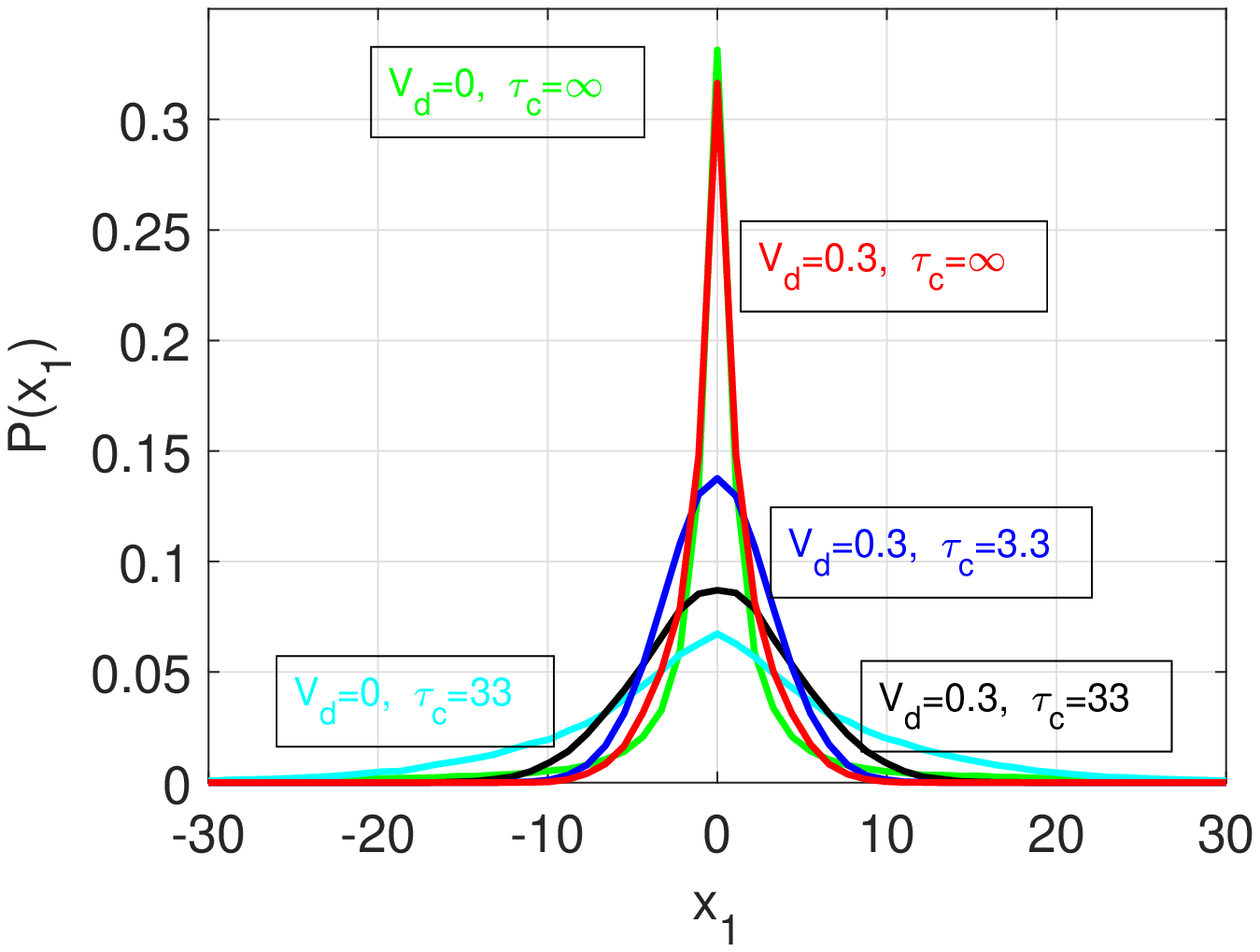} 
\hspace{0.5cm}
\includegraphics[height=5.5cm]{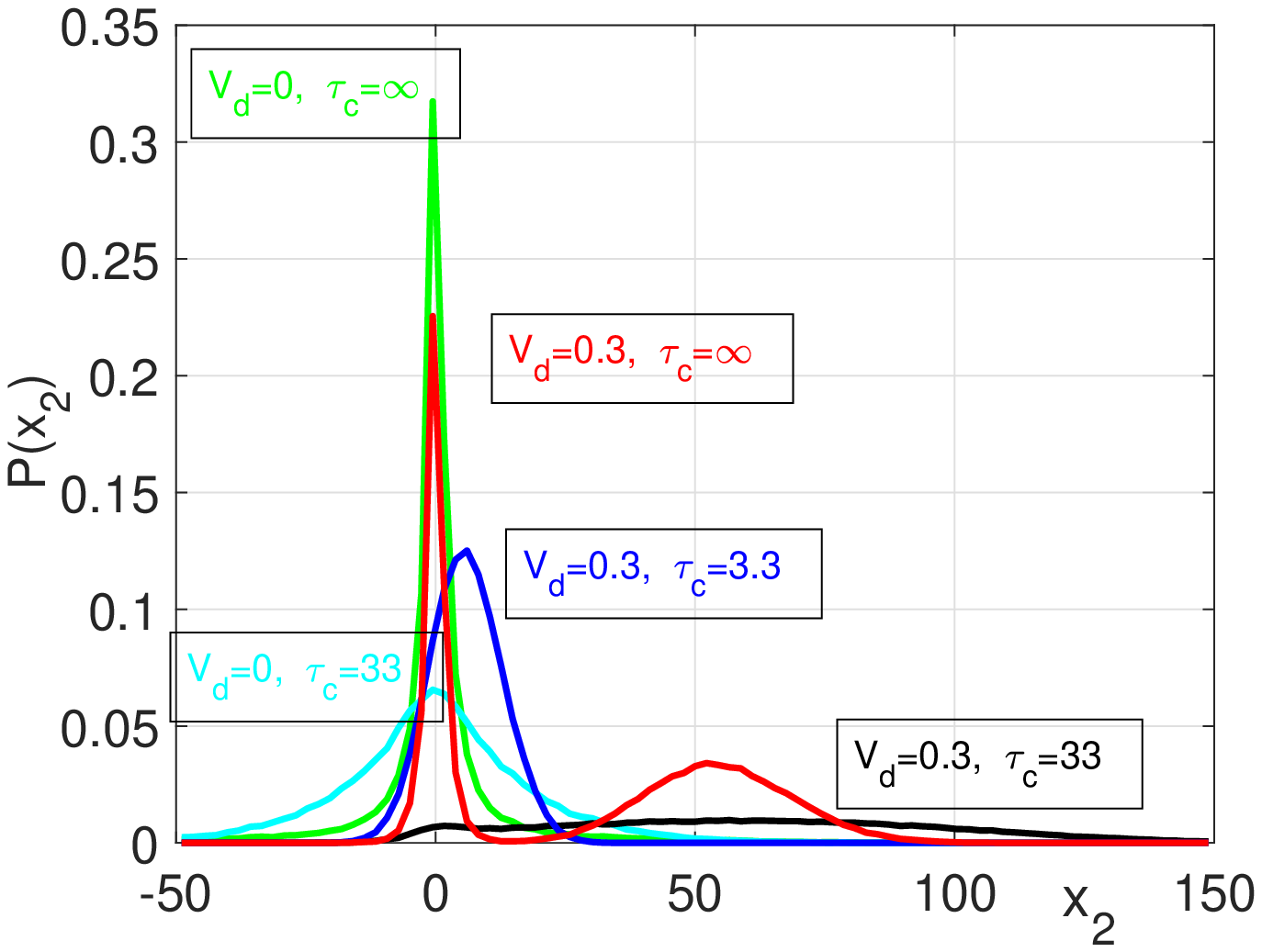}
}
\caption{The probabilities of the displacement along $x_1$ (left panel) and $%
x_2$ (right panel) directions, show the effect of the time variation of the
potential for typical cases with the values of $V_d$ and $\protect\tau_c$
that label the curves.}
\label{Pxycomp}
\end{figure}

The dispersions of the trajectories $\left\langle \delta
x_{i}^{2}(t)\right\rangle $ and the probabilities $P(x_{i},t)$\ for a
typical case that illustrates the effects of weka time variation of the
potential are shown in Figs. \ref{MSD} and \ref{Pxycomp}, by the black lines
for $V_{d}=0.3$ and $\tau _{c}=33$. We also present, for comparison, two
examples with $V_{d}=0.3$ (for the static case $\tau _{c}=\infty $\ (red
lines) and for fast time variation $\tau _{c}=3.3$\ (blue)), and two
examples with $V_{d}=0$ (for $\tau _{c}=\infty $ (green) and $\tau _{c}=33$
(cyan)).

When $V_{d}=0,$ the subdiffusive transport at $\tau _{c}=\infty $\ with $%
\left\langle \delta x_{i}^{2}(t)\right\rangle \sim t^{0.68}$\ is transformed
into normal transport ($\left\langle \delta x_{i}^{2}(t)\right\rangle
\rightarrow 2D^{i}t$) at large $\tau _{c}$\ (Fig. \ref{MSD}, green and cyan
curves). The process appears for all finite values of $\tau _{c},$ which
only influence the diffusion coefficients $D^{i}.$ However, the
probabilities of displacements are Gaussian only for fast time variation. In
the static case, $P(x_{i},t)$\ has a steep peak in $x_{i}=0,$ which
corresponds to trapped trajectories, superposed on a large Gaussian
component, which yields from the trajectories that are not closed at time $%
t. $ The steep peak is flattened by time variation and the probabilities
have extended exponential shapes at large $\tau _{c}.$ When $\tau _{c}$\
decreases, $P(x_{i},t)$\ evolves to Gaussian distribution, which is attained
when $\tau _{c}<\tau _{fl}$.

The average velocity makes the transport strongly anisotropic. In frozen
potential $\tau _{c}=\infty ,$ the transport is ballistic in the parallel
direction and saturated perpendicular to $\mathbf{V}_{d},$ as discusses in
Section 4.2 and also shown in Figs. \ref{MSD} and \ref{Pxycomp} (red
curves). The normal transport and the Gaussian probability are reached only
for fast time variation of the potential ($\tau _{c}<\tau _{fl}$), as seen
in the example for $\tau _{c}=3.3$ (blue curves). In these conditions, the
motion along the contour lines of $\phi (\mathbf{x},t)$ is completely
hindered, which means that the quasi-coherent components are eliminated.

Compared to these cases, the trajectory statistics at slow time variation in
the presence of $V_{d}$ (the black curves) is strongly anomalous with
complex behavior. Trajectory dispersion at large time has nonstandard
time-dependence in both directions%
\begin{equation*}
\left\langle \delta x_{1}^{2}(t)\right\rangle \sim t^{\alpha _{1}},\ \
\left\langle \delta x_{2}^{2}(t)\right\rangle \sim t^{\alpha _{2}},
\end{equation*}%
where $\alpha _{1}<1$\ and $\alpha _{2}>1,$\ which corresponds to
subdiffusive perpendicular transport (but not saturated as in the static
case) and superdiffusive parallel transport (but not of ballistic type).
These powers are functions of $\tau _{c}$ and $V_{d},$ $\alpha _{i}(\tau
_{c},V_{d}).$\ When $\tau _{c}$ decreases, $\alpha _{1}$ increases and
saturates $\alpha _{1}\rightarrow 1,$ while\ $\alpha _{2}$ decreases and
saturates $\alpha _{2}\rightarrow 1.$ A similar effect is determined by the
increase of $V_{d},$ which leads to normal transport at $V_{d}\gtrsim 1.$ In
the example presented in Fig. \ref{MSD} (black curves), $\alpha
_{1}(33,0.3)=0.57,$ $\alpha _{2}(33,0.3)=1.35.$ The probabilities are very
large, especially in the parallel direction, and the peak in $\mathbf{x}=0$
persists for very long time ($t=300=6\tau _{c}$ in Fig. \ref{Pxycomp}). 


Thus, the transport and the statistics of displacements are non-standard
when $V_{d}\lesssim 1$ and $\tau _{c}>\tau _{fl}.$ In these conditions the
structure of the contour lines of the potential shows island of closed lines
in a network of open lines. Also, the trajectories approximately follow the
contour lines for distances of the order of the correlation length before
they are removed by the time variation of the potential.

\section{7. Enhanced coherence and long memory effects}

\bigskip

\subsection{7.1 Hidden ordered motion}

\begin{figure}[tbh]
\centerline{
\includegraphics[height=5.5cm]{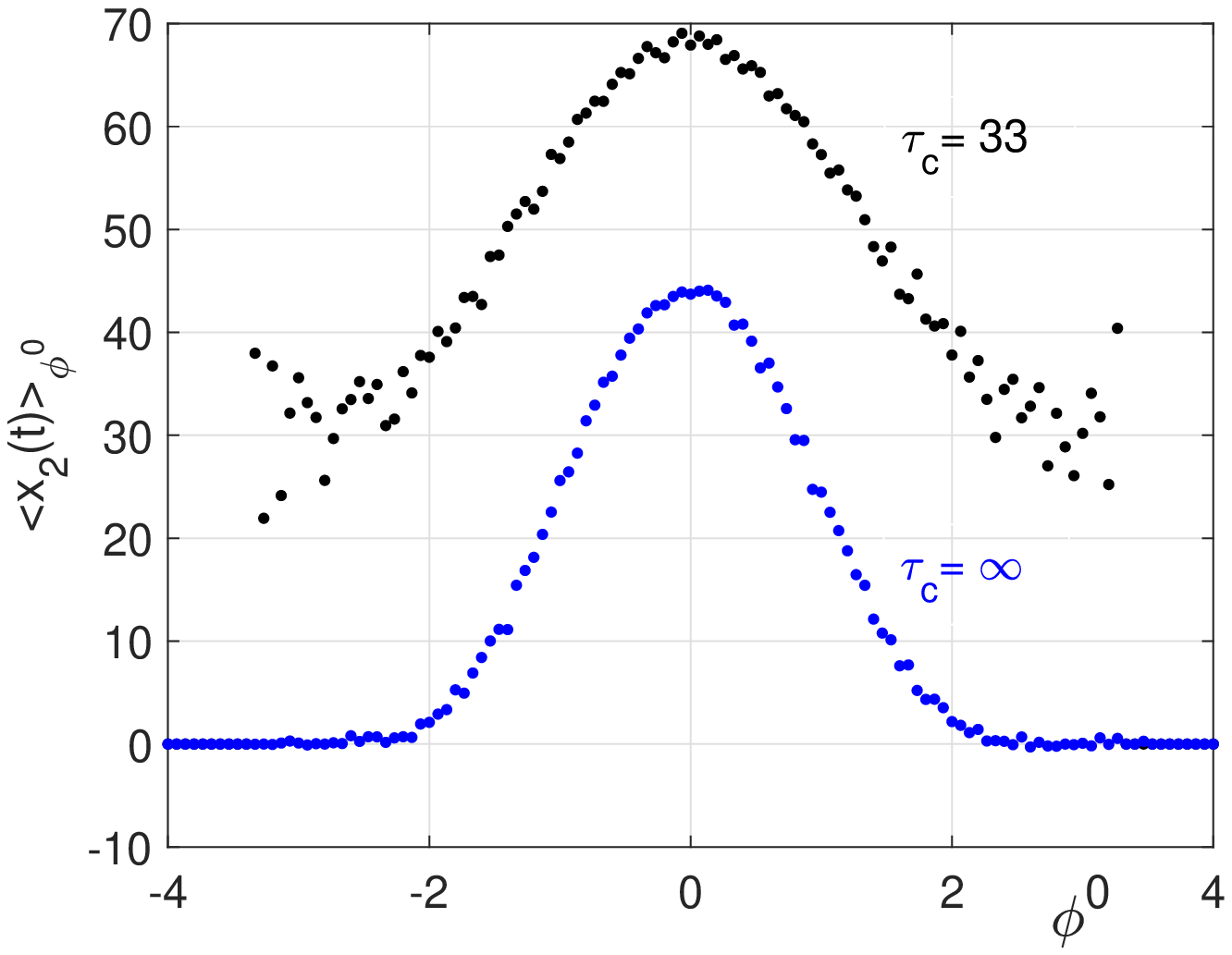}
\hspace{0.5cm}
\includegraphics[height=5.5cm]{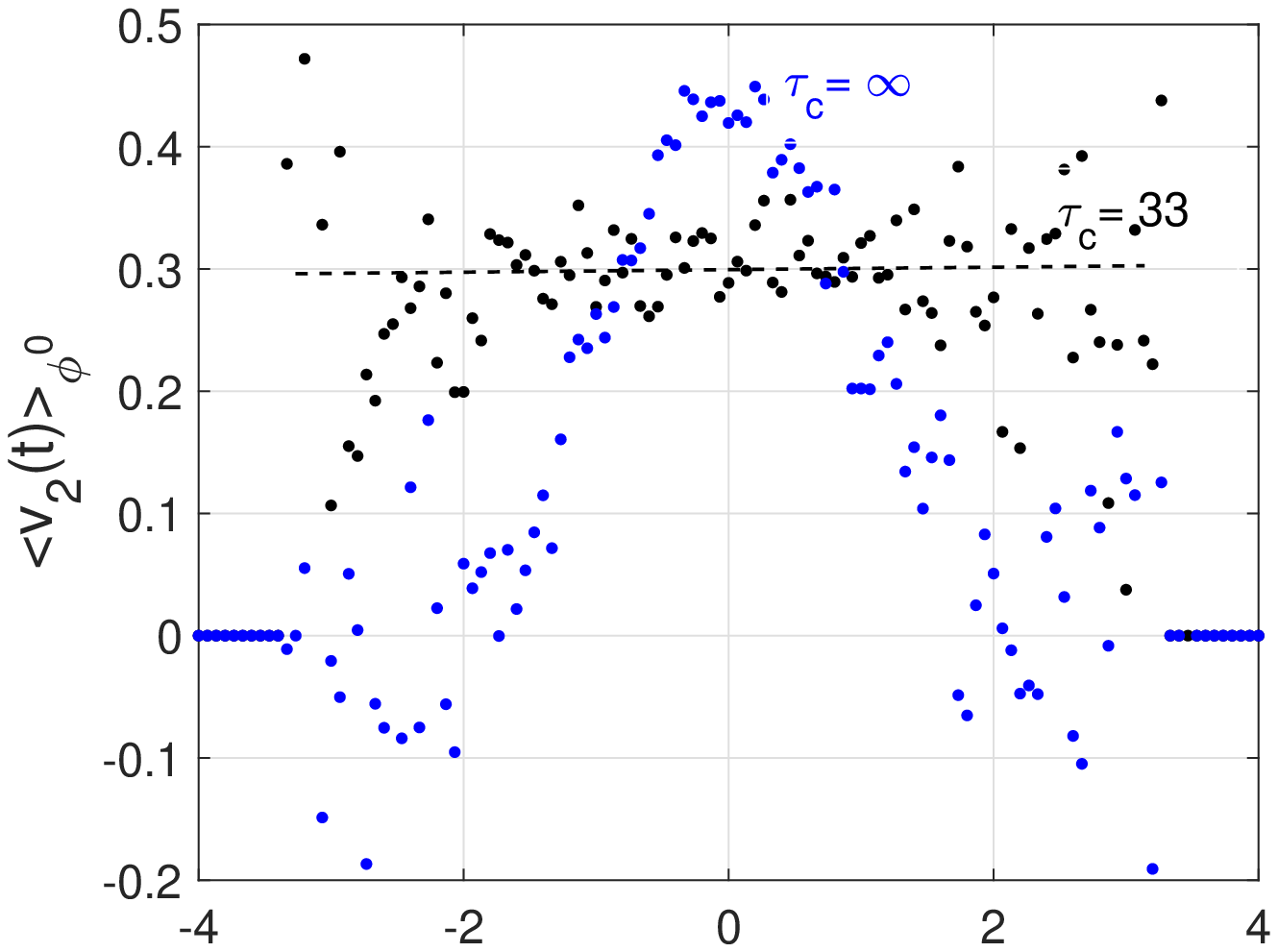}
}
\caption{Ordered parallel displacements $\left\langle x_{2}(t)\right\rangle
_{\protect\phi^{0}}$ (left panel) and average velocity $\left\langle
v_{2}(t)\right\rangle _{\protect\phi^{0}}$ (right panel) as functions of $%
\protect\phi^{0}$ at $t=200$ for a time-dependent potential with $\protect%
\tau_{c}=33$ (black points) compared to the static case (dashed blue lines)}
\label{taucymed}
\end{figure}

Ordered motion conditioned by the initial potential $\phi ^{0}$ was found in
the presence of an average velocity $V_{d}$ for the free trajectories. It is
represented by the average displacements $\left\langle x_{i}(t)\right\rangle
_{\phi ^{0},fr}$ that are conditioned by the initial potential and by the
category $c=fr$ [Eq. (\ref{medcondxy})]. These quantities obtained in a
time-dependent potential (with $\tau _{c}=33)$ are shown in Fig. (\ref%
{taucxmed}) for $x_{1}$ and in Fig. (\ref{taucymed}) for $x_{2}$, compared
to the static case. Significant differences appear for both directions.

\begin{figure}[tbh]
\centerline{
\includegraphics[height=5.5cm]{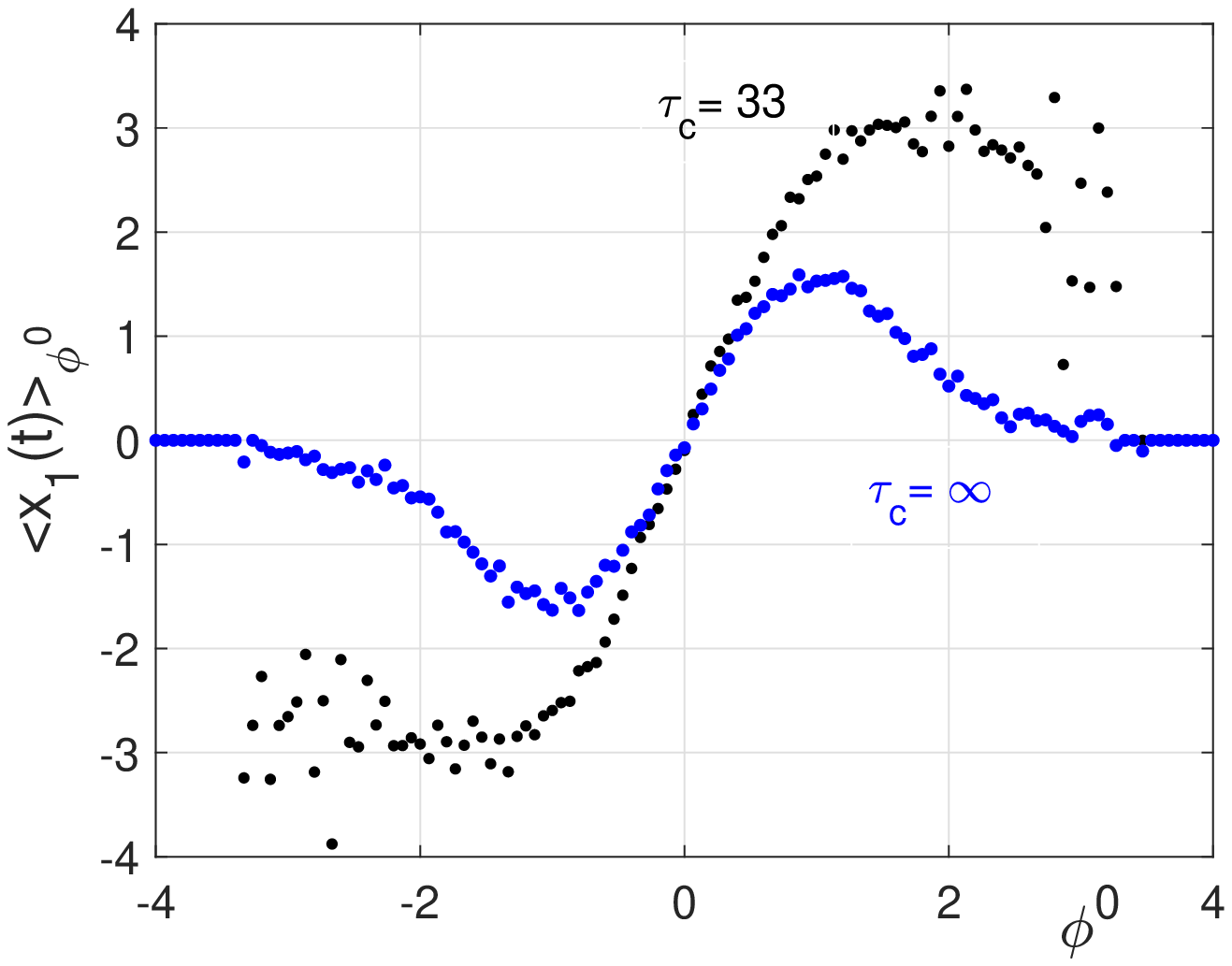}
\hspace{0.5cm}
\includegraphics[height=5.5cm]{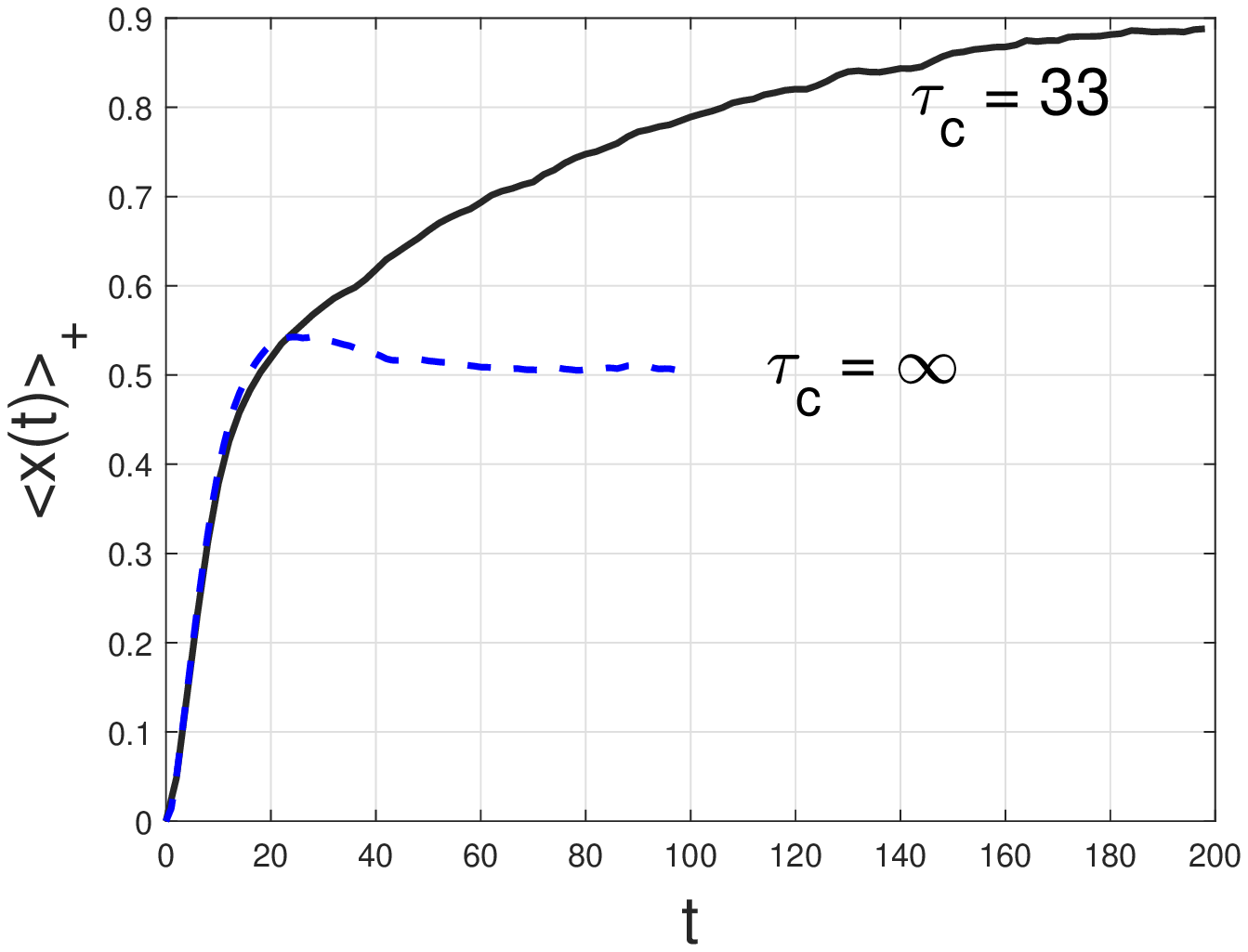}
}
\caption{Ordered perpendicular displacements $\left\langle
x_{1}(t)\right\rangle _{\protect\phi^{0}}$ as functions of $\protect\phi^{0}$
at $t=200$ (left panel) and $\left\langle x_{1}(t)\right\rangle _{+}$ as
function of time (right panel) for a time-dependent potential with $\protect%
\tau_{c}=33$ (black points) compared to the static case (dashed blue lines)}
\label{taucxmed}
\end{figure}

One can see in Fig. (\ref{taucxmed}, left panel) that the perpendicular
displacements $\left\langle x_{1}(t)\right\rangle _{\phi ^{0}}$ are larger
in time-depending potential, although the calculations are at a very large
time, $t=6\tau _{c}$ (where the EC of the potential (\ref{EC}) is
negligible, $E(t)=10^{-8}$). The main contribution comes from large values
of the potential $\left\vert \phi ^{0}\right\vert ,$ which is negligible in
the static potential. The amplitude of the ordered motion is represented by
the average displacements conditioned by the sign of $\phi ^{0},$ $%
\left\langle x_{1}(t)\right\rangle _{+}$\ and $\left\langle
x_{1}(t)\right\rangle _{-},$\ which determine by time derivative the HDs.
Surprisingly, the time evolution of these quantities shows a continuous
increase in time-dependent potential, while it saturates in the static case,
as seen in Fig. (\ref{taucxmed}, right panel) for $\left\langle
x_{1}(t)\right\rangle _{+}$. This means that the hidden drifts are
transitory in static potentials, but they are long-life statistical
quantities in time dependent potential. Their amplitude decays on a long
time scale, much longer than the decorrelation time of the potential.


The time variation of the potential modifies the parallel displacements $%
\left\langle x_{2}(t)\right\rangle _{\phi ^{0}}$\ by determining the
extension of the contribution on the whole range of $\phi ^{0}$ and the
dependence on these quantities of the initial potential (Fig. (\ref{taucymed}%
, left panel). It is peaked on $\phi ^{0}=0$ and has a weak (algebraic)
decay at large $\left\vert \phi ^{0}\right\vert ,$ instead of the
concentration on the domain of free trajectories with uniform average
displacement Eq. (\ref{medcondy}). However, the average Lagrangian velocity
is uniform on the whole range of $\phi ^{0}$ and it has the Eulerian value $%
V_{d},$ as seen in Fig. (\ref{taucymed}, right panel). The process of
concentration of the Lagrangian average velocity on the domain of free
trajectories found in static potentials is eliminated by the time-variation.

The fluctuations of the trajectories $\left\langle \delta
x_{i}^{2}(t)\right\rangle _{\phi ^{0}}$ and the transport $\left\langle
\delta x_{i}(t)\delta v_{i}(t)\right\rangle _{\phi ^{0}}$ conditioned by the
initial potential are all asymptotically uniform on the whole domain of $%
\phi ^{0}$. They reach this stage in a long time compared to the correlation
time of the potential ($t\gg \tau _{c}$) staring from the values
corresponding to static potential that are maintained at small time $t<\tau
_{c}$.

These results show that the trajectories are statistically identical with
the exception of the perpendicular average displacement $\left\langle
x_{i}(t)\right\rangle _{\phi ^{0}},$ which depend on $\phi ^{0}.$\ 

\subsection{7.2 Long memory}

The persistent Lagrangian order and the non-standard characteristics of the
trajectories in the time dependent case can be understood by analyzing the
statistics of the Lagrangian potential $\phi (t)\equiv \phi (\mathbf{x}%
(t),t).$

\begin{figure}[tbh]
\centerline{
\includegraphics[height=5.5cm]{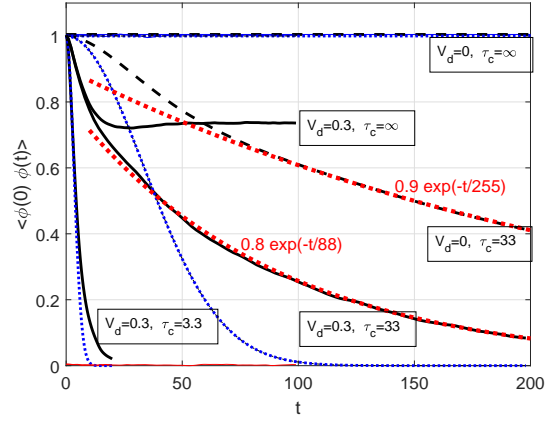}
}
\caption{Correlation of the Lagrangian potential for $V_d=0$ (dashed lines)
and $V_d=0.3$ (continuous lines) for static ($\protect\tau_c=\infty$), slow (%
$\protect\tau_c=33$) and fast time-variation ($\protect\tau_c=3.3$),
compared to the Eulerian correlations (dotted blue lines). Long tails with
exponential decay appear in time-dependent potentials with large $\protect%
\tau_c$. }
\label{CorfiL}
\end{figure}

\begin{figure}[tbh]
\centerline{
\includegraphics[height=5.5cm]{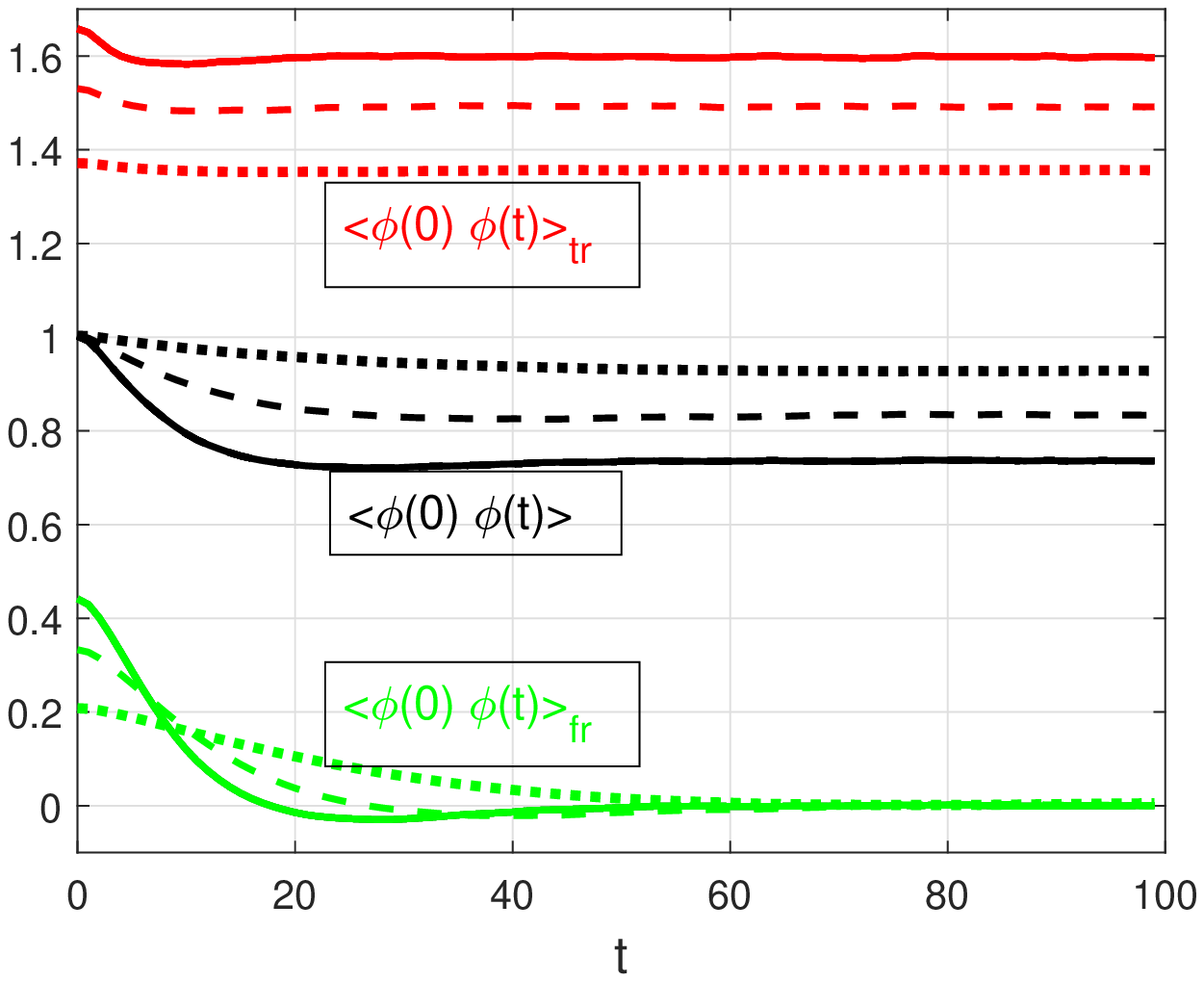}
\hspace{0.5cm}
\includegraphics[height=5.5cm]{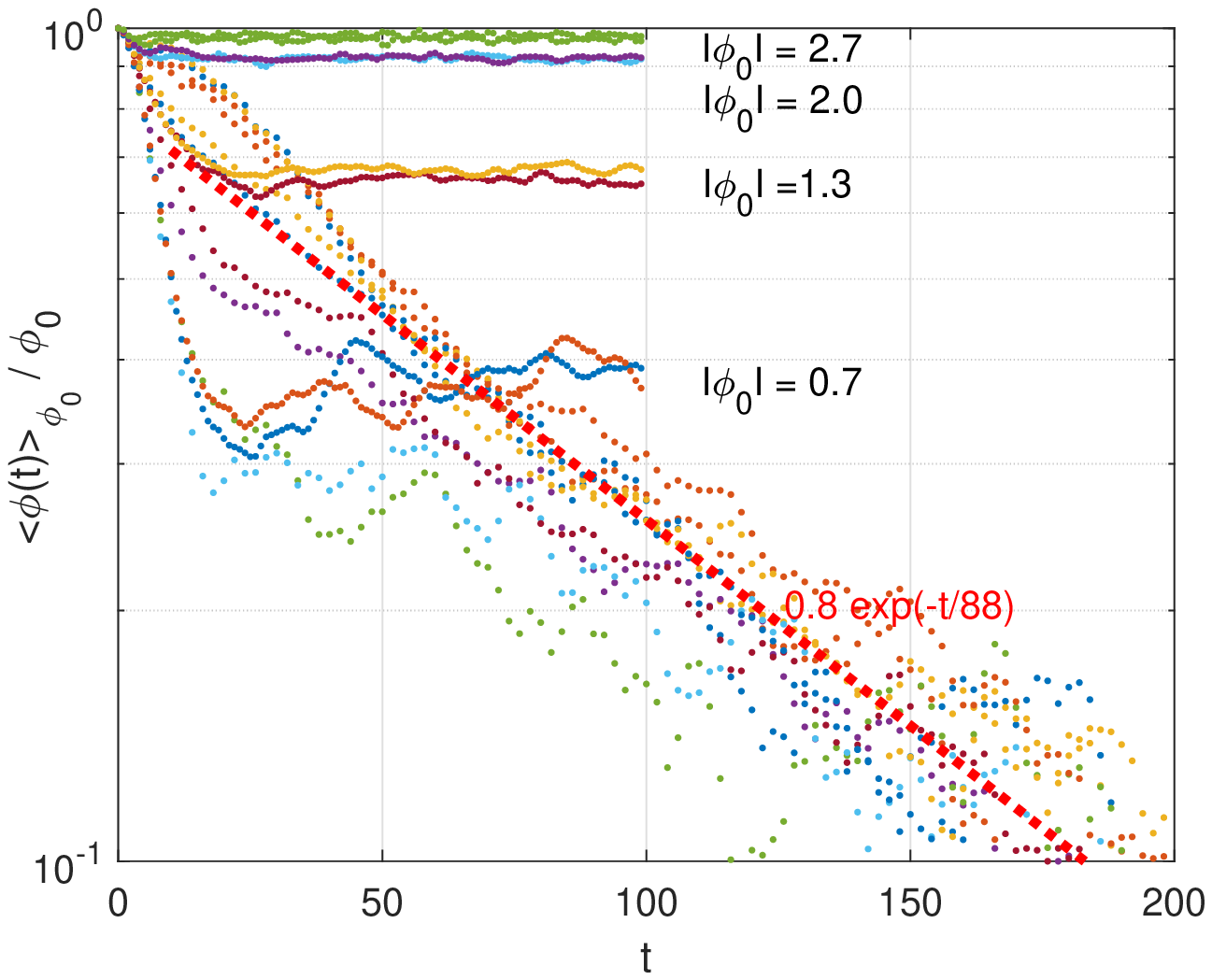}
}
\caption{Caracterization of the memory of the Lagrangian potential. Left:
The correlations conditioned by the category $<\protect\phi(0) \protect\phi%
(t)>_c$ for $V_d=0.1$ (dotted), $V_d=0.2$ (dashed), $V_d=0.3$ (continuous).
Right: The normalized average potential conditioned by the initial potential
for $V_d=0.3$, for $\protect\tau_c=\infty$ (continuous), $\protect\tau_c=33$
(dotted), and the values of $\protect\phi_0$ that label the curves. }
\label{condfil}
\end{figure}

The distribution of the Lagrangian potential has the same invariance
property as the Lagrangian velocity. It has the Gaussian probability of the
Eulerian potential at any time $t,$ for both the static and the
time-dependent cases, at any value of the average velocity $V_{d}.$\
However, significant differences appear between these cases concerning the
correlation and the average conditioned by the initial potential, as seen in
Figs. (\ref{CorfiL}) and \ref{condfil}.

The correlation of the Lagrangian potential $L_{\phi }(t)=\left\langle \phi
(0)~\phi (t)\right\rangle $ is far from the Eulerian time-correlation, $E(%
\mathbf{0},t).$\ Starting from the trivial case of static potential with $%
V_{d}=0,$\ where the invariance of the potential implies $L_{\phi }(t)=$\ $E(%
\mathbf{0})=1,$ in all the other cases shown in Fig. (\ref{CorfiL}), the
Lagrangian correlation is stronger than the Eulerian one, as it has a long
tail with much slower decay. It demonstrates that the Lagrangian potential
has a long-time memory.


The memory effect is strongest in almost static potentials (very large $\tau
_{c})$ with average velocity $V_{d}=0.$ The Lagrangian correlation decreases
much slower than the Eulerian one, and it is larger than than $E(\mathbf{0}%
,t)$ at any time (Fig. \ref{CorfiL}, the curve for $V_{d}=0,~\tau _{c}=33)$.
In this example, at $t=200\cong 6\tau _{c},$ $L_{\phi }$\ decreases only at $%
0.4,$ while $E(\mathbf{0},t)=1.5~10^{-8}.$ \ 

The average velocity ($V_{d}\neq 0)$ determines a faster decrease of $%
L_{\phi }(t)$ at small time that leads to smaller values compared to the
case $V_{d}=0$ (Fig. \ref{CorfiL}, the curve for $V_{d}=0.3,~\tau _{c}=33)$.
The decorrelation takes place on two time-scales. There is a fast decay at
small time that is followed by a slaw decrease of $L_{\phi }(t).$ The fast
decay is the same for $\tau _{c}=\infty $ and $\tau _{c}=33$ at $V_{d}=0.3$,
which shows that this process is not a consequence of the potential time
variation, but rather of the presence of $V_{d}.$

In the static case, the memory of the Lagrangian potential is infinite ($%
L_{\phi }(t)$ saturates). The asymptotic value is positive and it is a
decreasing function of $V_{d}$. The time-dependence of $L_{\phi }(t)$ is the
result of a selective decorrelation mechanism determined by the average
velocity. This process can be understood by analyzing the correlation of $%
\phi (t)$ conditioned by the category $c=tr,~fr$ in the static potential ($%
\tau _{c}=\infty ).$ As seen in Fig. \ref{condfil}, left panel, $%
\left\langle \phi (0)~\phi (t)\right\rangle _{c}$ decays to zero for the
free trajectories, while it saturates for the trapped trajectories at a
value that is comparable to the conditioned amplitude $\left\langle \phi
^{2}(0)\right\rangle _{tr}$. This demonstrates that in static potential the
decorrelation affects only the free trajectories and that the memory effect
is determined by the trapped trajectories, which approximately maintain the
initial potential $\phi (0)$. The asymptotic value of the average Lagrangian
potential is thus%
\begin{equation}
\left\langle \phi (0)~\phi (t)\right\rangle =\left\langle \phi (0)~\phi
(t)\right\rangle _{tr}n_{tr}=\phi ^{0}n_{tr}.  \label{ficorst}
\end{equation}

It is interesting to note that at finite $\tau _{c},$ the significant
decrease of the correction appears at any time, although it seems that the
process determined by $V_{d}$\ is transitory in static potentials.\ This is
caused by the interaction of the effects of $V_{d}$\ with the influence
produced by the time variation, which determine the two time-scale evolution
of $L_{\phi }(t).$ It is clearly evidenced by the average Lagrangian
potential conditioned by the initial value $\phi ^{0}$ normalized by this
value, $\left\langle \phi (t)\right\rangle _{\phi ^{0}}/\phi ^{0}$
represented in Fig. \ref{condfil}, right panel for the static case (lines)
and for $\tau _{c}=33$ (points)\ for several values of $\phi _{0}.$

An important property of the Lagrangian potential in slow time-dependent
potential is that its correlation and the average conditioned by $\phi ^{0}$
have the same time-decay (as seen in Fig. \ref{CorfiL} for the case $%
V_{d}=0.3,$ $\tau _{c}=33$ and in the right panel of Fig. \ref{condfil}, all
the curves have the same behavior $\exp \left( -t/88\right) $).

\bigskip

The long memory of the potential and the increase of the average
displacements $\left\langle x_{1}(t)\right\rangle _{\phi ^{0}}$ (Fig. \ref%
{taucxmed}) are the result of the same process. It consists of the
liberation of the trapped trajectories with large $\left\vert \phi
^{0}\right\vert $ followed by repeated stochastic events of capture and
release combined with the constraint of the total potential invariance (\ref%
{fitinv}) that approximately holds for small time intervals. Considering the
case of the peaks of the potential, the liberation of the trapped
trajectories with large $\phi ^{0}$ is produced when the time variation
determines the decrease of the potential to $\Delta $. The contour lines of
the potential that are open have the average perpendicular displacement $%
\Delta /V_{d}$ and the average potential along them equal to zero (as
imposed by Eq. (\ref{fitinv})). The stochastic recapture is uniformly
distributed over the potential and has the average perpendicular location $%
\Delta /V_{d}$. This cancels asymptotically the average of the potentials on
the trapping events and leads to the average $\Delta /V_{d}$ of the
positions of the trapping events. This happens on a time scale that is much
larger than $\tau _{c}$. These released trajectories with large $\phi ^{0}$
determine the slow decay of their initial average potential and the increase
of their average displacement from zero to the largest possible value $%
\Delta /V_{d}$. Thus, the memory of the Lagrangian potential and the
strengthening of the coherence of the trajectories are both determined by
the slow evolution toward uniform distribution of the trapping events on the
trajectories caused by the time-variation of the potential.

\section{8. Summary and conclusions}

A detailed study of the Lagrangian coherence of the trajectories in
2-dimensional incompressible velocity fields is presented. The strong order
that appear in this kind of velocity fields is determined by the Hamiltonian
structure of the advection equations (\ref{eqm}), (\ref{pot}) and by the
space correlation of the stochastic potential. The trajectories follow the
contour lines of the potential in the static case, and, for slowly varying
potentials, they remain close to them for long time intervals. This study is
focused on the identification and understanding of the order generated by an
average velocity $V_{d}$ superposed on the stochastic field. It determines
an average potential, which strongly modifies the structure of contour lines
of the total potential $\phi _{t}(\mathbf{x},t)=\phi (\mathbf{x},t)+x_{1}Vd$
\ by generation of a network of open lines between islands of closed lines.
As a result, two categories of trajectories are found in static (frozen)
potential: trapped (closed, periodic) and free (with unlimited displacement
in the parallel direction to $\mathbf{V}_{d}$). \ 

The results presented here are based on the numerical simulation of the
trajectories and on a complex statistical analysis that includes conditional
averages and correlations connected to the topology of the contour lines of
the potential $\phi _{t}.$ The statistics of displacements and of the
Lagrangian velocity are determined for the whole ensemble $R$, for the
categories trapped ($tr$) and free ($fr$), and also on sets of contour lines
of the potential conditioned by the value $\phi ^{0}$ at the starting point
of the trajectories. This analysis reveals the origin of coherence and
provides explanations for the nonstandard statistics, transport and memory
of this motion.

In the case of frozen potentials, we have found that the statistical
properties determined for the two categories $tr,$ $fr$ are completely
different compared to those obtained in the whole space $R$. The average
velocity $V_{d}$ generates coherence in the Lagrangian velocity, which
acquires average components from he stochastic ones for both categories. The
supplementary coherent velocity cancels the average velocity $V_{d}$ of the
trapped trajectories, and it determines larger velocity for the free
trajectories that compensate the missing contribution of the trapped ones ($%
\left\langle v_{2}(t)\right\rangle _{fr}=V_{d}/n_{fr}$). Thus, the
statistical invariance of the Lagrangian velocity (Lumley theorem) is
ensured in a rather non-trivial manner that involves hidden coherent
parallel velocity.

The statistical analysis conditioned by the initial potential $\phi ^{0}$
reveals additional important aspects. The free trajectories have Gaussian
distribution of $\phi ^{0}$ with a width $\Delta $ that is smaller than the
dispersion of the Eulerian potential. It shows the existence of ordered
perpendicular motion (average displacements across $\mathbf{V}_{d}$ (\ref%
{medcondx})) that appear for the free trajectories and are proportional with 
$\phi ^{0}.$ These averages $\left\langle x_{1}(t)\right\rangle _{\phi
^{0},fr}$ increase in time from zero and saturate at $\phi ^{0}/V_{d}.$\
They generate the hidden drifts, a pair of transitory average velocities
perpendicular on $\mathbf{V}_{d}$ conditioned by the sign of $\phi ^{0},$
which have opposite directions and exactly compensate each other. We have
also found that the Lagrangian statistics of the free trajectories
conditioned by $\phi ^{0}$ depends on the value $\phi ^{0}$ only through the
average perpendicular displacement. This means that the trajectories in the
category $fr$ for different values of $\phi ^{0}$ are statistically
identical, and are organized in strips with limited perpendicular extensions.

The probability of the displacements is non-Gaussian with separated
contributions of the categories: a steep peak for $tr$ and Gaussian
distribution that move along $\mathbf{V}_{d},$ but with larger velocity $%
V_{d}/n_{fr}$ for the $fr$ subensemble. The time-invariant Gaussian
distribution of the Lagrangian velocity is the sum of the non-Gaussian
contributions of the two categories, which are both non-Gaussian, but
invariant.

The transport is produced only by the free trajectories. A paradoxical
behavior was found: the statistics of the trajectories is strongly
non-Gaussian, but the transport is produced by Gaussian trajectories, which
in fact yield from the non-Gaussian velocity distribution of the free
trajectories. The transport is anomalous, subdiffusive across $\mathbf{V}%
_{d} $ and superdiffusive of ballistic type along $\mathbf{V}_{d}.$ The
latter results from the ordered parallel motion. category.

The free trajectories can be associated to a geometrical locus on the
two-dimensional space $\left( x_{1},x_{2}\right) $, in the sense that each
point is the initial condition for such trajectory and all trajectories are
confined in this domain $fr$. The complement of $fr$ is the geometric locus
of the trapped trajectories $tr$, which is composed of the islands of closed
contour lines of the potentials. The (Eulerian) statistical characteristics
of each geometrical locus were identified.

The time-dependence of the stochastic potential produces anomalous increase
of the Lagrangian coherence, instead of the expected decay after the
decorrelation time. In particular, the perpendicular average displacements
conditioned by $\phi ^{0}$ significantly increase and the transitory hidden
drifts become long-life structure that survive at $t\gg \tau _{c}.$ The
enhanced coherence is found to be associated to a long memory of the
Lagrangian potential. Also,\ the trajectories conditioned by the initial
potential become statistically identical for all values of $\phi ^{0},$ not
only on the domain of small potential with width $\Delta ,$ as in frozen
potentials. These effects are caused by the stochastic liberation by the
time-variation of the potential of the trajectories that initially are
trapped, followed by repeated stochastic captures that are constraint by of
the approximate invariance of the total potential.

\bigskip

\end{document}